\documentclass[11pt,a4paper]{article}
\pdfoutput=1

\usepackage{amsfonts}
\usepackage{subfig}
\usepackage{jheppub}
\usepackage{empheq}

\newcommand{\e}{\mathrm{e}}
\newcommand{\ud}{\,\mathrm{d}}
\newcommand{\udel}{\,\mathrm{\delta}}
\newcommand{\gz}{\tilde{g}(z)}
\newcommand{\pdz}{\partial_z}
\newcommand{\zt}{\tilde{z}}
\newcommand{\pdzt}{\partial_{\tilde{z}}}
\DeclareMathOperator{\im}{Im}
\DeclareMathOperator{\re}{Re}

\title{The Sound of Topology in the AdS/CFT Correspondence}

\author{Igal Arav}
\author{and Yaron Oz}

\affiliation{Raymond and Beverly Sackler School of Physics and Astronomy,\\
Tel-Aviv University, Tel-Aviv 69978, Israel}

\emailAdd{aravigal@post.tau.ac.il}
\emailAdd{yaronoz@post.tau.ac.il}

\abstract{
Using the gauge/gravity correspondence, we study the properties of 2-point correlation functions
of finite-temperature strongly coupled gauge field theories, defined on a curved space of general spatial topology with a dual
black hole description. We derive approximate asymptotic expressions for the correlation
functions and their poles, supported by exact numerical calculations, and study their dependence on
the dimension of spacetime and the spatial topology. The asymptotic structure of the correlation functions depends
on the relation between the spatial curvature and the temperature, and
is noticeable when they are of the same order.
In the case of a hyperbolic topology, a specific temperature is identified for which exact
analytical solutions exist for all types of perturbations. The asymptotic structure of the correlation
functions poles is found to behave in a non-smooth manner when approaching this temperature.
}


\arxivnumber{}

\begin{document}

\maketitle

\section{Introduction}
\label{sec:intro}

The aim of this paper is to analyze the properties of 2-point correlation functions
of finite-temperature strongly coupled (deconfined) gauge field theories defined on spacetimes with general spatial topology,
and admit a dual black hole description via the AdS/CFT correspondence.
Of specific interest is the dependence of these
properties on the topology of the space on which the gauge theory is defined, and its temperature.
We will see that the topology is indeed encoded
in the properties of the correlation function and its poles, corresponding to quasinormal modes in the
dual bulk spacetime: the sound of topology. In
particular, the asymptotic ``slope'' of the poles encodes information about the spatial
topology and the temperature, but is ``universal'' otherwise, i.e. it doesn't depend on the type of operators
considered.

One motivation for this work is the fact that important aspects of the strong coupling dynamics of
gauge field theories can depend crucially on the topology of space-time.
For instance, the confinement-deconfinement phase transition is a property of strongly coupled thermal conformal gauge theories defined on a space-time with
spherical spatial topology. The theory is always in the deconfined phase in the planar and hyperbolic cases. 
Thus, one may hope that insights 
to the  confinement-deconfinement phase transition mechanism can be gained by analyzing
the strong coupling dynamics on different  spatial topologies. This can be done 
by using the gravitational dual description.
A second motivation is the recently studied correspondence between gravitational dynamics and strongly coupled condensed matter systems. One can naturally envision in and out of equilibrium condensed matter systems realized on different spatial
topologies, where our study can be of much relevance.
A third motivation, perhaps a more remote one, is the relevance of the interplay between topology and field theory dynamics to
astrophysical/cosmological setups.

We distinguish three classes of spatial topology \footnote{In the spherical case, a Hawking-Page phase transition
exists, which is interpreted in terms of the AdS/CFT correspondence as a transition between a higher
temperature deconfined phase and a lower temperature confined phase (see~\cite{Witten:1998qj},
~\cite{Witten:1998zw}). Phase transitions might exist also for the flat and hyperbolic cases, e.g. the
transition between the black hole solution at a higher temperature and the Horowitz-Myers ``AdS
soliton'' at a lower temperature (see \cite{Surya:2001vj}). }: The spherical,
the flat and
the hyperbolic (The relevant thermodynamic quantities and their dependence on the topology are given in
Appendix~\ref{app:thermodynamicquant}).
Consider a gauge theory
defined on a $(d-1)$-dimensional spacetime with the topology of
$ \mathbb{R}\times\Omega_{d-2}^{FT} $,
where $ \Omega_{d-2}^{FT} $  is a constant curvature $ (d-2) $-dimensional Riemannian manifold of
arbitrary topology\footnote{We assume that $ d\ge 4 $.}.
The two basic scales that characterize the theory are the temperature $ T $ and the scalar curvature
of the $ \Omega_{d-2}^{FT} $ manifold $ R_\Omega^{FT} $.
In the analysis we will consider two different regimes:
\begin{enumerate}
\item The regime of $ \frac{\sqrt{|R_\Omega^{FT}|}}{T} \to 0 $: In this case the curvature of the
$ \Omega_{d-2}^{FT} $ manifold may essentially be neglected, so we expect all physical
quantities to behave as they do in the case of a flat ($ R_\Omega^{FT} = 0 $) space. Therefore,
in this limit one does not distinguish between different topologies (aside from the known changes
of the Laplace operator spectra between different topologies). This limit includes
the hydrodynamic limit\footnote{The hydrodynamic condition $ \frac{L_s^{FT}}{T} \to 0$
requires that $ \frac{\sqrt{|R_\Omega^{FT}|}}{T} \to 0 $ since the minimal eigenvalue
$ L_{s,min}^{FT} $ is of the same order as $ \sqrt{|R_\Omega^{FT}|} $.}:
$ \frac{\omega}{T},\frac{L_s^{FT}}{T} \to 0 $, where $ \omega $ denotes the frequency and  $ L_s^{FT} $ is the spatial Laplace
operator eigenvalue.
\item The regime of $ \frac{\sqrt{|R_\Omega^{FT}|}}{T} = \text{fixed} $: In this case it will be
possible to distinguish between the different topologies. In particular we will study the
limit of $ \frac{L_s^{FT}}{T} = \text{fixed} $ and $ \frac{\omega}{T} \gg 1 $, where analytical
asymptotic expressions will be derived for the quasinormal modes and the 2-point correlation functions.

\end{enumerate}
Note that the $ \frac{\sqrt{|R_\Omega^{FT}|}}{T} \gg 1 $ regime does not exist in the black hole 
description in the case of spherical topology (since the temperature has a finite minimal value) - it
corresponds to a different (confining) phase on the field theory side. Nor does it exist in the case of 
flat topology (for which $ R_\Omega^{FT} = 0 $).
While this regime does exist for the hyperbolic case, there is nothing separating it qualitatively 
from the $ \frac{\sqrt{|R_\Omega^{FT}|}}{T} = \text{fixed} $ regime in the context of this paper.
For these reasons, we won't consider this regime separately here.

The main new results of this work are (see Appendix~\ref{app:notations} for a list of notations
and definitions):
\begin{itemize}
\item In the asymptotic limit, where $ \frac{L_s^{FT}}{T} \sim \text{fixed} $ and
$ \frac{\omega}{T} \gg 1 $, the following holds:
\begin{itemize}
\item For all types of topologies and all the temperatures , the quasinormal
frequencies (and therefore the poles of the field theory correlators) are of the form
\begin{equation}
\lambda_n = \lambda_0 + n\Delta\lambda \ ,
\end{equation}
where $ \Delta\lambda $ depends on the topology and temperature.
\item In this limit we derive asymptotic expressions for the field theory
2-point retarded correlation functions of the operators dual to the different perturbation
types. These are given
in Equations~\ref{eq:asymptcftcorrelatorscalarevend} and~\ref{eq:asymptcftcorrelatorscalaroddd} for the
scalar perturbation modes,
in Equations~\ref{eq:asymptcftcorrelatorlongvectorevend} and~\ref{eq:asymptcftcorrelatorlongvectoroddd}
for the longitudinal vector perturbation modes,
and in Equations~\ref{eq:asymptcftcorrelatortransvectorevend}
and~\ref{eq:asymptcftcorrelatortransvectoroddd} for the transverse vector perturbation modes.

\end{itemize}

\item In the hyperbolic case there is
a specific temperature $ T_c $, where the bulk ``black hole'' solution has no singularity, and is isometric to
AdS. In this special case the following holds:
\begin{itemize}
\item At $ T=T_c $ we derive exact expressions for the field theory 2-point retarded correlation
functions of the operators dual to the different perturbation types. These are given
in Equations~\ref{eq:scalarexactcorrelatorintceven} and~\ref{eq:scalarexactcorrelatorintcodd}
for the scalar perturbation modes,
in Equations~\ref{eq:longvectorexactcorrelatorintceven} and~\ref{eq:longvectorexactcorrelatorintcodd}
for the longitudinal vector perturbation modes
and in Equations~\ref{eq:transvectorexactcorrelatorintceven}
and~\ref{eq:transvectorexactcorrelatorintcodd}
for the transverse vector perturbation modes.

\item For $ T>T_c $, the asymptotic $ \Delta\lambda $ has a real component $ \re(\Delta\lambda)>0 $.
As $ T\to T_c $, $ \Delta\lambda $ becomes imaginary according to
Equation~\ref{eq:asymptparamsaroundtcde4b} for $ d=4 $ (as previously calculated in \cite{Koutsoumbas:2006xj}),
and according to Equation~\ref{eq:asymptparamsaroundtcdg4b} for $ d>4 $.
\end{itemize}

\end{itemize}

The paper is organized as follows.
In Section~\ref{sec:quasinormalmodes}, we will review some known results
regarding quasinormal modes (formulated in our notation).
We will calculate the QNM equations for massless scalar and
vector perturbations, the exact solutions for the ``special'' case mentioned above (the hyperbolic case
with the specific temperature $ T_c $), the approximate quasinormal frequencies in the asymptotic limit
($ \frac{L_s^{FT}}{T} = \text{fixed} $ and $ \frac{\omega}{T} \gg 1 $) and the diffusion mode in the
hydrodynamic limit. We will discuss the dependence of the asymptotic quasinormal modes on the topology
and the temperature, in particular in the hyperbolic case around the temperature $ T_c $, and demonstrate
it using some numerical results.
In Section~\ref{sec:cftcorrelators}, we will use AdS/CFT dictionary to calculate the 2-point correlation
functions of the dual field theory. We will obtain exact expressions for the hyperbolic case with
$ T=T_c $ and approximate expressions in the asymptotic limit for general topology and temperature.
We will compare the asymptotic expressions with exact numerical solutions.
Section~\ref{sec:discussion} is devoted to a discussion. Details of calculations are outlined in the appendices.

\section{Quasi-Normal Modes}
\label{sec:quasinormalmodes}

\subsection{QNM Definition}
\label{subsec:qnmdefinition}

Quasinormal modes of black holes or black branes are defined as the late-time oscillation modes of the
black hole metric and the fields coupled to it, satisfying certain boundary conditions. Put
differently, these are the eigenmodes of the linearized equations of motion over the black hole
background. The boundary conditions are specified at the black hole horizon and at spatial infinity. The boundary condition
at the horizon is chosen so that the solution corresponds to  a wave ingoing into the horizon, while
the boundary condition choice at spatial infinity depends on the asymptotic nature of the spacetime.

Choosing the incoming wave solution gives the boundary condition:
\begin{equation}
\left.\psi\right|_{z=0} \sim z^{-\frac{i\omega}{C}} \ ,
\end{equation}
where $ z=0 $ corresponds to the BH horizon and $ z=1 $ corresponds to spatial infinity.
As for the boundary condition at spatial infinity, different boundary conditions have been
investigated for different asymptotic spacetime geometries. In the context of AdS/CFT, the relevant
boundary condition is the one that corresponds to the poles of the retarded correlator of the
CFT operators dual to the investigated field. For the fields discussed here, this condition
amounts to the Dirichlet boundary condition (see~\cite{Son:2002sd},~\cite{Berti:2009kk}
and~\cite{Nunez:2003eq}):
\begin{equation}
\left.\psi\right|_{z=1}=0 \ .
\end{equation}

\subsection{QNM Equations}
\label{subsec:qnmequations}

We consider the d-dimensional black hole metric given by a metric of the form:
\begin{equation}
\label{eq:rmetric}
\ud s_{bulk}^2 = -f(r)\ud t^2 + \frac{1}{f(r)} \ud r^2 + r^2 \ud\Omega_{d-2}^2 \ ,
\end{equation}
where
$
\ud\Omega_{d-2}^2 = \sum_{i,j} (g_\Omega)_{ij} \ud x^i \ud x^j
$
is the inner metric of a constant curvature, (d-2)-dimensions Riemannian manifold with of same
topology as the $ \Omega_{d-2}^{FT} $ manifold.

From the vacuum Einstein equation with a cosmological constant
one can deduce the form of $ f(r) $ to be
\begin{equation}
f(r) = k + \frac{r^2}{R^2} - \frac{r_0^{d-3}}{r^{d-3}} \ ,
\end{equation}
where $ R $ is related to the cosmological constant by
$
R^2 = -\frac{(d-2)(d-1)}{2\Lambda}
$
and $ k $ is related to the scalar curvature $ R_\Omega $ of the manifold $ \Omega_{d-2} $
by
(see, for example, \cite{Birmingham:1998nr})
$
k = \frac{R_\Omega}{(d-2)(d-3)}
$.
In terms of these coordinates, the horizon of the black hole is located at
$ r=r_+ $ where $ f(r_+)=0 $, while the boundary is located at $ r\to\infty $.

After making the transformation to the coordinate $ z = 1-\frac{r_+}{r} $,
the metric takes the form:
\begin{equation}
\label{eq:zmetric}
\ud s^2 = -\frac{\rho^2 \tilde{g}(z)}{(1-z)^2} \ud t^2 + \frac{r_+^2}{\rho^2\tilde{g}(z)(1-z)^2}\ud z^2 + \frac{r_+^2}{(1-z)^2}\ud \Omega_{d-2}^2 \ ,
\end{equation}
where $ \rho \equiv \frac{r_+}{R} $, $ K \equiv \frac{k}{\rho^2} $ and
\begin{equation}
\label{eq:gzexp}
\tilde{g}(z) \equiv 1 + K(1-z)^2 - (1+K)(1-z)^{d-1} \ .
\end{equation}
In terms of the new coordinate $ z $, the horizon is located at $ z=0 $ while the boundary is located at $ z=1 $.

Let us assume that the spectrum of the scalar and vector Laplace operator $ \Delta_\Omega $\footnote{In
the context of this work we shall define the Laplace operator as: $ \Delta \equiv \udel\ud $ where
$ \ud $ and $ \udel $ are the exterior derivative and codifferential operators respectively.}
defined on the manifold $ \Omega_{d-2} $ is given by
\begin{align}
\Delta_\Omega H_{L^2}(x) & = L_s^2 H_{L^2}(x) \\
\Delta_\Omega \boldsymbol{A}_{L^2}(x) & = L_v^2 \boldsymbol{A}_{L^2}(x)
\end{align}
respectively.
In order for a QNM  corresponding to the frequency $ \omega $ and the above Laplace operator eigenvalue to exist, the corresponding boundary conditions problem should have a non-trivial solution. Introducing the dimensionless parameters
$
\lambda \equiv \frac{\omega r_+}{\rho^2}
$, $
q_s \equiv \frac{L_s}{\rho}
$, $
q_v \equiv \frac{L_v}{\rho}
$ and $
C \equiv \tilde{g}'(0)
$,
we find the following equations (see Appendix~\ref{app:qnmequationsderivation} for a detailed
derivation):
\begin{itemize}
\item For a (massless, minimally coupled) scalar QNM, the equation is
\begin{equation}
\label{eq:scalareq}
(1-z)^{d-2}\partial_z\left[ \frac{\tilde{g}(z)}{(1-z)^{d-2}} \partial_z \psi \right] +
\left[ \frac{\lambda^2}{\tilde{g}(z)} - q_s^2 \right] \psi = 0 \ ,
\end{equation}
along with the boundary conditions:
\begin{equation}
\label{eq:scalarbc}
\left.\psi\right|_{z=0} \sim z^{-\frac{i\lambda}{C}}
\qquad
\left.\psi\right|_{z=1} = 0 \ .
\end{equation}
\item For a ``longitudinal'' vector QNM, the equation is
\begin{equation}
\label{eq:longvectoreq}
\partial_z \left[ \tilde{g}(z)(1-z)^{d-4} \partial_z \left( \frac{1}{(1-z)^{d-4}} \psi \right) \right] + \left[ \frac{\lambda^2}{\tilde{g}(z)} - q_s^2 \right] \psi = 0 \ ,
\end{equation}
along with the boundary conditions:
\begin{multline}
\label{eq:longvectorbc}
\left.\psi\right|_{z=0} \sim z^{-\frac{i\lambda}{C}}\\
\left. (1-z)^{d-4}\partial_z\left[ \frac{1}{(1-z)^{d-4}} \psi \right] \right|_{z=1} = \left.\left( \partial_z \psi + \frac{d-4}{1-z}\psi \right)\right|_{z=1} = 0 \ ,
\end{multline}
or \emph{equivalently} the equation
\begin{multline}
\label{eq:longvectoreq2}
(1-z)^{d-4}\partial_z\left[ \frac{\tilde{g}(z)}{(1-z)^{d-4}} \partial_z \psi \right]\\
+ \frac{q_s^2}{\lambda^2-q_s^2 \tilde{g}(z)} \tilde{g}(z)\partial_z\tilde{g}(z) \partial_z\psi + \left[ \frac{\lambda^2}{\tilde{g}(z)} - q_s^2 \right] \psi =0 \ ,
\end{multline}
along with the boundary conditions:
\begin{equation}
\label{eq:longvectorbc2}
\left.\psi\right|_{z=0} \sim z^{-\frac{i\lambda}{C}}
\qquad
\left.\psi\right|_{z=1} = 0 \ .
\end{equation}
\item For a ``transverse'' vector QNM, the equation is
\begin{equation}
\label{eq:transvectoreq}
(1-z)^{d-4}\partial_z\left[ \frac{\tilde{g}(z)}{(1-z)^{d-4}} \partial_z \psi \right] +
\left[ \frac{\lambda^2}{\tilde{g}(z)} - q_v^2 \right] \psi = 0 \ ,
\end{equation}
along with the boundary conditions:
\begin{equation}
\label{eq:transvectorbc}
\left.\psi\right|_{z=0} \sim z^{-\frac{i\lambda}{C}}
\qquad
\left.\psi\right|_{z=1} = 0 \ .
\end{equation}
\end{itemize}

It can be easily seen from the above equations that if $ \psi(z) $ is as a solution for
any of the equations (satisfying the corresponding boundary conditions) with frequency
$ \lambda $, then $ \psi^*(z) $  is a solution for the same equation (with the same
boundary conditions) with frequency $ -\lambda^* $. This means that the QNM frequencies
are always symmetric with respect to the imaginary axis.

A detailed derivation of the QNM Equations can be found in Appendix~\ref{app:qnmequationsderivation}.

\subsection{Exact solutions for the \texorpdfstring{$K=-1$}{K=-1} case}
\label{subsec:qnmexactsolutionsfortc}

In the case where the $ \Omega_{d-2} $ manifold is hyperbolic, we consider the special state
corresponding to the temperature:
\begin{equation}
T_C \equiv \frac{\sqrt{|k|}}{2\pi R} = \frac{1}{2\pi}\sqrt{\frac{|R_\Omega^{FT}|}{(d-2)(d-3)}} \ ,
\end{equation}
for which $ K=-1 $, and the horizon radius is given by $ \rho_C = \sqrt{|k|} $.
In this case, the QNM equations
(Equations~\ref{eq:scalareq}, \ref{eq:longvectoreq} and~\ref{eq:transvectoreq}) can be solved
analytically. This has been done in~\cite{Birmingham:2006zx},~\cite{Aros:2002te}
and~\cite{LopezOrtega:2007vu}. Here we summarize these solutions in our notation.

Making the transformation:
\begin{align}
w &\equiv\gz = 1-(1-z)^2 \\
\psi &\equiv w^\gamma (1-w)^\delta \phi
\end{align}
(where $ \gamma $ and $ \delta $ are defined in Appendix~\ref{app:exactsolutionsfortcderivation}
for each perturbation type), one obtains the hypergeometric equation. The solution to the
equation satisfying the incoming-wave boundary condition at the horizon is
$\sb{2}F_1\left(a,b;c;w\right)$, where $ a $, $ b $ and $ c $ are given in
Appendix~\ref{app:exactsolutionsfortcderivation} for each case.

Applying the QNM boundary condition at the AdS boundary we get for the scalar and transverse vector
cases:
\begin{equation}
\sb{2}F_1\left(a,b;c;1\right)=\frac{\Gamma(c)\Gamma(c-a-b)}{\Gamma(c-a)\Gamma(c-b)}=0 \ ,
\end{equation}
and for the longitudinal vector case (and using known formula fot the hypergeometric
function, taking into account $ c-a-b = \frac{d-5}{2} $) we get:
\begin{multline}
\left.(1-w)^{\frac{d-3}{2}}\partial_w\left[\frac{1}{(1-w)^\frac{d-5}
{2}}\sb{2}F_1(a,b;c;w)\right]\right|_{w=1} \\
= \frac{(c-a)(c-b)}{c}\sb{2}F_1(a,b;c+1;1) \\
= \frac{(c-a)(c-b)}{c} \frac{\Gamma(c+1)\Gamma(c-a-b+1)}{\Gamma(c-a+1)\Gamma(c-b+1)}
= 0  \ .
\end{multline}
The solution for which is given by $c-a=-n$ or $c-b=-n$ where $n\in\mathbb{Z}$ and $n\geq0$,
so that the quasi-normal frequencies are:
\begin{itemize}
\item For the massless scalar case:
\begin{equation}
\lambda_n = -\frac{i}{2}(d+1) \pm \sqrt{q^2-\frac{1}{4}(d-3)^2} -2ni \ .
\end{equation}
\item For the longitudinal vector case:
\begin{equation}
\lambda_n=-\frac{i}{2}(d-3)\pm\sqrt{q^2-\frac{1}{4}(d-3)^2}-2ni \ .
\end{equation}
\item For the transverse vector case:
\begin{equation}
\lambda_n = -\frac{i}{2}(d-1) \pm \sqrt{q^2-\frac{1}{4}(d-5)^2} -2ni \ .
\end{equation}
\end{itemize}

\subsection{QNM Asymptotics (for \texorpdfstring{$ K>-1 $}{K>-1})}
\label{subsec:qnmasymoptotics}

For the case $ K>-1 $, one may find an approximate analytical expression for the $n$-th QNM frequency
for $ n\to\infty $ - that is, in the asymptotic limit of $ \frac{L_s^{FT}}{T} = \text{fixed} $ and
$ \frac{\omega}{T} \gg 1 $. This has been done in~\cite{Natario:2004jd} using the ``monodromy'' method.
Here we summarize this calculation using our notations.
The general stages of the calculation are as follows:
\begin{enumerate}

\item We bring each of the QNM equations to a Schr\"odinger form, with some effective potentials,
using appropriate coordinate and function transformations.
\item We find the poles of these effective potentials on the complex plane, and write approximate forms
for the equations in the vicinity of the poles.
\item We solve the approximate equations around each pole using Bessel functions.
\item Using the asymptotic limit of $ |\lambda| \gg 1 $, we replace the Bessel functions with their
asymptotic forms.
\item We match the solutions on each region of the complex plane, and apply the appropriate boundary
conditions, to get equations for values of $ \lambda $ that allow for non-trivial solutions.
\item We solve the equations analytically and obtain an asymptotic expression of the form:
\begin{equation}
\lambda_n = \lambda_0 + n\Delta\lambda \ .
\end{equation}

\end{enumerate}

\subsubsection{Calculation of Asymptotic QNMs}

Starting from the QNM Equations, we perform the following transformation:
\begin{equation}
\psi(z) = (1-z)^{\frac{\alpha}{2}} \phi(z)
\end{equation}
\begin{equation}
\zt \equiv \int \frac{1}{\gz} \ud z = \sum_{k=1}^{d-1} \gamma_k \ln(z-z_k) + \zt_0 \ ,
\end{equation}
where:
\begin{equation}
\alpha=
\begin{cases}
d-2 & \text{for scalar},\\
d-4 & \text{for vector}
\end{cases} \ ,
\end{equation}
$ z_k $ ($k=1,\ldots, d-1 $) are the zeros of the polynomial $ \gz $ and\footnote{The branch cuts are chosen here so that $ 0\leq\arg(z)<2\pi $.}
\begin{equation}
\gamma_k = \lim_{z\to z_k} \frac{z-z_k}{\gz} = \frac{1}{\tilde{g}'(z_k)}
\end{equation}
\begin{equation}
\zt_0 = - \sum_{k=1}^{d-1} \gamma_k \ln(1-z_k)
\end{equation}
(so that $ \zt(z=1) = 0 $).
After this transformation the equations take the Schr\"odinger form:
\begin{equation}
\partial_{\zt}^2\phi + \left[ \lambda^2-V(\zt) \right]\phi = 0 \ ,
\end{equation}
where the effective potential $ V(z) $ is given for each case in Appendix~\ref{app:effectivepotentials}.

Next we turn to the calculation of the asymptotic QNMs using the monodromy method. We first note that
the asymptotic QNM frequencies satisfy (see~\cite{Cardoso:2004up} and~\cite{Natario:2004jd})
\begin{equation}
\im(\lambda\zt_0)=0 \ .
\end{equation}

\begin{figure}[htbp]
\begin{center}
\subfloat[$z$ plane contour]{%
\label{fig:zcontour}\includegraphics[width=8cm]{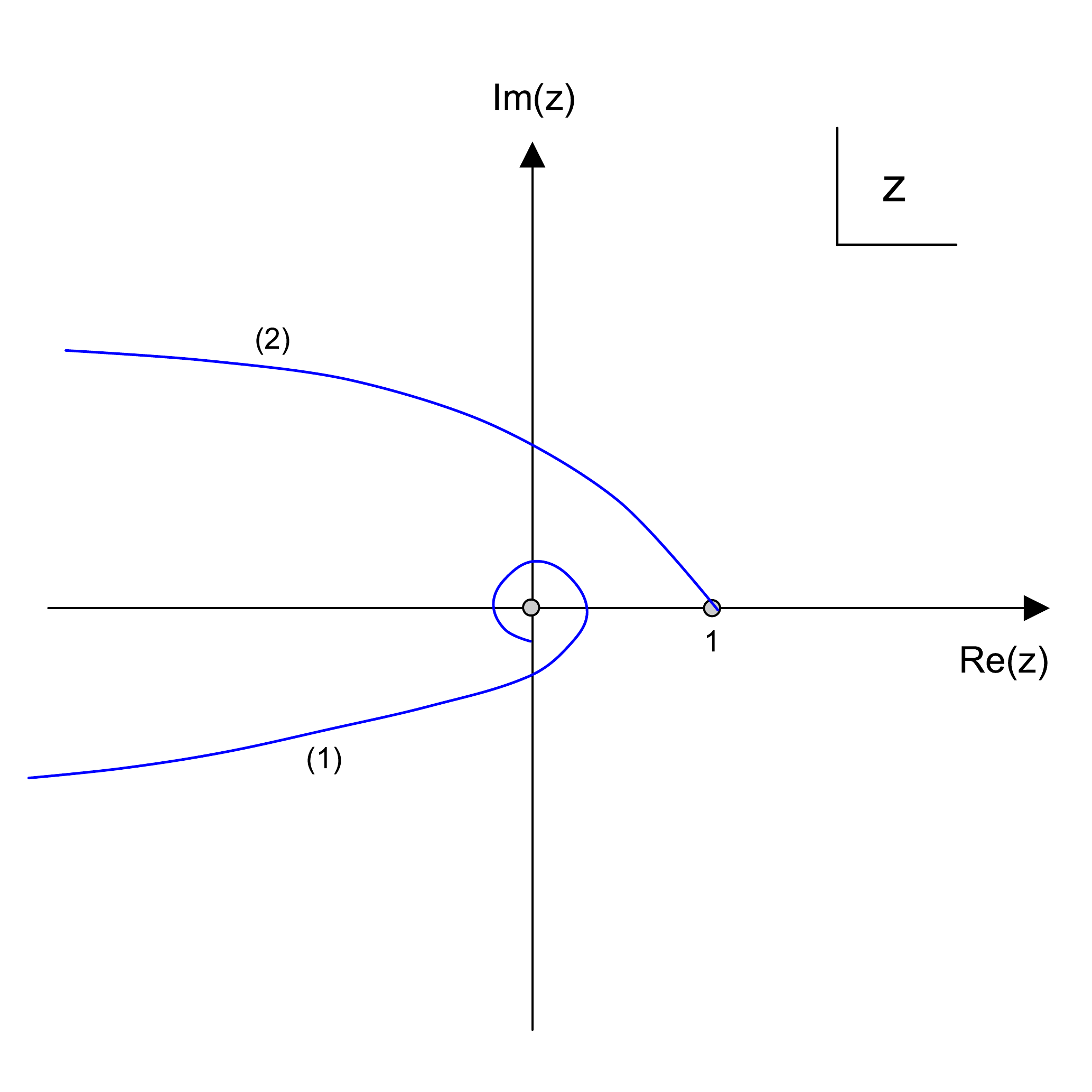}}
\hspace{1cm}
\subfloat[$\zt$ plane contour]{%
\label{fig:ztcontour}\includegraphics[width=8cm]{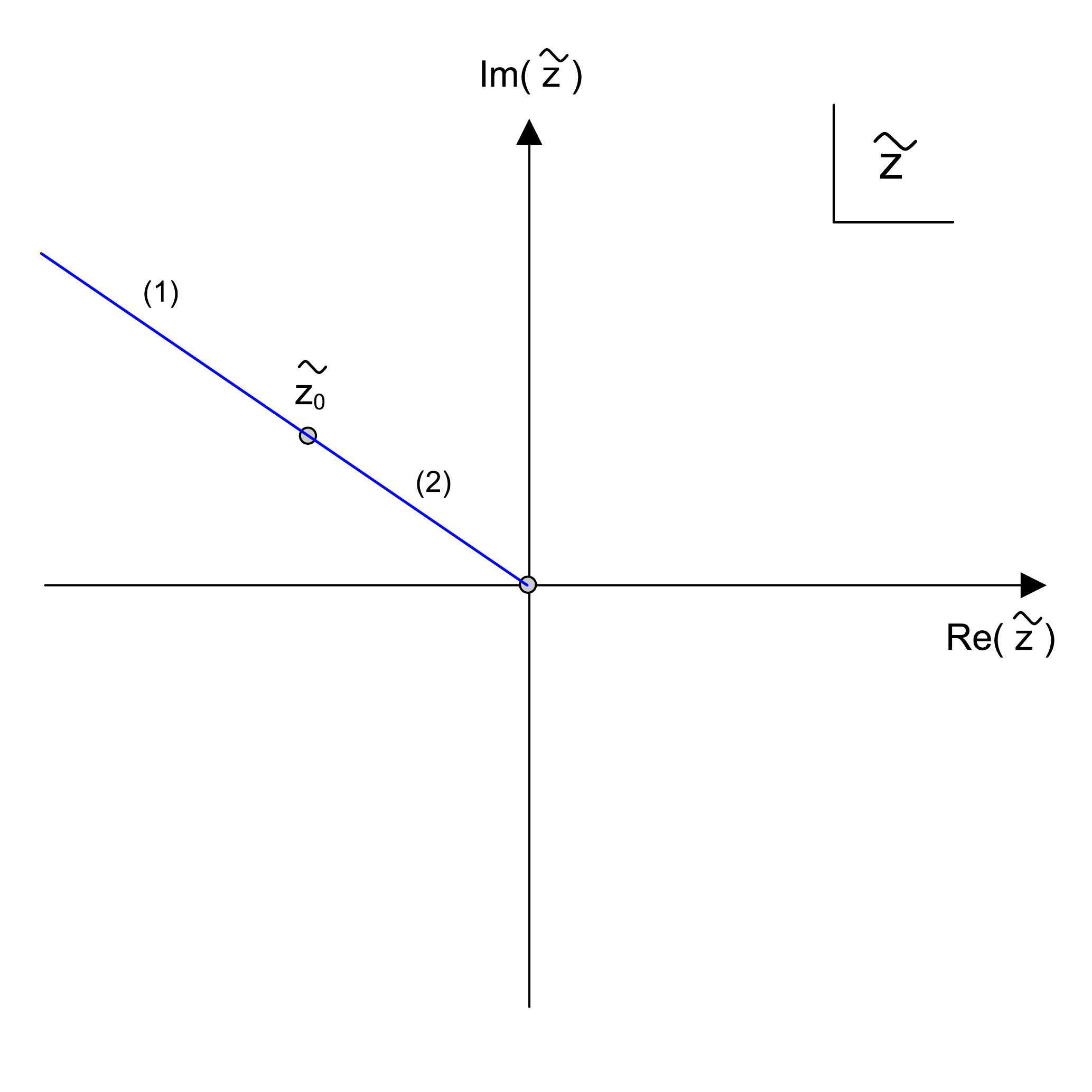}}
\end{center}
\caption{Chosen contour for the monodromy calculation: \protect\subref{fig:zcontour} In the $z$ plane
and \protect\subref{fig:ztcontour} In the $ \zt $ plane.}
\label{fig:contour}
\end{figure}

Looking at the anti-Stokes lines, defined by
\begin{equation}
\im(\lambda\zt) = 0 \ ,
\end{equation}
we choose two of them to be the contour for the calculation\footnote{The choice of branch cuts so
that $ 0\leq\arg(z)<2\pi $ makes sure that the contours don't pass through a branch cut. Choosing
$ 0<\arg(z)\leq 2\pi $ instead would give the complex conjugates of these contours.}.
We then proceed by developing the equation, and finding its solutions, around points along the
chosen contour (the boundary and $ z\to\infty $ ) and ``matching'' the solutions. The chosen
contour is shown in Figure~\ref{fig:contour}\footnote{The assumed shape of the anti-Stokes lines
is only true in the case of $ K>-1 $.}.

We first look at the equation at $ z\to\infty $ or $ \zt\to\zt_0 $. As shown in
Appendix~\ref{app:effectivepotentials}, the equation at $ \zt\to\zt_0 $ is approximately
\begin{equation}
\pdzt^2\phi + \left[\lambda^2-\frac{j_\infty^2-1}{4(\zt-\zt_0)^2}\right] = 0 \ .
\end{equation}
The solution to this equation can be written as:
\begin{equation}
\phi \approx A_+ P_+(\zt-\zt_0) + A_- P_-(\zt-\zt_0) \ .
\end{equation}
In the non-scalar case where $ j_\infty \neq 0 $ (and more generally $ j_\infty $ isn't an even integer),
\begin{equation}
P_\pm(\zt-\zt_0) \equiv \sqrt{2\pi\lambda(\zt-\zt_0)}J_{\pm\frac{j_\infty}{2}}\left(\lambda(\zt-\zt_0)\right) \ ,
\end{equation}
where $ J_{\pm\frac{j_\infty}{2}} $ are the Bessel functions.
The fact that asymptotically\footnote{Another assumption is that
$ |\lambda| \gg |q|$, so that the effect of the effective potential can be neglected in the regions
between the the ``special'' points ($ z\rightarrow0 $, $ z\rightarrow1 $, $ z\rightarrow\infty $)}
$ |\lambda| \gg 1 $ allows us to replace the Bessel functions with their corresponding asymptotic
forms.
For direction (1) (as appears in Figure~\ref{fig:contour}), we have $ \lambda(\zt-\zt_0) > 0 $, so
that
\begin{multline}
\label{eq:monoasympinftyplus}
\phi \approx
\left[A_+ \e^{i(-\lambda\zt_0-\alpha_+)} + A_- \e^{i(-\lambda\zt_0-\alpha_-)}\right] \e^{i\lambda\zt}\\
+ \left[A_+ \e^{i(\lambda\zt_0+\alpha_+)} + A_- \e^{i(\lambda\zt_0+\alpha_-)} \right] \e^{-i\lambda\zt}
\ ,
\end{multline}
where
\begin{equation}
\alpha_\pm = \frac{\pi}{4}(1 \pm j_\infty) \ .
\end{equation}
For direction (2) we have $ \lambda(\zt-\zt_0) < 0 $, so that
\begin{multline}
\label{eq:monoasympinftymin}
\phi \approx
\left[A_+\e^{i(-\lambda\zt_0+3\alpha_+)} + A_-\e^{i(-\lambda\zt_0+3\alpha_-)}\right]\e^{i\lambda\zt}\\
+\left[A_+\e^{i(\lambda\zt_0+\alpha_+)} + A_-\e^{i(\lambda\zt_0+\alpha_-)}\right]\e^{-i\lambda\zt} \ .
\end{multline}

Next we look at the equation at the boundary. As shown in Appendix~\ref{app:effectivepotentials}, the
equation at $ \zt\to 0 $ is approximately
\begin{equation}
\pdzt^2\phi + \left[\lambda^2-\frac{j_1^2-1}{4(\zt-\zt_0)^2}\right] = 0 \ .
\end{equation}
The solution is therefore:
\begin{equation}
\phi \approx B_+ P_+(\zt) + B_- P_-(\zt)\\
= B_+\sqrt{2\pi\lambda\zt} J_{\frac{j_1}{2}}(\lambda\zt)
+ B_-\sqrt{2\pi\lambda\zt} J_{-\frac{j_1}{2}}(\lambda\zt) \ .
\end{equation}
We now apply the boundary conditions at the boundary ($ z=1 $, $ \zt=0 $) to this solution. In all cases (the scalar, transverse vector and longitudinal vector) we end up with the condition
\begin{equation}
B_- = 0 \ .
\end{equation}
Next we again assume $ |\lambda| \gg 1 $ so that the Bessel function can be replaced with its asymptotic form. For direction (2), we have $ \lambda\zt > 0 $, so that
\begin{equation}
\label{eq:monoasympboundary}
\phi \approx 2B_+\cos(\lambda\zt-\beta_+)
= \left[B_+ \e^{-i\beta_+}\right]\e^{i\lambda\zt} + \left[B_+ \e^{i\beta_+}\right]\e^{-i\lambda\zt} \ ,
\end{equation}
where
\begin{equation}
\beta_+ = \frac{\pi}{4}(1+j_1) \ .
\end{equation}

Finally from the boundary conditions at the horizon ($ z=0 $, $ \zt\to -\infty $) we have (for line (1) as it appears in Figure~\ref{fig:contour})
\begin{equation}
\label{eq:monoasymphorizon}
\phi \sim \e^{-i\lambda\zt} \ .
\end{equation}

Now we ``match'' the solutions on lines (1) and (2).
In the non-scalar case, in which $ j_\infty \ne 0 $ so that $ \alpha_+ \ne \alpha_- $.
In this case we have for line (1) (from Equations~\ref{eq:monoasympinftyplus} and~\ref{eq:monoasymphorizon})
\begin{equation}
\label{eq:monoeq1}
A_+\e^{i(-\lambda\zt_0-\alpha_+)} + A_-\e^{i(-\lambda\zt_0-\alpha_-)} = 0 \ .
\end{equation}
For line (2) we have (from Equations~\ref{eq:monoasympinftymin} and~\ref{eq:monoasympboundary})
\begin{align}
A_+\e^{i(-\lambda\zt_0+3\alpha_+)} + A_-\e^{i(-\lambda\zt_0+3\alpha_-)} &= B_+\e^{-i\beta_+}\\
A_+\e^{i(\lambda\zt_0+\alpha_+)} + A_-\e^{i(\lambda\zt_0+\alpha_-)} &= B_+\e^{i\beta_+} \ .
\end{align}

The condition for having a non-trivial solution for this system of equations is:
\begin{equation}
\tan(\lambda\zt_0-\beta_+)
= \frac{\e^{i3\alpha_+}\sin(\alpha_+) - \e^{i3\alpha_-}\sin(\alpha_-)}{\e^{i3\alpha_+}\cos(\alpha_+) - \e^{i3\alpha_-}\cos(\alpha_-)} \ .
\end{equation}
Finally, we obtain the result\footnote{Obviously, $ -\lambda_n^* $ would also be a solution.
Choosing the branch cuts so that $ 0<\arg(z)\leq 2\pi $ would give this solution instead.}:
\begin{equation}
\label{eq:asymptoticlambda}
\lambda_n = \lambda_0 + n\Delta\lambda \ ,
\end{equation}
where
\begin{equation}
\lambda_0 = \frac{\beta_+ + \arctan\left[\frac{\e^{i3\alpha_+}\sin(\alpha_+) - \e^{i3\alpha_-}\sin(\alpha_-)}{\e^{i3\alpha_+}\cos(\alpha_+) - \e^{i3\alpha_-}\cos(\alpha_-)}\right]}{\zt_0}
\qquad
\Delta\lambda = \frac{\pi}{\zt_0} \ .
\end{equation}

In the scalar case, in which $ j_\infty = 0 $, we have:
\begin{align}
P_+(\zt-\zt_0) &\equiv \sqrt{2\pi\lambda(\zt-\zt_0)}J_0\left(\lambda(\zt-\zt_0)\right)\\
P_-(\zt-\zt_0) &\equiv \sqrt{2\pi\lambda(\zt-\zt_0)}Y_0\left(\lambda(\zt-\zt_0)\right) \ .
\end{align}
In a calculation similar to the non-scalar case, we again obtain the asymptotic form
in Equation~\ref{eq:asymptoticlambda},
where
\begin{equation}
\lambda_0 = \frac{\beta_+ + \frac{\pi}{4}-\frac{i}{2}\ln(2)}{\zt_0}
\qquad
\Delta\lambda = \frac{\pi}{\zt_0} \ .
\end{equation}

\subsubsection{Asymptotic Dependence on Topology and Temperature}

\begin{figure}[htbp]
\centering
\includegraphics[width=10cm]{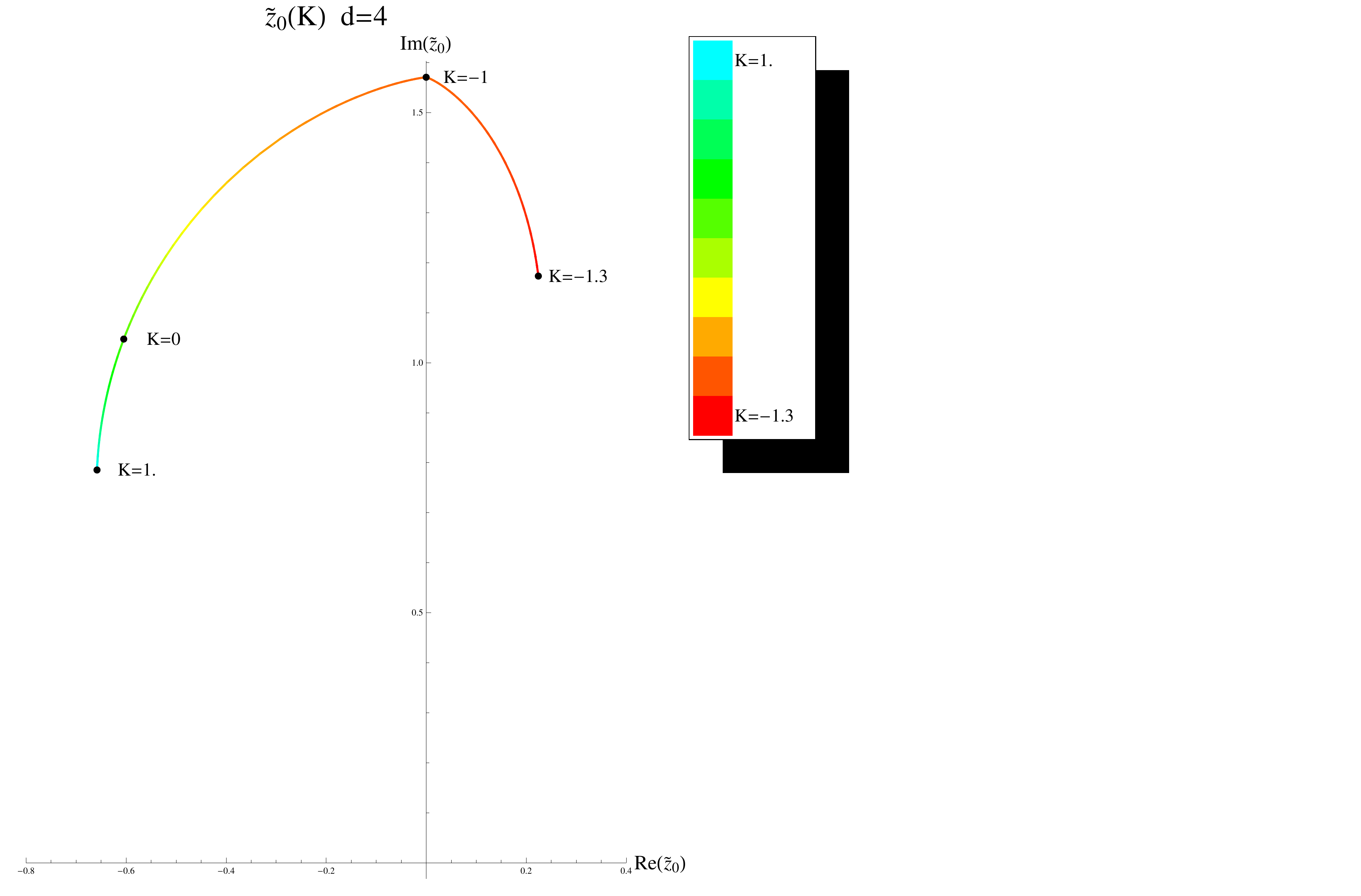}
\caption{The dependence of the parameter $ \zt_0 $ on the parameter $ K $ on the $ \zt_0 $ complex
plane, in $ d=4 $ dimensions.}
\label{fig:zt0valuesd4}
\end{figure}

\begin{figure}[htbp]
\centering
\includegraphics[width=10cm]{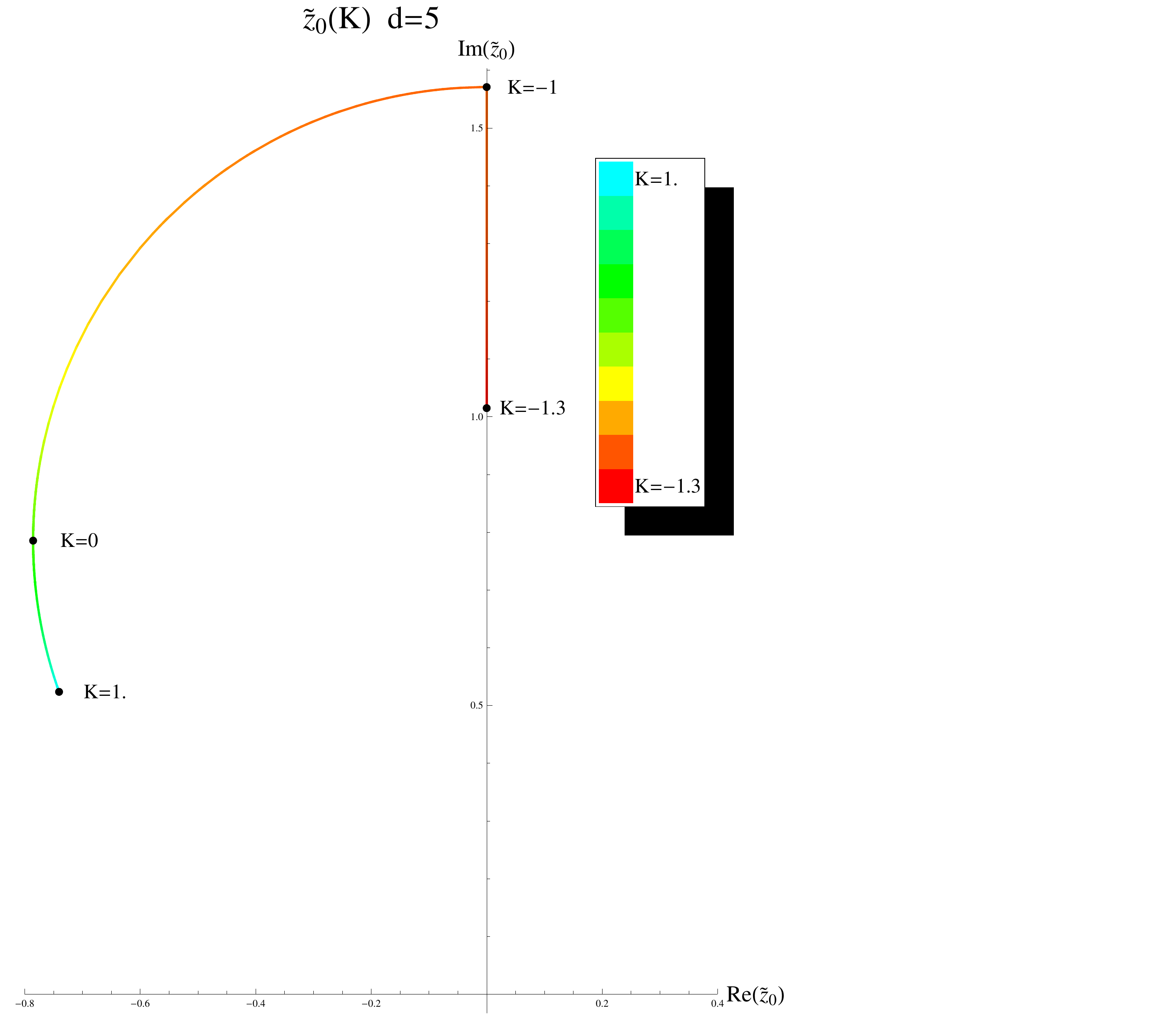}
\caption{The dependence of the parameter $ \zt_0 $ on the parameter $ K $ on the $ \zt_0 $ complex
plane, in $ d=4 $ dimensions.}
\label{fig:zt0valuesd5}
\end{figure}

From the results of the calculation in this section we see that the asymptotic parameters of the
QNM frequencies (namely $ \lambda_0 $ and $ \Delta\lambda $) of each perturbation type depend on
the $ \Omega_{d-2} $ manifold topology and the temperature through the parameter $ \zt_0 $,
which depends only on $ d $ and $ K $ (where $ K $ contains the dependency on both the topology and
the temperature). In fact, the asymptotic slope of the frequencies ($ \Delta\lambda $) depends only
on the spatial topology and the temperature - it is the same for all types of perturbations
(even for types that aren't discussed here, such as tensorial perturbations):
\begin{equation}
\Delta\lambda = \frac{\pi}{\zt_0} \ .
\end{equation}
The dependence of $ \zt_0 $ on $ K $ is plotted in Figure~\ref{fig:zt0valuesd4} for $ d=4 $ and in
Figure~\ref{fig:zt0valuesd5} for $ d=5 $.
In the case of spherical ($ K>0 $) and flat ($ K=0 $) topologies, the parameter $ \zt_0 $, and therefore
the slope $ \Delta\lambda $, always has an imaginary part and a real part (for all temperature values).
However in the hyperbolic case, at the temperature $ T_c $ (as defined in
Section~\ref{sec:intro}) where $ K=-1 $, the slope becomes completely
imaginary\footnote{In dimensions $ d=4 $ and $ d=5 $, $ \re\zt_0 \ge 0 $ below the temperature $ T_c $,
so that the asymptotic slope of the QNM frequencies remains imaginary below this temperature as well.}
(as seen in the exact results given in Subsection~\ref{subsec:qnmexactsolutionsfortc} for this case).
Moreover, it can be clearly seen in the Figures~\ref{fig:zt0valuesd4} and~\ref{fig:zt0valuesd5} that
$ \zt_0 (K) $ (and therefore $ \Delta\lambda(T) $) is not smooth at this temperature.

In the following we study two specific cases: the case of $ K=0 $, and the case of $ K\to -1$
($ k<0 $,$ T\to T_c $).

\paragraph{The case of $ K=0 $}

Let us first review the case of $ K=0 $ (where the $ \Omega_{d-2} $ manifold is ``flat''),
described for example in \cite{Natario:2004jd} as the limiting case of large spherical black holes.
In this case, the roots of $ \gz $ are analytically known and we can get an accurate expression for the
asymptotic QNM frequencies (and most importantly the gap $ \Delta\lambda $).

Assuming that $ K=0 $, we get for $ \gz $:
\begin{equation}
\gz = 1-(1-z)^{d-1} \ .
\end{equation}
Its roots are given by:
\begin{equation}
z_k = 1 - \e^{i2\pi\frac{k-1}{d-1}} \qquad k=1,\ldots,(d-1) \ .
\end{equation}

After calculating the residues and using the definition of $ \zt_0 $, we obtain the
results:
\begin{equation}
\zt_0 = -\frac{\pi}{d-1}\e^{-i\frac{\pi}{d-1}}\frac{1}{\sin\left(\frac{\pi}{d-1}\right)}
\end{equation}
\begin{equation}
\Delta\lambda = \frac{\pi}{\zt_0} = -(d-1)\e^{i\frac{\pi}{d-1}}\sin\left(\frac{\pi}{d-1}\right) \ .
\end{equation}
For example, for $ d=4 $ we get $ \Delta\lambda = -\frac{3\sqrt{3}}{4}-i\frac{9}{4} $,
and for $ d=5 $ we get $ \Delta\lambda = -2(1+i) $.

\paragraph{The case of $ K\to -1 $}

Here we consider the case of $ K\to -1 $ ($ T\to T_c $) from above (meaning $ K>-1 $). Define:
\begin{equation}
K\equiv -1+\epsilon \qquad \epsilon>0 \ ,
\end{equation}
so that
\begin{equation}
\gz = 1-(1-\epsilon)(1-z)^2-\epsilon(1-z)^{d-1} \ .
\end{equation}
Notice also that from the discussion in Section~\ref{sec:intro} and Appendix~\ref{app:thermodynamicquant}
we can relate $ \epsilon $ to the temperature difference by:
\begin{equation}
\epsilon = \frac{2}{d-2} \frac{\Delta T}{T_c} \ .
\end{equation}

We first find an approximation for the roots of $ \gz $. For $ K=-1 $ ($ \epsilon=0 $) we have:
\begin{equation}
\gz = 1-(1-z)^2 = 0
\qquad\Rightarrow\qquad
z_1=0, z_2=2 \ .
\end{equation}
In this case there are only \emph{two} roots. When $ K>-1 $, there are $ d-1 $ roots. Two of them will
be close to the two roots of the $ K=-1 $ case, while the others go to $ \infty $ as $ K\to -1 $.

Developing the roots and residues around $ z=0 $ and $ z=2 $ to first order in $ \epsilon $,
we get:
\begin{align}
z_1 & =0 \\
\gamma_1 &\approx \frac{1}{2+(d-3)\epsilon} \approx \frac{1}{2}-\frac{d-3}{4}\epsilon \\
z_2 &\approx 2+\frac{1+(-1)^d}{2}\epsilon \\
\gamma_2 &\approx -\frac{1}{2}-\frac{(-1)^d(d-2)+1}{4}\epsilon \ .
\end{align}

Approximating the other roots we get:
\begin{align}
1-z_k &\approx \epsilon^{-\frac{1}{d-3}} \e^{i\left(\frac{\pi}{d-3}+\frac{2\pi}{d-3}j\right)} \\
\gamma_k &\approx -\frac{1}{d-3}\epsilon^{\frac{1}{d-3}}\e^{-i\left(\frac{\pi}{d-3}+\frac{2\pi}{d-3}j\right)} \ ,
\end{align}
where $ j \equiv d-k-1 = 0,\ldots,d-4 $.

We put these values into the definition for $ \zt_0 $, and after some calculations we obtain the following results:\\
For the case $ d=4 $:
\begin{empheq}[box=\fbox]{align}
\label{eq:asymptparamsaroundtcde4a}
\zt_0 &\approx \frac{i\pi}{2} + \epsilon\ln\epsilon\\
\label{eq:asymptparamsaroundtcde4b}
\Delta\lambda &= \frac{\pi}{\zt_0} \approx -2i + \frac{4}{\pi}\epsilon\ln\epsilon \ ,
\end{empheq}
And for the case $ d>4 $:
\begin{empheq}[box=\fbox]{align}
\label{eq:asymptparamsaroundtcdg4a}
\zt_0 &\approx \frac{i\pi}{2} - \frac{\pi}{d-3}\frac{1}{\sin\left(\frac{\pi}{d-3}\right)}\epsilon^{\frac{1}{d-3}}\\
\label{eq:asymptparamsaroundtcdg4b}
\Delta\lambda &= \frac{\pi}{\zt_0} \approx -2i - \frac{4}{d-3}\frac{1}{\sin\left(\frac{\pi}{d-3}\right)}\epsilon^{\frac{1}{d-3}} \ .
\end{empheq}
For example, for $ d=5 $ we get
\begin{equation}
\Delta\lambda \approx -2i-2\sqrt{\epsilon} \ .
\end{equation}

\subsection{Hydrodynamic Approximation}

Taking the hydrodynamic limit of $ \lambda,q \to 0 $ ($ \frac{\omega}{T},\frac{L_s^{FT}}{T} \to 0 $,
$ \frac{\sqrt{|R_\Omega^{FT}|}}{T} \to 0 $ is required) in the appropriate QNM equations, one may
calculate the hydrodynamic constants of the dual gauge theory (see
~\cite{Son:2007vk},~\cite{Kovtun:2005ev}).
Since the longitudinal vector mode is coupled to charge density fluctuations in the gauge theory, its
QNM spectrum in the hydrodynamic limit has to contain a mode corresponding to the charge diffusion mode
in the gauge theory. Here we review the derivation of the diffusion mode from the QNM Equation
in our notation.

Starting from Equation~\ref{eq:longvectoreq} for a longitudinal vector QNM, let us
define a new function $ \psi' = z^{-\frac{i\lambda}{C}}\psi $, and get a new ``shifted''
equation:
\begin{multline}
\label{eq:longvectoreqshifted}
\left( \pdz-\frac{i\lambda}{Cz} \right)
\left[ \gz (1-z)^{d-4} \left(\pdz-\frac{i\lambda}{Cz}\right) \left( \frac{1}{(1-z)^{d-4}}\psi \right) \right]\\
+ \left[ \frac{\lambda^2}{\gz}-q^2 \right] \psi = 0 \ ,
\end{multline}
with boundary conditions
\begin{equation}
\left. \psi \right|_{z=0} \sim \text{const.}
\qquad
\left. (1-z)^{d-4} \left(\pdz-\frac{i\lambda}{Cz}\right) \left(\frac{1}{(1-z)^{d-4}}\psi\right) \right|_{z=1} = 0 \ .
\end{equation}

For the hydrodynamic approximation we shall assume $ \lambda \sim q^2 \ll 1 $, and develop $ \psi $ in
orders of $ \lambda $ :
\begin{equation}
\psi = \psi^{(0)} + \psi^{(1)} + \psi^{(2)} + \ldots \ .
\end{equation}
We put this into Equation~\ref{eq:longvectoreqshifted}, equate each order to 0 and solve the equation
with the corresponding boundary conditions in terms of the higher order solutions.

For the 0-order we get the general solution:
\begin{equation}
\psi^{(0)}
\sim C_0\left[ \frac{1}{C}\ln z + \text{const.} \right] + D_0 (1-z)^{d-4} \ ,
\end{equation}
since $ \left. \gz \right|_{z=0} \sim Cz $.
From the boundary condition at $ z=0 $ we get
\begin{equation}
C_0 = 0 \ ,
\end{equation}
so that
\begin{equation}
\psi^{(0)} = D_0 (1-z)^{d-4} \ ,
\end{equation}
while the boundary conditions at $ z=1 $ are automatically fulfilled.

Next we turn to the $ \lambda^1 $ order, for which we get the solution:
\begin{multline}
\psi^{(1)} = \frac{i\lambda D_0}{C}(1-z)^{d-4} \ln z
-(1-z)^{d-4} \frac{q^2 D_0}{d-3} \int \frac{1-z}{\gz} \ud z\\
+C_1 (1-z)^{d-4} \int \frac{1}{\gz (1-z)^{d-4}} \ud z
+D_1 (1-z)^{d-4} \ .
\end{multline}
Applying the boundary condition at $ z=0 $ we get
\begin{equation}
\left.\psi^{(1)}\right|_{z=0} \sim \frac{i\lambda D_0}{C}\ln z
-\frac{q^2 D_0}{(d-3)C}\ln z
+\frac{C_1}{C}\ln z + D_1 \sim \text{const.} \ ,
\end{equation}
from which we get the condition
\begin{equation}
\label{eq:hydrocond1}
\left( i\lambda - \frac{q^2}{d-3} \right)D_0 + C_1 = 0 \ .
\end{equation}
From the boundary condition at $ z=1 $ we have
\begin{multline}
\left. (1-z)^{d-4}\pdz\left( \frac{1}{(1-z)^{d-4}}\psi^{(1)} \right) \right|_{z=1}
-\frac{i\lambda}{C}\left. \psi^{(0)} \right|_{z=1} \\
= \frac{i\lambda D_0}{C}\delta_{d,4} + C_1 - \frac{i\lambda D_0}{C}\delta_{d,4}
= C_1 = 0 \ .
\end{multline}
Putting this into Equation~\ref{eq:hydrocond1} we get to the conclusion
\begin{equation}
\lambda = -\frac{i}{d-3} q^2 \ ,
\end{equation}
which is a (normalized) diffusion relation, with the diffusion constant equal to $ \frac{1}{d-3} $,
or, in terms of $ \omega $ and $ L_\mathbf{s}^{FT} $:
\begin{equation}
\omega \approx -iD \left(L_\mathbf{s}^{FT}\right)^2
\qquad
D = \frac{d-1}{d-3} \frac{1}{4\pi T} \ .
\end{equation}
This is true for any dimension $ d \ge 4 $ and any $ \Omega_{d-2} $ topology and temperature for whom
the hydrodynamic condition can be fulfilled. As expected, the leading term in this approximation
doesn't depend on the topology of $ \Omega_{d-2} $.

\subsection{Numerical Calculation of QNMs}

In the following we present some results of exact numerical calculations of the QNM spectra in several
cases in order to illustrate the dependence on topology and temperature. The spectra have been
calculated using the method outlined in Appendix~\ref{app:numericalmethods}.

Figures~\ref{fig:qnmresultsforseveralkscalard4q0K10m05}
and~\ref{fig:qnmresultsforseveralklongvectord5q141K10m05} contain the exact QNM spectra for the cases
of spherical ($ K=1 $), flat ($ K=0 $) and hyperbolic ($ K=-0.5 $) topologies, for the scalar and
longitudinal vector cases. The dependence of the asymptotic slope of the spectrum on the parameter $ K $
can be easily seen in these figures (as well as its independence of other parameters such as the spatial
mode or perturbation type). The (pure imaginary) hydrodynamic mode can also be seen in the spectra of
the longitudinal vector perturbations.

\begin{figure}[htbp]
\begin{minipage}{0.49\textwidth}
\centering
\includegraphics[width=\textwidth]{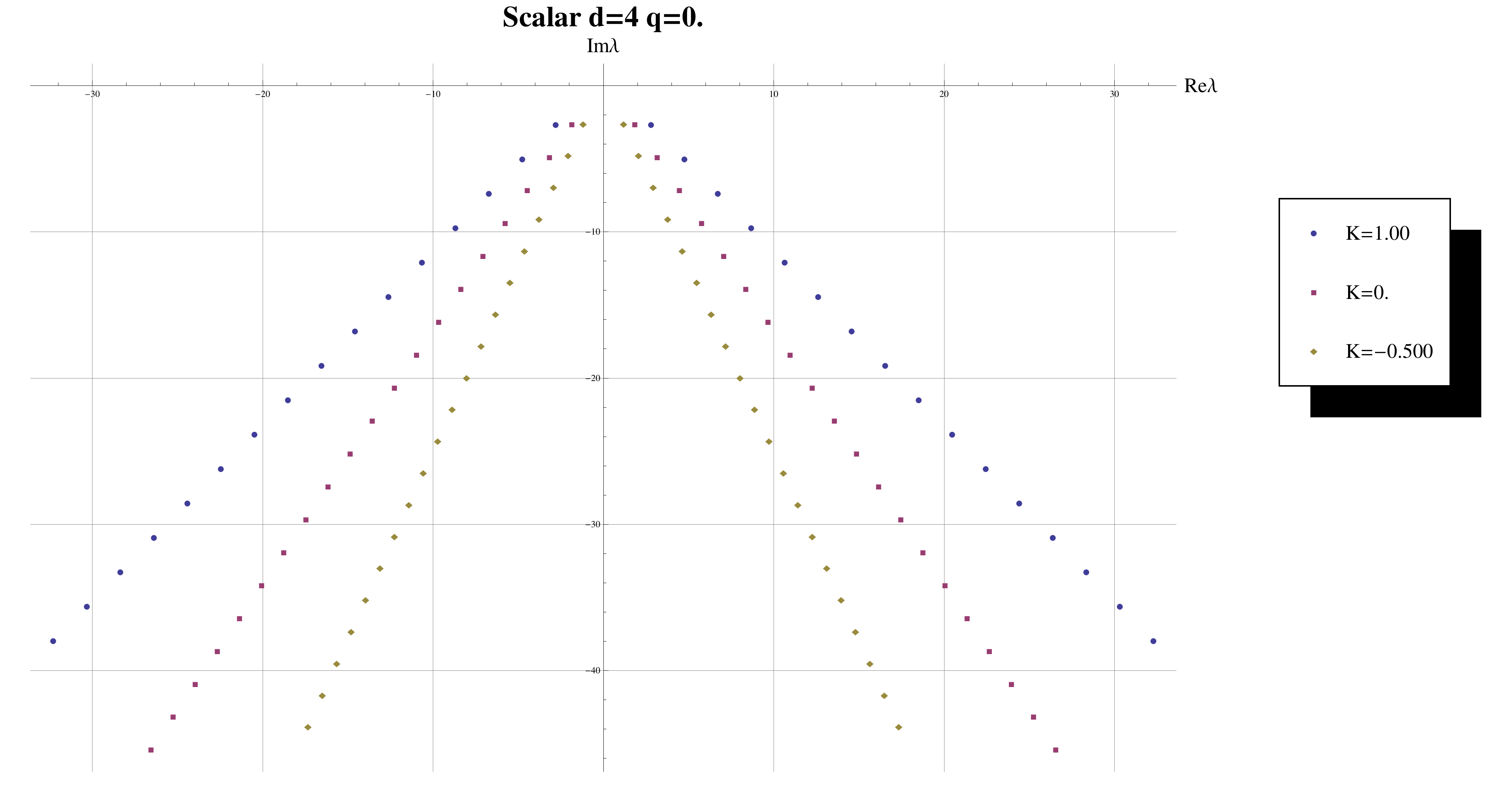}
\caption{Numerically calculated QNM frequencies for the case of scalar perturbations with
$ d=4 $,$ K = 1 , 0 , -0.5 $ and $ q_\mathbf{s} = 0 $.}
\label{fig:qnmresultsforseveralkscalard4q0K10m05}
\end{minipage}
\hfill
\begin{minipage}{0.49\textwidth}
\includegraphics[width=\textwidth]{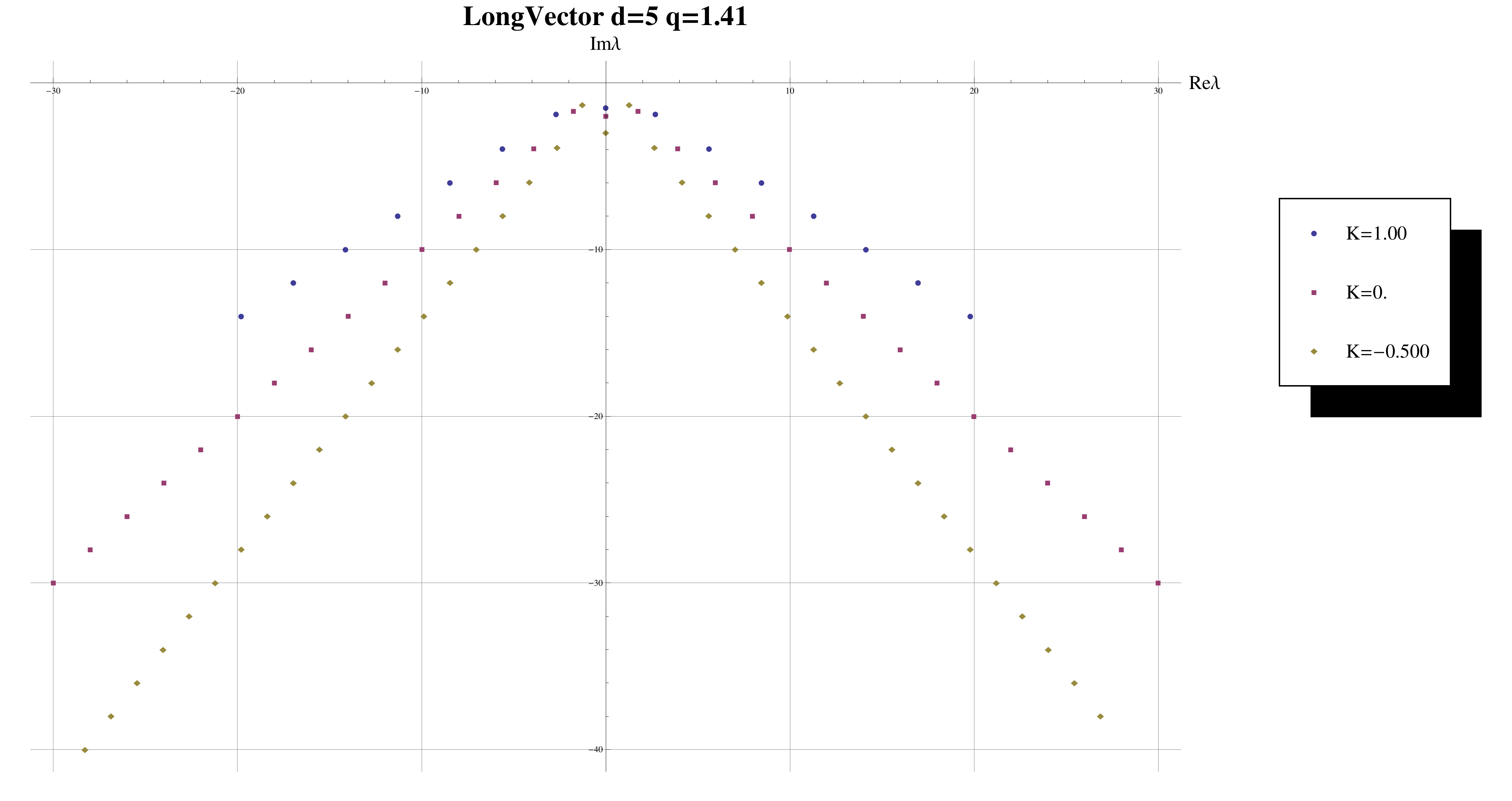}
\caption{Numerically calculated QNM frequencies for the case of longitudinal vector perturbations with
$ d=5 $,$ K = 1 , 0 , -0.5 $ and $ q_\mathbf{s} = \sqrt{2} $.}
\label{fig:qnmresultsforseveralklongvectord5q141K10m05}
\end{minipage}
\end{figure}

\begin{figure}[htbp]
\begin{minipage}{0.49\textwidth}
\centering
\includegraphics[width=\textwidth]{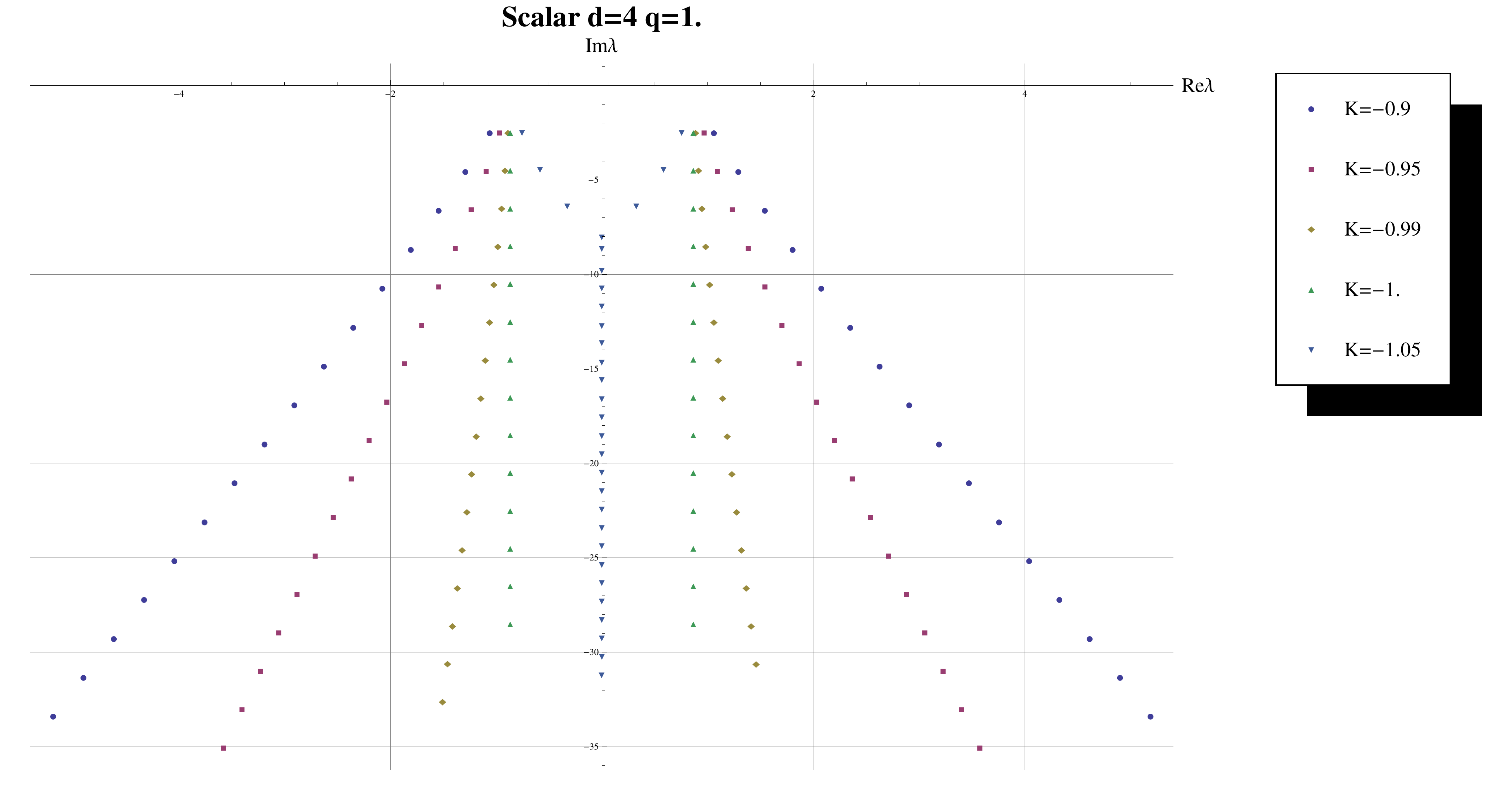}
\caption{Numerically calculated QNM frequencies for the case of scalar perturbations with
$ d=4 $,$ -1.05 < K < -0.9 $ and $ q_\mathbf{s} = 1 $.}
\label{fig:qnmresultsforseveralkscalard4q1}
\end{minipage}
\hfill
\begin{minipage}{0.49\textwidth}
\centering
\includegraphics[width=\textwidth]{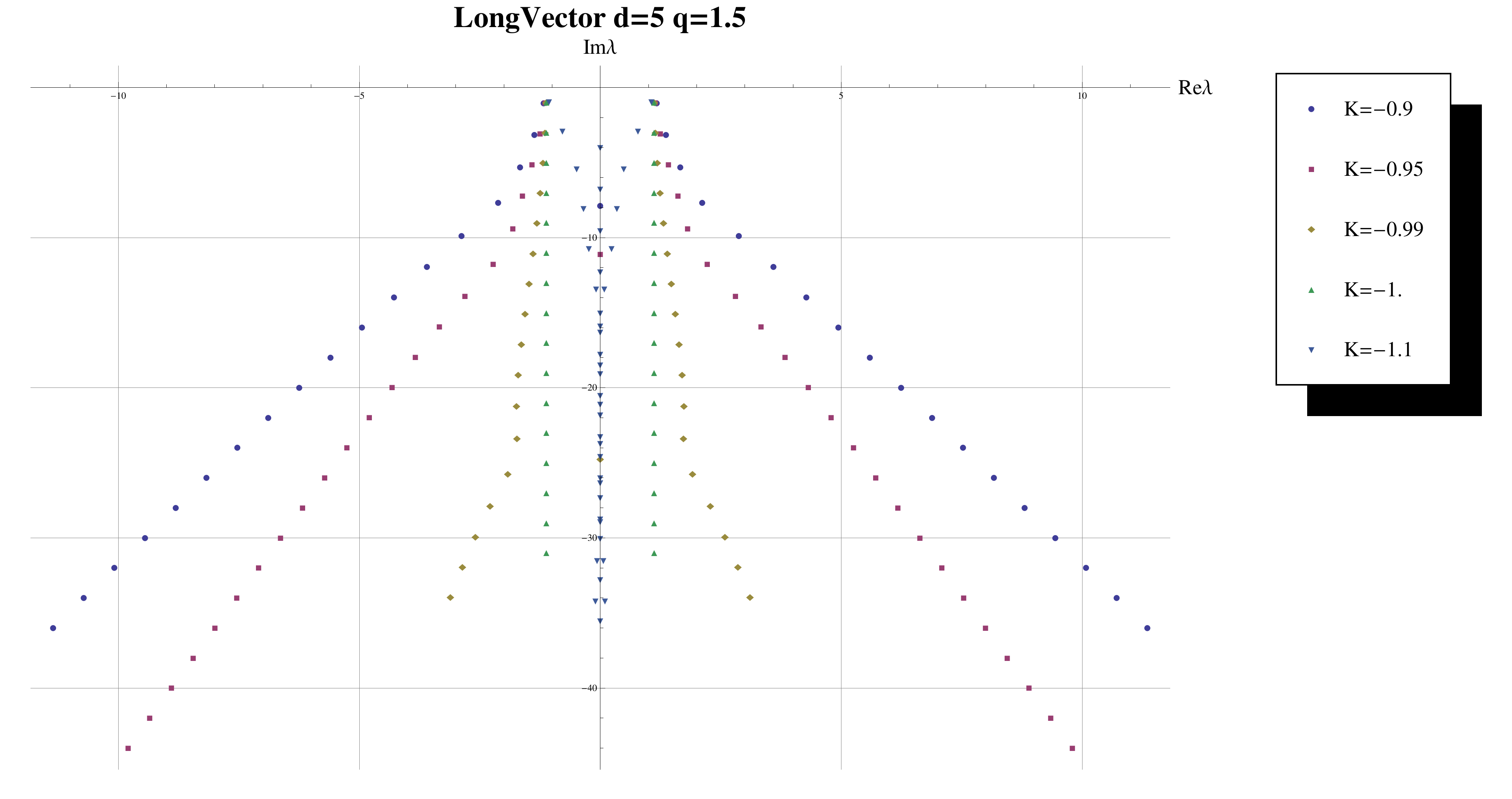}
\caption{Numerically calculated QNM frequencies for the case of longitudinal vector perturbations with
$ d=5 $,$ -1.1 < K < -0.9 $ and $ q_\mathbf{s} = 1.5 $.}
\label{fig:qnmresultsforseveralklongvectord5q1p5}
\end{minipage}
\end{figure}

The numerical results for $ -1.1 < K < -0.9 $, as illustrated in
Figures~\ref{fig:qnmresultsforseveralkscalard4q1} and~\ref{fig:qnmresultsforseveralklongvectord5q1p5},
demonstrate the phenomenon discussed in Sections~\ref{subsec:qnmexactsolutionsfortc}
and~\ref{subsec:qnmasymoptotics} for an hyperbolic $ \Omega_{d-2} $ manifold: At $ T=T_c $ ($ K=-1 $) the
asymptotic QNM frequency gap ($ \Delta\lambda $) becomes imaginary, and for $ d=4,5 $ it remains
imaginary for $ T<T_c $ ($ K<-1 $).

\clearpage

\section{CFT Correlators}
\label{sec:cftcorrelators}

\subsection{CFT Correlators From Holographic Principle}
\label{subsec:cftcorrelatorsfromhol}

Via the AdS/CFT corresondence, a black hole (or black brane) background in
the bulk spacetime, such as the ones discussed here, corresponds to a dual CFT at finite temperature in the deconfined phase.
The correspondence  allows one to calculate the 2-point correlation function of CFT operators
by calculating the on-shell action of the bulk field dual to that operator.

The general prescription for calculating the dual CFT correlators from the AdS/CFT correspondence
is discussed
in~\cite{Maldacena:1997re},~\cite{Witten:1998qj},~\cite{Gubser:1998bc}, and~\cite{Aharony:1999ti}.
In the case of Minkowski spacetime, the calculation of the Minkowski correlators from AdS/CFT
requires some additional subtleties, as explained in~\cite{Son:2002sd}.
In particular, the retarded green function defined as:
\begin{equation}
G^R\left(t,x;t',x'\right) = -i\theta(t-t')\left\langle\left[O(t,x),O(t',x')\right]\right\rangle
\end{equation}
can be calculated by choosing the incoming-wave boundary condition at the black hole horizon for
the solution of the classical EOM in the bulk spacetime (see
Subsection~\ref{subsec:qnmdefinition}).

One consequence of the above prescription is the fact that the QNM frequency spectrum of the
bulk field (as defined in Subsection~\ref{subsec:qnmdefinition}) comprises the poles of the retarded
correlator of the dual CFT operator. Therefore all of the results of
Section~\ref{sec:quasinormalmodes} serve to teach about the poles of the retarded correlators of
the theories dual to the discussed black holes. The purpose of this section is the expansion of
those results from knowledge of the poles to expressions for the correlators themselves.

\subsection{General Formulae For Correlators}
\label{subsec:generalformulaeforcorrelators}

The background spacetime metric on which the field theory dual to the bulk black hole
is defined will be given by:
\begin{equation}
\label{eq:cftmetricnoR}
\ud s_{FT}^2 = -\ud t^2 + \ud\Omega_{FT,d-2}^{2} \ ,
\end{equation}
where
$
\ud\Omega_{FT,d-2}^2 = \sum_{i,j}(g_\Omega^{FT})_{ij}\ud x_{FT}^i \ud x_{FT}^j
$
is the inner metric of the $ \Omega_{d-2}^{FT} $ manifold.
This metric is conformally related
to the bulk background metric at its boundary, with the conformal factor chosen to cancel the
divergence in the bulk metric. The relation between the metrics is then given by
(see~\cite{Emparan:1999gf})
\begin{equation}
g_{\mu\nu}^{FT} = \lim_{r\to\infty}\left(\frac{R^2}{r^2}g_{\mu\nu}^{bulk}\right) \ ,
\end{equation}
or:
\begin{equation}
\label{eq:cftmetric}
\ud s_{FT}^2 = -\ud t^2 + R^2 \ud\Omega_{d-2}^2 \ .
\end{equation}
Comparing this with Equation~(\ref{eq:cftmetricnoR}) we get the relation between the field theory
manifold and the bulk manifold $ \ud\Omega_{FT,d-2}^2 = R^2\ud\Omega_{d-2}^2 $, and therefore
the relation between the corresponding curvatures -
$
R_\Omega^{FT} = \frac{R_\Omega}{R^2}
$
and
$
k_{FT} \equiv \frac{R_\Omega^{FT}}{(d-2)(d-3)} = \frac{k}{R^2}
$,
and the relation between the corresponding Laplace operator eigenvalues
$
L_s^{FT} = \frac{L_s}{R}
$.

Using the AdS/CFT prescription for a massless scalar bulk field and a gauge vector bulk field, we
may find expressions for the retarded correlation functions of the dual field theory operators.
This is done by integrating by parts the expressions for the action of these fields,
using the EOMs they satisfy to get an expression that depends only on the boundary values of these
fields and taking the derivative of the expression with respect to the the boundary values.
We find the following expressions:
\begin{itemize}

\item For a massless scalar field, the correlation function is given by:
\begin{equation}
\label{eq:scalarcftcorrelator}
G^R(\omega,\mathbf{s})
= \left.-2C_s\frac{r_+^{d-1}}{R^d}\frac{1}{(1-z)^{d-2}}
\pdz\widehat{\psi}_{\omega,\mathbf{s}}(z)\right|_{z\to 1} \ ,
\end{equation}
where $ \widehat{\psi}_{\omega,\mathbf{s}} $ satisfies the scalar QNM equation
(Equation~\ref{eq:scalareq}) along with the boundary conditions:
\begin{equation}
\left.\widehat{\psi}_{\omega,\mathbf{s}}\right|_{z=0} \sim z^{-\frac{i\lambda}{C}}
\qquad
\left.\widehat{\psi}_{\omega,\mathbf{s}}\right|_{z\to 1} = 1 \ .
\end{equation}
$ C_s $ is the appropriate normalization constant for the bulk scalar field (see Appendix C).
\item For the longitudinal component of a vector field, the correlation function is given
by:
\begin{equation}
\label{eq:longvectorcftcorrelator}
G_{tt}^R(\omega,\mathbf{s})
= \left.4C_v\frac{r_+^{d-3}}{R^{d-2}}\frac{1}{(1-z)^{d-4}}
\widehat{\psi}_{\omega,\mathbf{s}}\right|_{z\to 1}
\end{equation}
\begin{align}
\label{eq:longvectorcftcorrelator2}
G_{t\|}^R(\omega,\mathbf{s}) &= i\frac{\lambda}{q_\mathbf{s}}G_{tt}^R(\omega,\mathbf{s}) \\
\label{eq:longvectorcftcorrelator3}
G_{\|\|}^R(\omega,\mathbf{s}) &= \frac{\lambda^2}{q_\mathbf{s}^2}G_{tt}^R(\omega,\mathbf{s}) \ ,
\end{align}
where $ \widehat{\psi}_{\omega,\mathbf{s}} $ satisfies the longitudinal vector QNM equation
(Equation~\ref{eq:longvectoreq}) along with the boundary conditions:
\begin{equation}
\left.\widehat{\psi}_{\omega,\mathbf{s}}\right|_{z=0} \sim z^{-\frac{i\lambda}{C}}
\quad
\left.\widehat{\chi}_{\omega,\mathbf{s}}\right|_{z\to 1}
= \left.\frac{1}{q_\mathbf{s}^2}\gz(1-z)^{d-4}\pdz
\left[\frac{1}{(1-z)^{d-4}}\widehat{\psi}_{\omega,\mathbf{s}}\right]\right|_{z\to 1} = 1 \ .
\end{equation}
$ C_v $ is the appropriate normalization constant for the bulk vector field (see Appendix C).
\item For the transverse component of a vector field, the correlation function is given by:
\begin{equation}
\label{eq:transvectorcftcorrelator}
G_{\bot\bot}^R(\omega,\mathbf{v})\\
= \left.-4C_v\frac{r_+^{d-3}}{R^{d-2}}\frac{1}{(1-z)^{d-4}}
\pdz\widehat{\psi}_{\omega,\mathbf{v}}(z)\right|_{z\to 1} \ ,
\end{equation}
where $ \widehat{\psi}_{\omega,\mathbf{v}} $ satisfies the transverse vector QNM equation
(Equation~\ref{eq:transvectoreq}) along with the boundary conditions:
\begin{equation}
\left.\widehat{\psi}_{\omega,\mathbf{v}}\right|_{z=0} \sim z^{-\frac{i\lambda}{C}}
\qquad
\left.\widehat{\psi}_{\omega,\mathbf{v}}\right|_{z\to 1} = 1 \ .
\end{equation}

\end{itemize}

Several notes about taking the limit $ z\to 1 $:
\begin{enumerate}
\item The expression $ \pdz\widehat{\psi}_{\omega,\mathbf{s}}(z) $ in the limit $ z\to 1 $ is
calculated by normalizing $ \psi $ at $ z=1-\epsilon $, differentiating it and then taking
$ \epsilon\to 1 $, so that:
\begin{equation}
\left.\pdz\widehat{\psi}_{\omega,\mathbf{s}}(z)\right|_{z\to 1}
= \left.\frac{\pdz\psi_{\omega,\mathbf{s}}(z)}{\psi_{\omega,\mathbf{s}}(z)}\right|_{z\to 1} \ .
\end{equation}
\item When taking the limit $ z\to 1 $ , one must drop the contact terms - the terms that are polynomials in $ \omega $ and $ L_\mathbf{s} $ and diverge as $ z\to 1 $ (These terms are
removed by process of renormalization). This can be done by developing
$ \pdz\widehat{\psi}_{\omega,\mathbf{s}}(z) $ in orders of $ 1-z $, and then dropping all terms
up to the lowest non-contact-term order.
\end{enumerate}

Note also that the spacetime coordinates correlators for the vector perturbations are related to the
above longitudinal and transverse components by the following relations:
\begin{eqnarray}
G_{tt}^R(t,x;t',x') &=& \int\frac{\ud\omega}{2\pi}\sum_\mathbf{s}
G_{tt}^R(\omega,\mathbf{s})H_\mathbf{s}^{FT*}(x)H_\mathbf{s}^{FT}(x')\e^{-i\omega(t'-t)}\\
G_{ti}^R(t,x;t',x') &=&  \int\frac{\ud\omega}{2\pi}\sum_\mathbf{s}
G_{t\|}^R(\omega,\mathbf{s})H_\mathbf{s}^{FT*}(x)\frac{\partial_i H_\mathbf{s}^{FT}(x')}
{L_\mathbf{s}^{FT}}\e^{-i\omega(t'-t)} \\
G_{ij}^R(t,x;t',x') &=&  \int\frac{\ud\omega}{2\pi}\sum_\mathbf{s}
G_{\|\|}^R(\omega,\mathbf{s})\frac{\partial_i H_\mathbf{s}^{FT*}(x)}
{L_\mathbf{s}^{FT}}\frac{\partial_j H_\mathbf{s}^{FT}(x')}
{L_\mathbf{s}^{FT}}\e^{-i\omega(t'-t)} \nonumber\\
&& + \int\frac{\ud\omega}{2\pi}\sum_\mathbf{v}
G_{\bot\bot}^R(\omega,\mathbf{v})\tilde{A}_{\mathbf{v},i}^{FT*}(x)
\tilde{A}_{\mathbf{v},j}^{FT}(x')\e^{-i\omega(t'-t)} \ .
\end{eqnarray}

A detailed derivation of these expressions can be found in Appendix~\ref{app:corrformulaederivation}.

\subsection{Exact Correlators for the \texorpdfstring{$K=-1$}{K=-1} Case}
\label{subsec:cftexactcorrelatorsfortc}

As in Subsection~\ref{subsec:qnmexactsolutionsfortc}, in the $K=-1$ case (where the $\Omega_{d-2}$
manifold is hyperbolic and $T=T_c$), analytical expressions for the CFT correlators may be found.
This section is dedicated to the calculation of the correlators in this case.

\subsubsection{Scalar Correlator}

Continuing from the transformation defined in
Appendix~\ref{app:exactsolutionsfortcderivation}, the expression in
Equation~\ref{eq:scalarcftcorrelator} can be written in terms of $ \phi $ and $ w $:
\begin{equation}
\label{eq:scalarcftcorrelatorinw}
G^R(\omega,\mathbf{s})
= \left.-4C_s\frac{r_+^{d-1}}{R^d}\frac{1}{(1-w)^\frac{d-3}{2}}
\frac{\partial_w\phi_{\omega,\mathbf{s}}}{\phi_{\omega,\mathbf{s}}}\right|_{w\to 1}
\end{equation}
(where a contact term has been dropped).

As explained in Appendix~\ref{app:exactsolutionsfortcderivation}, the solution to
the EOM with an incoming-wave boundary condition at the horizon is
$ \phi_{\omega,\mathbf{s}} = {}_2F_1\left(a,b;c;w\right)  $,
where $ a,b $ and $ c $ are given by Equations~\ref{eq:scalartransexactsolutionfortcparams1}
and~\ref{eq:scalartransexactsolutionfortcparams2}. Define:
\begin{equation}
\Delta \equiv c-a-b = \frac{d-1}{2} \ .
\end{equation}

There are now two possible cases:
\begin{enumerate}
\item $ d $ is even, so that $ \Delta $ is non-integer. In this case, the hypergeometric
function satisfies the following connection formula:
\begin{multline}
\label{eq:hypergeometricevenconnectionformula}
\phi_{\omega,\mathbf{s}}(w) = {}_2F_1\left(a,b;c;w\right) \\
= \frac{\Gamma(c)\Gamma(\Delta)}{\Gamma(c-a)\Gamma(c-b)}\,{}_2F_1\left(a,b;1-\Delta;1-w\right)\\
+ \frac{\Gamma(c)\Gamma(-\Delta)}{\Gamma(a)\Gamma(b)}(1-w)^\Delta
\,{}_2F_1\left(c-a,c-b;\Delta+1;1-w\right) \ .
\end{multline}
Defining the coefficients:
\begin{align}
A(\omega,\mathbf{s}) &=  \frac{\Gamma(c)\Gamma(\Delta)}{\Gamma(c-a)\Gamma(c-b)} \\
B(\omega,\mathbf{s}) &=  \frac{\Gamma(c)\Gamma(-\Delta)}{\Gamma(a)\Gamma(b)} \ ,
\end{align}
we get:
\begin{equation}
\left.\frac{\partial_w\phi_{\omega,\mathbf{s}}}{\phi_{\omega,\mathbf{s}}}\right|_{w=1-\epsilon}
\approx -\Delta\frac{B(\omega,\mathbf{s})}{A(\omega,
\mathbf{s})}\epsilon^{\Delta-1}\\
= -\Delta \frac{\Gamma(-\Delta)}{\Gamma(\Delta)} \frac{\Gamma(c-a)\Gamma(c-b))}
{\Gamma(a)\Gamma(b)} \epsilon^{\Delta-1} \ .
\end{equation}
Putting this into Equation~\ref{eq:scalarcftcorrelatorinw} we get:
\begin{equation}
\label{eq:scalarexactcorrelatorintceven}
\boxed{
G^R(\omega,\mathbf{s})
= 2(d-1)C_s \frac{r_+^{d-1}}{R^d}\frac{\Gamma(-\Delta)}{\Gamma(\Delta)}
\frac{\Gamma(a+\Delta)}{\Gamma(a)}\frac{\Gamma(b+\Delta)}{\Gamma(b)} \ .
}
\end{equation}

\item $ d $ is odd, so that $ \Delta $ is an integer. In this case, the hypergeometric
function satisfies\footnote{Here and in related formulae, $ \psi(z) $ represents the Digamma function,
defined by : $ \psi(z) \equiv \frac{\Gamma'(z)}{\Gamma(z)}$.}:
\begin{multline}
\label{eq:hypergeometricoddconnectionformula}
\phi_{\omega,\mathbf{s}}(w) = {}_2F_1\left(a,b;c;w\right) \\
= \frac{\Gamma(c)\Gamma(\Delta)}{\Gamma(a+\Delta)\Gamma(b+\Delta)}
\sum_{n=0}^{\Delta-1}\frac{(a)_n (b)_n}{n!(1-\Delta)_n}(1-w)^n \\
- \frac{\Gamma(c)}{\Gamma(a)\Gamma(b)}(w-1)^\Delta\sum_{n=0}^{\infty}
\frac{(a+\Delta)_n(b+\Delta)_n}{n!(n+\Delta)!}(1-w)^n [\ln(1-w)\\
+\psi(a+\Delta+n)+\psi(b+\Delta+n)-\psi(n+1)-\psi(n+\Delta+1)] \ ,
\end{multline}
so that\footnote{The terms $ \psi(1) $ and $ \psi(\Delta+1) $ have been dropped from $ B $ because
after dividing by A they amount to a contact term.}:
\begin{align}
A(\omega,\mathbf{s}) &=  \frac{\Gamma(c)\Gamma(\Delta)}{\Gamma(a+\Delta)\Gamma(b+\Delta)} \\
B(\omega,\mathbf{s}) &=  \frac{\Gamma(c)}{\Gamma(a)\Gamma(b)}\frac{(-1)^{\Delta+1}}{(\Delta)!}
\left[\psi(a+\Delta)+\psi(b+\Delta)\right]
\end{align}

\begin{equation}
\left.\frac{\partial_w\phi_{\omega,\mathbf{s}}}{\phi_{\omega,\mathbf{s}}}\right|_{w=1-\epsilon}
\approx -\frac{(-1)^{\Delta+1}}{\Gamma^2(\Delta)}(a)_\Delta(b)_\Delta
\left[\psi(a+\Delta)+\psi(b+\Delta)\right]\epsilon^{\Delta-1} \ .
\end{equation}

Putting this into Equation~\ref{eq:scalarcftcorrelatorinw} we get:
\begin{equation}
\label{eq:scalarexactcorrelatorintcodd}
\boxed{
G^R(\omega,\mathbf{s})
= 4C_s\frac{r_+^{d-1}}{R^d}\frac{(-1)^{\Delta+1}}{\Gamma^2(\Delta)}(a)_\Delta(b)_\Delta
\left[\psi(a+\Delta)+\psi(b+\Delta)\right] \ .
}
\end{equation}

\end{enumerate}

\subsubsection{Vector Correlators}

The calculation of the exact vector correlators in the $ K=-1 $ case is similar to the scalar case:
We write the expressions for the correlators in Equations~\ref{eq:longvectorcftcorrelator}
and~\ref{eq:transvectorcftcorrelator} in terms of the variables defined in
Appendix~\ref{app:exactsolutionsfortcderivation} and calculate the limits using the exact solutions.
A detailed derivation is given in Appendix~\ref{app:exactvectorcorrelatorsderivationfortc}, and here
we quote the final results.

For the longitudinal vector mode, we define:
\begin{equation}
\Delta \equiv \frac{d-5}{2} \ .
\end{equation}
For the case of even $ d $ ($ \Delta $ is non-integer), we obtain the expression:
\begin{equation}
\label{eq:longvectorexactcorrelatorintceven}
\boxed{
G_{tt}^R(\omega,\mathbf{s})
= \frac{4C_v}{d-5}\frac{r_+^{d-3}}{R^{d-2}}q_\mathbf{s}^2
\frac{\Gamma(-\Delta)}{\Gamma(\Delta)}\frac{\Gamma(a+\Delta)}{\Gamma(a)}
\frac{\Gamma(b+\Delta)}{\Gamma(b)} \ .
}
\end{equation}
For the case of odd $ d $ ($ \Delta $ is an integer), we obtain the expression:
\begin{equation}
\label{eq:longvectorexactcorrelatorintcodd}
\boxed{
G_{tt}^R(\omega,\mathbf{s})
= 2C_v\frac{r_+^{d-3}}{R^{d-2}}q_\mathbf{s}^2\frac{(-1)^{\Delta+1}}{\Gamma^2(\Delta+1)}
(a)_\Delta(b)_\Delta\left[\psi(a+\Delta)+\psi(b+\Delta)\right] \ .
}
\end{equation}
The other components of the longitudinal mode correlator can be calculated from
Equations~\ref{eq:longvectorcftcorrelator2} and~\ref{eq:longvectorcftcorrelator3}.

For the transverse vector mode, we define:
\begin{equation}
\Delta \equiv \frac{d-3}{2} \ .
\end{equation}
The caluclation gives for even $ d $:
\begin{equation}
\label{eq:transvectorexactcorrelatorintceven}
\boxed{
G_{\bot\bot}^R(\omega,\mathbf{v})
= 4(d-3)C_v \frac{r_+^{d-3}}{R^{d-2}}\frac{\Gamma(-\Delta)}{\Gamma(\Delta)}
\frac{\Gamma(a+\Delta)}{\Gamma(a)}\frac{\Gamma(b+\Delta)}{\Gamma(b)} \ ,
}
\end{equation}
and for odd $ d $:
\begin{equation}
\label{eq:transvectorexactcorrelatorintcodd}
\boxed{
G_{\bot\bot}^R(\omega,\mathbf{v})
= 8C_v\frac{r_+^{d-3}}{R^{d-2}}\frac{(-1)^{\Delta+1}}{\Gamma^2(\Delta)}(a)_\Delta(b)_\Delta
\left[\psi(a+\Delta)+\psi(b+\Delta)\right] \ .
}
\end{equation}

\subsection{Asymptotic Correlator Expressions (for \texorpdfstring{$ K>-1 $}{K>-1} )}
\label{subsec:cftasymptoticcorrelatorexpressions}

Continuing Subsection~\ref{subsec:qnmasymoptotics}, the goal of this subsection is to find an approximate
expression for the correlators of the CFT operators coupled to the scalar and vector fields in the
bulk spacetime for large values of the frequency, where $ |\lambda| \gg q $ (corresponding to
$ \frac{L_s^{FT}}{T} = \text{fixed} $ and $ \frac{\omega}{T} \gg 1 $)
and $ \im\lambda < 0 $. The general stages of the calculation are as follows:
\begin{enumerate}

\item We start from the effective potentials and their poles calculated in
Subsection~\ref{subsec:qnmasymoptotics}.
\item We solve the approximate QNM equations around each pole using Bessel functions.
\item Using the asymptotic limit of $ |\lambda| \gg 1 $, we replace the Bessel functions with their
asymptotic forms.
\item We match the solutions on each region of the complex plane, and apply the boundary condition at
the horizon, to get an expression for the solution near the boundary up to a multiplicative constant.
\item We use the approximate solution near the boundary and the expressions for the retarded correlators
from Subsection~\ref{subsec:generalformulaeforcorrelators} to obtain an asymptotic expression for the
correlators.

\end{enumerate}

\subsubsection{Calculation of Asymptotic Correlators for the Scalar Case}
\label{subsubsec:cftasymptoticcorrelatorexpressionsscalar}

Proceeding from the definitions in Subsection~\ref{subsec:qnmasymoptotics}, the expression in
Equation~\ref{eq:scalarcftcorrelator} can be written in terms of $ \phi $ and $ \zt $
(after dropping the contact term and using the fact that $ \zt\approx z-1 $ for
$ z\to 1 $):
\begin{equation}
\label{eq:scalarcftcorrelatorinzt}
G^R(\omega,\mathbf{s}) =  \left.-2C_s\frac{r_+^{d-1}}{R^d}\frac{(-1)^d}{\zt^{d-2}}
\frac{\pdzt\phi_{\omega,\mathbf{s}}(\zt)}{\phi_{\omega,\mathbf{s}}(\zt)}\right|_{\zt\to 0} \ .
\end{equation}

We continue in a method similar to the one applied in Subsection~\ref{subsec:qnmasymoptotics}:
find asymptotic solutions around the poles of the effective potential and ``match'' the
solutions\footnote{
While in the general case, where $ \im(\lambda\zt_0)\neq 0 $, the anti-Stokes lines
emanating from $ \zt=0 $ and $ \zt_0 $ don't align, the ``matching'' of the solutions in
the region (2) is still possible since there are no Stokes lines between these two
anti-Stokes lines in this region, so that the asymptotic solution remains
the same between them.
}. As before, we assume that $ |\lambda|\gg 1 $ and $ |\lambda|\gg q $.
We also assume that $ \lambda $ is in the 3rd quadrant of the complex plane
($ \re\lambda<0 $, $ \im\lambda<0 $).
The solutions around $ \zt\to\zt_0 $ ($ z\to\infty $) are given by:
\begin{align}
P_+(\zt-\zt_0) &\equiv \sqrt{2\pi\lambda(\zt-\zt_0)}J_0\left(\lambda(\zt-\zt_0)\right)\\
P_-(\zt-\zt_0) &\equiv \sqrt{2\pi\lambda(\zt-\zt_0)}Y_0\left(\lambda(\zt-\zt_0)\right) \ .
\end{align}
Replacing the Bessel functions with their asymptotic forms gives for direction (1):
\begin{equation}
\label{eq:monoasympinftyplus2}
\phi \approx
\left[A_+\e^{i(-\lambda\zt_0-\frac{\pi}{4})}+A_-\e^{i(-\lambda\zt_0-\frac{3\pi}{4})}\right]\e^{i\lambda\zt}\\
+\left[A_+\e^{i(\lambda\zt_0+\frac{\pi}{4})}-A_-\e^{i(\lambda\zt_0-\frac{\pi}{4})}\right]\e^{-i\lambda\zt} \ ,
\end{equation}
and for direction (2):
\begin{equation}
\label{eq:monoasympinftymin2}
\phi \approx
\left[A_+\e^{i(-\lambda\zt_0+\frac{3\pi}{4})}-3A_-\e^{i(-\lambda\zt_0+\frac{\pi}{4})}\right]\e^{i\lambda\zt}\\
+\left[A_+\e^{i(\lambda\zt_0+\frac{\pi}{4})}-A_-\e^{i(\lambda\zt_0-\frac{\pi}{4})}\right]\e^{-i\lambda\zt} \ .
\end{equation}
From the boundary condition at the horizon ($ z=0 $, $ \zt\to -\infty $) we have:
\begin{equation}
\phi \approx \e^{-i\lambda\zt} \ .
\end{equation}

There are now two possible cases:
\begin{enumerate}
\item $ d $ is even. In this case the solution around $ \zt\to 0 $ is approximately:
\begin{equation}
\phi \approx B_+ P_+(\zt) + B_- P_-(\zt)\\
= B_+\sqrt{2\pi\lambda\zt} J_{\frac{j_1}{2}}(\lambda\zt)
+ B_-\sqrt{2\pi\lambda\zt} J_{-\frac{j_1}{2}}(\lambda\zt) \ ,
\end{equation}
where $ j_1 = d-1 $ (so that $ \frac{j_1}{2} $ is non-integer). Replacing the Bessel functions
with their asymptotic form, we have:
\begin{equation}
\phi
\approx \left[B_+\e^{-i\beta_+} + B_-\e^{-i\beta_-}\right]\e^{i\lambda\zt} +
 \left[B_+\e^{i\beta_+} + B_-\e^{i\beta_-}\right]\e^{-i\lambda\zt} \ ,
\end{equation}
where $ \beta_\pm \equiv \frac{\pi}{4}(1 \pm j_1)$.
We now match the solutions on lines (1) and (2). We have for line (1):
\begin{equation}
\label{eq:cftmonoeq3}
A_+\e^{i(-\lambda\zt_0-\frac{\pi}{4})}+A_-\e^{i(-\lambda\zt_0-\frac{3\pi}{4})} = 0
\qquad\Rightarrow\qquad
A_- = -iA_+ \ .
\end{equation}
For the section (2) we have:
\begin{align}
\label{eq:cftmonoeq4}
A_+\e^{i(-\lambda\zt_0+\frac{3\pi}{4})}-3A_-\e^{i(-\lambda\zt_0+\frac{\pi}{4})}
&= B_+\e^{-i\beta_+} + B_-\e^{-i\beta_-}\\
\label{eq:cftmonoeq5}
A_+\e^{i(\lambda\zt_0+\frac{\pi}{4})}-A_-\e^{i(\lambda\zt_0-\frac{\pi}{4})}
&= B_+\e^{i\beta_+} + B_-\e^{i\beta_-} \ .
\end{align}
Putting Equation~\ref{eq:cftmonoeq3} into Equations~\ref{eq:cftmonoeq4} and~\ref{eq:cftmonoeq5}
we get:
\begin{align}
B_+\e^{-i\theta_+} + B_-\e^{-i\theta_-} &= 2iA_+ \\
B_+\e^{i\theta_+} + B_-\e^{i\theta_-} &= 4iA_+ \ ,
\end{align}
where $ \theta_\pm \equiv \lambda\zt_0 - \frac{\pi}{4} - \beta_\pm $.
Solving for $ B_\pm $ we can get an expression for $ \frac{B_+}{B_-} $:
\begin{equation}
\frac{B_+}{B_-}
= \frac{\left|\begin{array}{cc}
2 & \e^{-i\theta_-} \\
4 & \e^{i\theta_-}
\end{array}\right|}
{\left|\begin{array}{cc}
\e^{-i\theta_+} & 2 \\
\e^{i\theta_+} & 4
\end{array}\right|}
= -i^{d-1}\frac{\e^{2i\theta_-}-2}{\e^{2i\theta_-}+2}
\end{equation}
(here we used the fact that $ d $ is even).

Next we evaluate the correlator. Developing $ \phi $ around $ \zt=0 $ we have:
\begin{multline}
\label{eq:asymptoticcorrelatorsphiaround0even}
\phi
\approx B_+\sqrt{2\pi\lambda\zt} J_{\frac{j_1}{2}}(\lambda\zt)
+ B_-\sqrt{2\pi\lambda\zt} J_{-\frac{j_1}{2}}(\lambda\zt) \\
\approx B_-\sqrt{2\pi\lambda\zt}(\frac{1}{2}\lambda\zt)^{-\frac{j_1}{2}}
\sum_{k=0}^\infty \frac{(-\frac{1}{4}\lambda^2\zt^2)^k}{k!\Gamma(-\frac{j_1}{2}+k+1)} \\
+ B_+\sqrt{2\pi\lambda\zt}(\frac{1}{2}\lambda\zt)^{\frac{j_1}{2}}
\sum_{k=0}^\infty \frac{(-\frac{1}{4}\lambda^2\zt^2)^k}{k!\Gamma(\frac{j_1}{2}+k+1)} \ .
\end{multline}
Defining the coefficients:
\begin{align}
A(\omega,\mathbf{s}) &= \frac{\sqrt{2\pi}\lambda^{-\Delta+\frac{1}{2}}}
{2^{-\Delta}\Gamma(-\Delta+1)} B_- \\
B(\omega,\mathbf{s}) &= \frac{\sqrt{2\pi}\lambda^{\Delta+\frac{1}{2}}}
{2^{\Delta}\Gamma(\Delta+1)} B_+
\end{align}
(where $ \Delta\equiv \frac{j_1}{2} = \frac{d-1}{2} $),
we get:
\begin{equation}
\left.\frac{\pdzt\phi_{\omega,\mathbf{s}}}{\phi_{\omega,\mathbf{s}}}\right|_{\zt=\epsilon}
= (2\Delta)\frac{B(\omega,\mathbf{s})}{A(\omega,\mathbf{s})}\epsilon^{2\Delta-1}
\approx (d-1)\left(\frac{i\lambda}{2}\right)^{d-1}\frac{\Gamma(-\Delta)}{\Gamma(\Delta)}
\frac{\e^{2i\theta_-}-2}{\e^{2i\theta_-}+2}\epsilon^{d-2} \ .
\end{equation}
And finally putting this into Equation~\ref{eq:scalarcftcorrelatorinzt}:
\begin{equation}
G^R(\omega,\mathbf{s}) \approx  -2(d-1)C_s\frac{r_+^{d-1}}{R^d}
\frac{\Gamma(-\Delta)}{\Gamma(\Delta)}\left(\frac{i\lambda}{2}\right)^{d-1}
\frac{\e^{2i\theta_-}-2}{\e^{2i\theta_-}+2}
\end{equation}
(Note that this expression is true up to the addition of contact terms).
This asymptotic formula is only true for $ \re\lambda<0 $. In order to get an expression for
$ \re\lambda>0 $, we can make use of the symmetry properties of the correlator\footnote{
This property can also be deduced from the QNM equations themselves.}:
\begin{equation}
G^R(\omega,\mathbf{s}) = G^{R*}(-\omega^*,\mathbf{s}) \ .
\end{equation}
We get for $ \re\lambda>0 $:
\begin{equation}
G^R(\omega,\mathbf{s}) \\
= -2(d-1)C_s\frac{r_+^{d-1}}{R^d}
\frac{\Gamma(-\Delta)}{\Gamma(\Delta)}\left(\frac{i\lambda}{2}\right)^{d-1}
\frac{\e^{2i\overline{\theta_-}}-2}{\e^{2i\overline{\theta_-}}+2} \ ,
\end{equation}
where we define:
\begin{equation}
\overline{\theta_-}(\lambda) \equiv -\theta_-^*(-\lambda*)
= \lambda\zt_0^*-\frac{\pi}{4}(d-3) \ .
\end{equation}
Looking at the asymptotic behaviour of the expressions for $ \re\lambda<0 $ and $ \re\lambda>0 $,
we may unify them into one asymptotic expression:
\begin{equation}
G^R(\omega,\mathbf{s}) \approx  2(d-1)C_s\frac{r_+^{d-1}}{R^d}
\frac{\Gamma(-\Delta)}{\Gamma(\Delta)}\left(\frac{i\lambda}{2}\right)^{d-1}
\frac{\e^{2i\theta_-}-2}{\e^{2i\theta_-}+2}\,
\frac{\e^{2i\overline{\theta_-}}-2}{\e^{2i\overline{\theta_-}}+2} \ .
\end{equation}
Finally, using $ \lambda=\frac{\omega R^2}{r_+} $, the asymptotic expression may be written:
\begin{equation}
\label{eq:asymptcftcorrelatorscalarevend}
\boxed{
G^R(\omega,\mathbf{s}) \approx  2(d-1)C_s R^{d-2}
\frac{\Gamma(-\Delta)}{\Gamma(\Delta)}\left(\frac{i\omega}{2}\right)^{d-1}
\frac{\e^{2i\theta_-}-2}{\e^{2i\theta_-}+2}\,
\frac{\e^{2i\overline{\theta_-}}-2}{\e^{2i\overline{\theta_-}}+2} \ .
}
\end{equation}

\item $ d $ is odd. In this case the solution around $ \zt\to 0 $ is approximately:
\begin{equation}
\phi \approx B_+ P_+(\zt) + B_- P_-(\zt)\\
= B_+\sqrt{2\pi\lambda\zt} J_{\frac{j_1}{2}}(\lambda\zt)
+ B_-\sqrt{2\pi\lambda\zt} Y_{\frac{j_1}{2}}(\lambda\zt) \ ,
\end{equation}
where $ j_1 = d-1 $ (so that $ \frac{j_1}{2} $ is an integer). Replacing the Bessel functions
with their asymptotic form, we have:
\begin{equation}
\phi
\approx \left[(B_+-iB_-)\e^{-i\beta_+}\right]\e^{i\lambda\zt} +
\left[(B_++iB_-)\e^{i\beta_+}\right]\e^{-i\lambda\zt} \ ,
\end{equation}
where $ \beta_+ = \frac{\pi}{4}(1+j_1) = \frac{\pi}{4}d $.
We again match the solutions on lines (1) and (2).
For (1) we again have:
\begin{equation}
\label{eq:cftmonoeq6}
A_+\e^{i(-\lambda\zt_0-\frac{\pi}{4})}+A_-\e^{i(-\lambda\zt_0-\frac{3\pi}{4})} = 0 \ .
\end{equation}
For (2):
\begin{align}
\label{eq:cftmonoeq7}
A_+\e^{i(-\lambda\zt_0+\frac{3\pi}{4})}-3A_-\e^{i(-\lambda\zt_0+\frac{\pi}{4})}
&= (B_+-iB_-)\e^{-i\beta_+}\\
\label{eq:cftmonoeq8}
A_+\e^{i(\lambda\zt_0+\frac{\pi}{4})}-A_-\e^{i(\lambda\zt_0-\frac{\pi}{4})}
&= (B_++iB_-)\e^{i\beta_+} \ .
\end{align}
Putting Equation~\ref{eq:cftmonoeq6} into Equations~\ref{eq:cftmonoeq7} and~\ref{eq:cftmonoeq8}
we get (defining $ \theta_+ \equiv \lambda\zt_0 - \frac{\pi}{4} - \beta_+ $):
\begin{align}
(B_+-iB_-)\e^{i\theta_+} &= 4iA_+ \\
(B_++iB_-)\e^{-i\theta_+} &= 2iA_+ \ .
\end{align}
Solving for $ B_\pm $ we can get an expression for $ \frac{B_+}{B_-} $:
\begin{equation}
\frac{B_+}{B_-}
= \frac{\left|\begin{array}{cc}
4 & -i\e^{i\theta_+} \\
2 & i\e^{-i\theta_+}
\end{array}\right|}
{\left|\begin{array}{cc}
\e^{i\theta_+} & 4 \\
\e^{-i\theta_+} & 2
\end{array}\right|}
= 2i\frac{\e^{2i\theta_+}}{\e^{2i\theta_+}-2}-i \ .
\end{equation}

Next we evaluate the correlator. Developing $ \phi $ around $ \zt=0 $ we have (we again
define $ \Delta\equiv \frac{j_1}{2} = \frac{d-1}{2} $):
\begin{equation}
\label{eq:asymptoticcorrelatorsphiaround0odd}
\phi
\approx B_+\sqrt{2\pi\lambda\zt} J_{\Delta}(\lambda\zt)
+ B_-\sqrt{2\pi\lambda\zt} Y_{\Delta}(\lambda\zt) \\
\end{equation}
Defining the coefficients:
\begin{equation}
A(\omega,\mathbf{s}) = -\frac{\sqrt{2\pi}\lambda^{-\Delta+\frac{1}{2}}}{\pi 2^{-\Delta}}
(\Delta-1)!\,B_-
\end{equation}
\begin{multline}
B(\omega,\mathbf{s}) = \frac{2\sqrt{2\pi}\lambda^{\Delta+\frac{1}{2}}}
{\pi 2^\Delta\Gamma(\Delta+1)}\ln\left(-\frac{\lambda}{2}\right)\,B_- \\
-\frac{\sqrt{2\pi}\lambda^{\Delta+\frac{1}{2}}}{\pi 2^\Delta(\Delta)!}
\left[\psi(1)+\psi(\Delta+1)\right]\,B_-
+\frac{\sqrt{2\pi}\lambda^{\Delta+\frac{1}{2}}}{2^\Delta \Gamma(\Delta+1)}\,B_+ \ ,
\end{multline}
we again have:
\begin{equation}
\left.\frac{\pdzt\phi_{\omega,\mathbf{s}}}{\phi_{\omega,\mathbf{s}}}\right|_{\zt=\epsilon}
\approx (2\Delta)\frac{B(\omega,\mathbf{s})}{A(\omega,\mathbf{s})}\epsilon^{2\Delta-1} \ .
\end{equation}
And finally after putting this into Equation~\ref{eq:scalarcftcorrelatorinzt}, and again making
use of the symmetry properties of the correlator we obtain:
\begin{empheq}[box=\fbox]{multline}
\label{eq:asymptcftcorrelatorscalaroddd}
G^R(\omega,\mathbf{s}) \approx
4C_s R^{d-2}\frac{(-1)^{\Delta+1}}{\Gamma^2(\Delta)}
\left(\frac{i\omega}{2}\right)^{d-1} \\
\left[2\pi i\frac{\e^{2i\theta_+}}{\e^{2i\theta_+}-2}
- 2\pi i\frac{\e^{2i\overline{\theta_+}}}{\e^{2i\overline{\theta_+}}-2}
+ 2\ln\left(\frac{i\lambda}{2}\right)\right] \ ,
\end{empheq}
where:
\begin{equation}
\overline{\theta_+}(\lambda) \equiv -\theta_+^*(-\lambda^*) = \lambda\zt_0^*
+\frac{\pi}{4}(d+1) \ .
\end{equation}

\end{enumerate}

\subsubsection{Calculation of Asymptotic Correlators for the Vector Case}
\label{subsubsec:cftasymptoticcorrelatorexpressionsvector}

The calculation of asymptotic expressions for the vector correlators is similar to the scalar case.
A detailed derivation is given in Appendix~\ref{app:cftasymptoticcorrelatorexpressionsvector}, and
here we quote the final results.

For the longitudinal vector mode, we define:
\begin{equation}
\Delta \equiv \frac{d-5}{2} \ .
\end{equation}
We distinguish between two possible cases\footnote{Note that for the case of vector gauge field
perturbations with $ d=4 $, the effective potential for both modes is $ V(z) = q^2 \gz $, and has a pole
of degree $ -\frac{3}{2} $ at $ \zt\to\zt_0 $. In this case the method applied here can't be used since
the equation is not of the Bessel type around the pole $ \zt_0 $, and since $ q $ obviously can't be
neglected (neglecting it results in an an expression with no poles). This case therefore requires
a separate and more complete treatment, which shall not be investigated here. This section will
therefore focus on the case of $ d\geq 5 $.}:
\begin{enumerate}

\item $ d $ is even. In this case we obtain the asymptotic expression:
\begin{empheq}[box=\fbox]{multline}
\label{eq:asymptcftcorrelatorlongvectorevend}
G_{tt}^R(\omega,\mathbf{s})
\approx \frac{4C_v}{d-5}R^{d-4}
(L_\mathbf{s}^{FT})^2\frac{\Gamma(-\Delta)}{\Gamma(\Delta)}\left(\frac{i\omega}{2}\right)^{d-5}\\
\frac{\e^{2i\theta_-}+2\cos\left(\frac{\pi}{d-2}\right)}
{\e^{2i\theta_-}-2\cos\left(\frac{\pi}{d-2}\right)}
\frac{\e^{2i\overline{\theta_-}}+2\cos\left(\frac{\pi}{d-2}\right)}
{\e^{2i\overline{\theta_-}}-2\cos\left(\frac{\pi}{d-2}\right)} \ ,
\end{empheq}
where:
\begin{align}
\theta_- &\equiv \lambda\zt_0 + \frac{\pi}{4}(d-7) \\
\overline{\theta_-}(\lambda) &\equiv -\theta_-^*(-\lambda^*)
= \lambda\zt_0^*-\frac{\pi}{4}(d-7) \ .
\end{align}

\item $ d $ is odd. In this case we obtain the asymptotic expression:
\begin{empheq}[box=\fbox]{multline}
\label{eq:asymptcftcorrelatorlongvectoroddd}
G_{tt}^R(\omega,\mathbf{s})
\approx 2C_v R^{d-4}(L_\mathbf{s}^{FT})^2\frac{(-1)^{\Delta+1}}{\Gamma^2(\Delta+1)}
\left(\frac{i\omega}{2}\right)^{d-5} \\
\left[2\pi i\frac{\e^{2i\theta_+}}{\e^{2i\theta_+} + 2\cos\left(\frac{\pi}{d-2}\right)}
- 2\pi i\frac{\e^{2i\overline{\theta_+}}}{\e^{2i\overline{\theta_+}} + 2\cos\left(\frac{\pi}{d-2}\right)}
+ 2\ln\left(\frac{i\lambda}{2}\right)\right] \ ,
\end{empheq}
where:
\begin{align}
\theta_+ &\equiv \lambda\zt_0 - \frac{\pi}{4}(d-3) \\
\overline{\theta_+}(\lambda) &\equiv -\theta_+^*(-\lambda*)
= \lambda\zt_0^*+\frac{\pi}{4}(d-3) \ .
\end{align}

\end{enumerate}
The other components of the longitudinal mode correlator can be calculated from
Equations~\ref{eq:longvectorcftcorrelator2} and~\ref{eq:longvectorcftcorrelator3}.

For the transverse vector mode, we define:
\begin{equation}
\Delta \equiv \frac{d-3}{2} \ .
\end{equation}
We again distinguish between two possible cases:
\begin{enumerate}

\item $ d $ is even. In this case we obtain the asymptotic expression:
\begin{empheq}[box=\fbox]{multline}
\label{eq:asymptcftcorrelatortransvectorevend}
G_{\bot\bot}^R(\omega,\mathbf{v})
\approx 4(d-3)C_v R^{d-4}
\frac{\Gamma(-\Delta)}{\Gamma(\Delta)}\left(\frac{i\omega}{2}\right)^{d-3}\\
\frac{\e^{2i\theta_-}-2\cos\left(\frac{\pi}{d-2}\right)}
{\e^{2i\theta_-}+2\cos\left(\frac{\pi}{d-2}\right)}
\frac{\e^{2i\overline{\theta_-}}-2\cos\left(\frac{\pi}{d-2}\right)}
{\e^{2i\overline{\theta_-}}+2\cos\left(\frac{\pi}{d-2}\right)} \ ,
\end{empheq}
where:
\begin{align}
\theta_- &\equiv \lambda\zt_0 + \frac{\pi}{4}(d-5) \\
\overline{\theta_-}(\lambda) &\equiv -\theta_-^*(-\lambda*)
= \lambda\zt_0^*-\frac{\pi}{4}(d-5) \ .
\end{align}

\item $ d $ is odd. In this case we obtain the asymptotic expression:
\begin{empheq}[box=\fbox]{multline}
\label{eq:asymptcftcorrelatortransvectoroddd}
G_{\bot\bot}^R(\omega,\mathbf{v})
\approx 8C_v R^{d-4}
\frac{(-1)^{\Delta+1}}{\Gamma^2(\Delta)}
\left(\frac{i\omega}{2}\right)^{d-3}\\
\left[2\pi i\frac{\e^{2i\theta_+}}{\e^{2i\theta_+} - 2\cos\left(\frac{\pi}{d-2}\right)}
- 2\pi i\frac{\e^{2i\overline{\theta_+}}}{\e^{2i\overline{\theta_+}} - 2\cos\left(\frac{\pi}{d-2}\right)}
+ 2\ln\left(\frac{i\lambda}{2}\right)\right] \ ,
\end{empheq}
where:
\begin{align}
\theta_+ &\equiv \lambda\zt_0 - \frac{\pi}{4}(d-1) \\
\overline{\theta_+}(\lambda) &\equiv -\theta_+^*(-\lambda*)
= \lambda\zt_0^*+\frac{\pi}{4}(d-1) \ .
\end{align}

\end{enumerate}

Notice that, taking into account the differences in the definition of $ \Delta $ and $ \theta_\pm $
between the longitudinal and the transverse perturbation modes:
\begin{align}
\Delta^{\text{long}} &= \Delta^{\text{trans}} - 1 \\
\theta_\pm^{\text{long}} &= \theta_\pm^{\text{trans}} \pm \frac{\pi}{2} \ ,
\end{align}
the asymptotic expressions for $ G_{\bot\bot}^R(\omega,\mathbf{v}) $ and
$ G_{\|\|}^R(\omega,\mathbf{s}) $ are identical. This is expected and serves as a check for our
calculations, since in this limit we have neglected the spatial mode $ L_{\mathbf{s}/\mathbf{v}}^{FT} $,
so that the differences between the longitudinal and the transverse modes are negligible.

\subsection{Numerical Calculation of Correlators}

In the following we present some results of exact numerical calculations of the gauge theory
retarded correlation functions for several cases, and compare them to the asymptotic analytical
expressions (as discussed in Subsection~\ref{subsec:cftasymptoticcorrelatorexpressions}).
The correlation functions have been calculated using the method outlined in Appendix~\ref{app:numericalmethods}.

The following calculations were made for a degree of $ N=200 $ and $ z_0 = 0.6 $ (see
Appendix~\ref{app:numericalmethods} for details):

\begin{itemize}

\item The case of scalar perturbation in $ d=4 $ bulk dimensions, with $ K=1 $ (spherical boundary
topology) and $ q_\mathbf{s} = 0 $ (s-wave). The results are given in
Figures~\ref{fig:correlationfunctionresultspolesscalard4K1q0},
~\ref{fig:correlationfunctionresultscirclescalard4K1q0R20},
~\ref{fig:correlationfunctionresultscontourscalard4K1q0}
and~\ref{fig:correlationfunctionresultsspectralscalard4K1q0}.

\begin{figure}[htbp]
\centering
\subfloat[
]{
\label{fig:correlationfunctionresultspolesscalard4K1q0}
\includegraphics[width=0.4\textwidth]{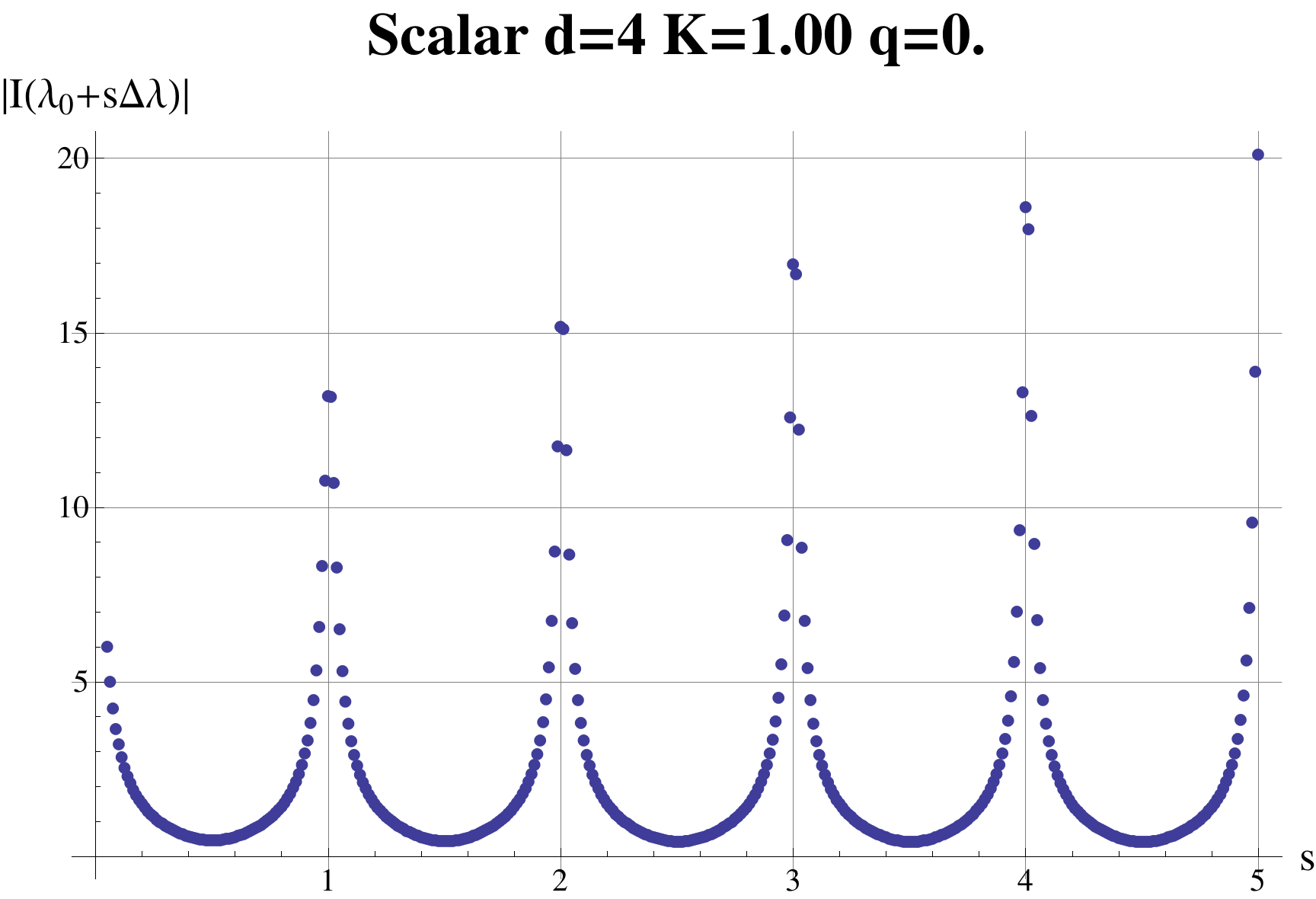}
}
\subfloat[
]{
\label{fig:correlationfunctionresultscirclescalard4K1q0R20}
\includegraphics[width=0.56\textwidth]{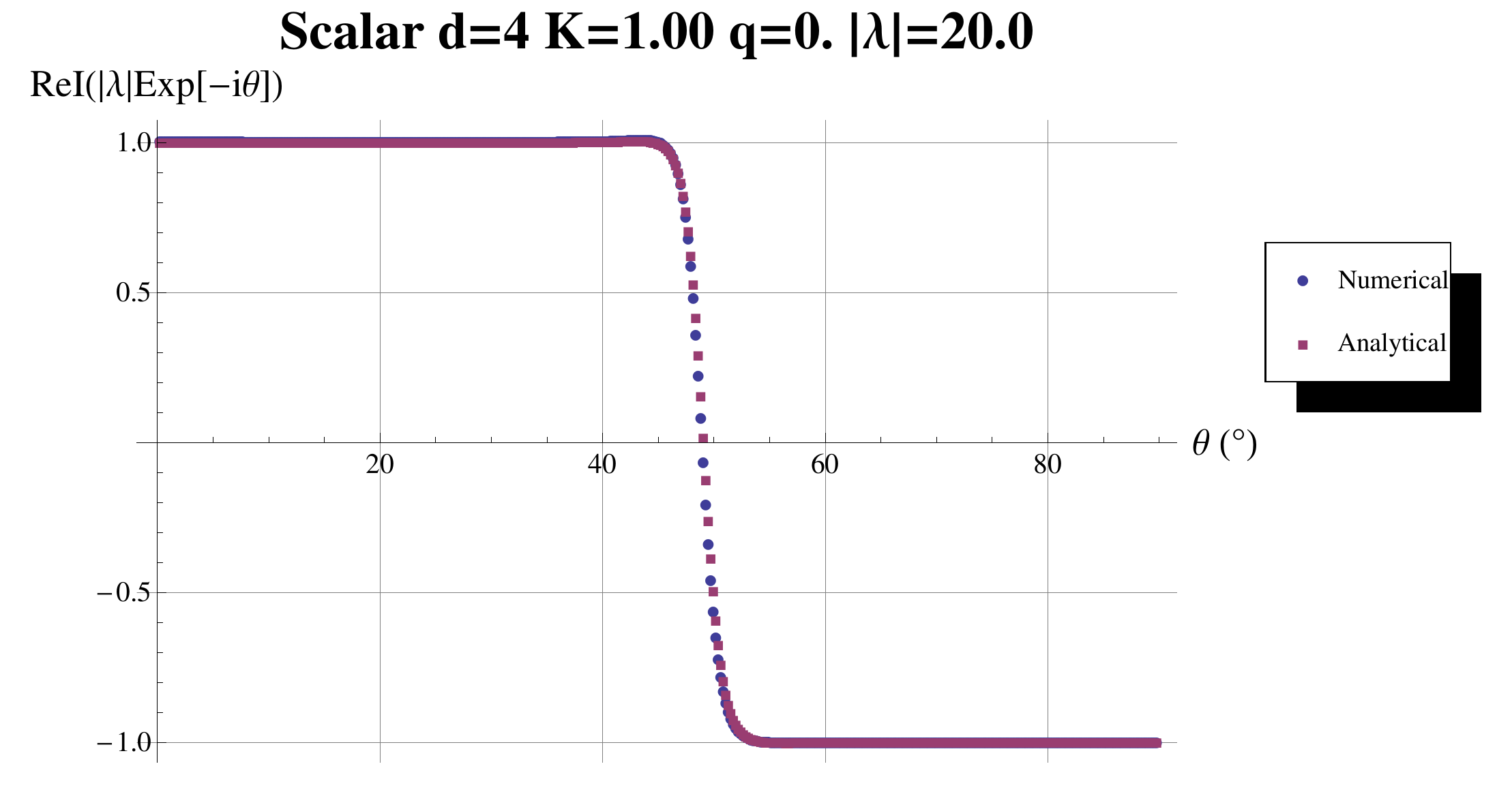}
}\\
\subfloat[
]{
\label{fig:correlationfunctionresultscontourscalard4K1q0}
\includegraphics[width=0.56\textwidth]{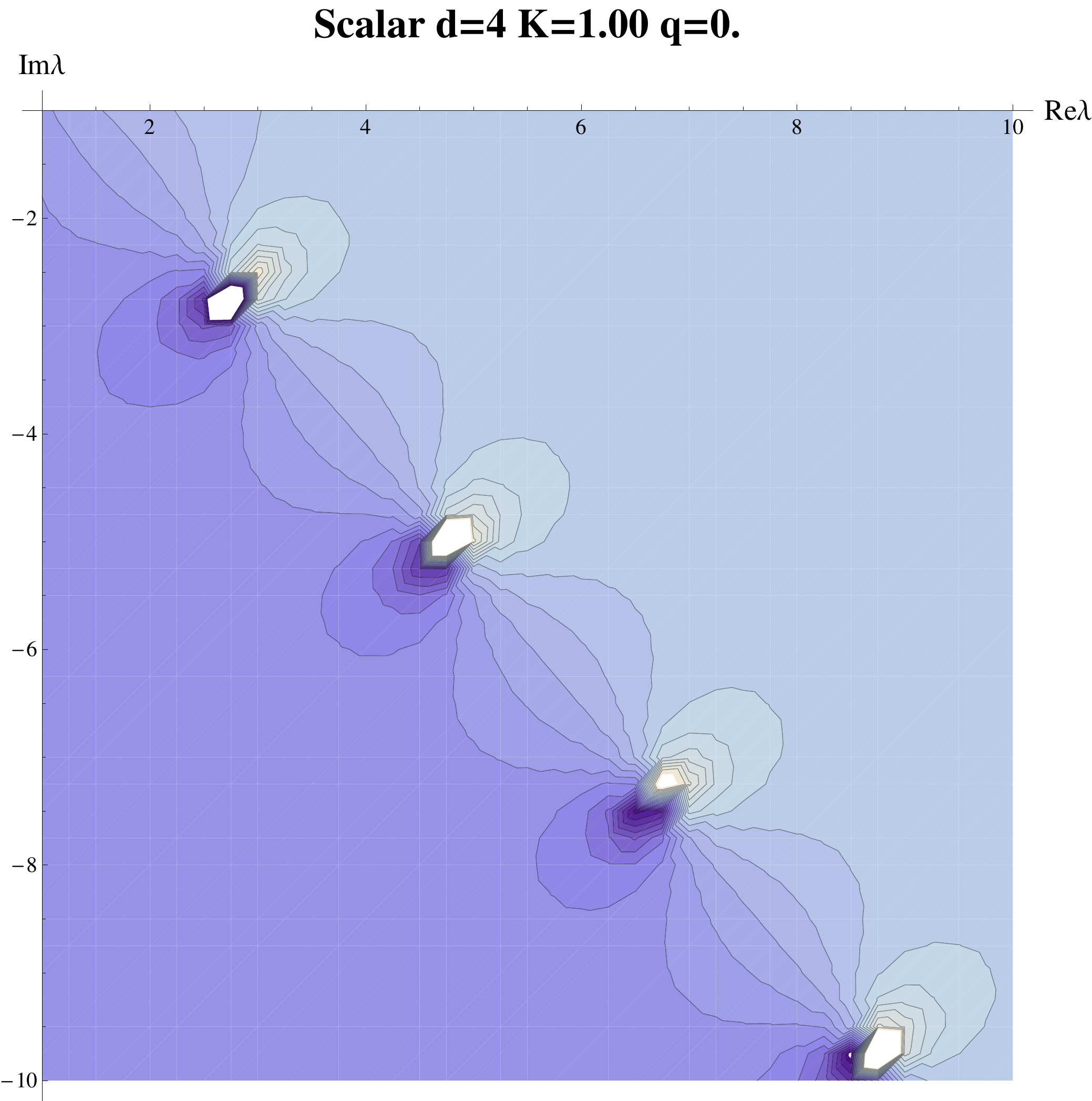}
}
\subfloat[
]{
\label{fig:correlationfunctionresultsspectralscalard4K1q0}
\includegraphics[width=0.4\textwidth]{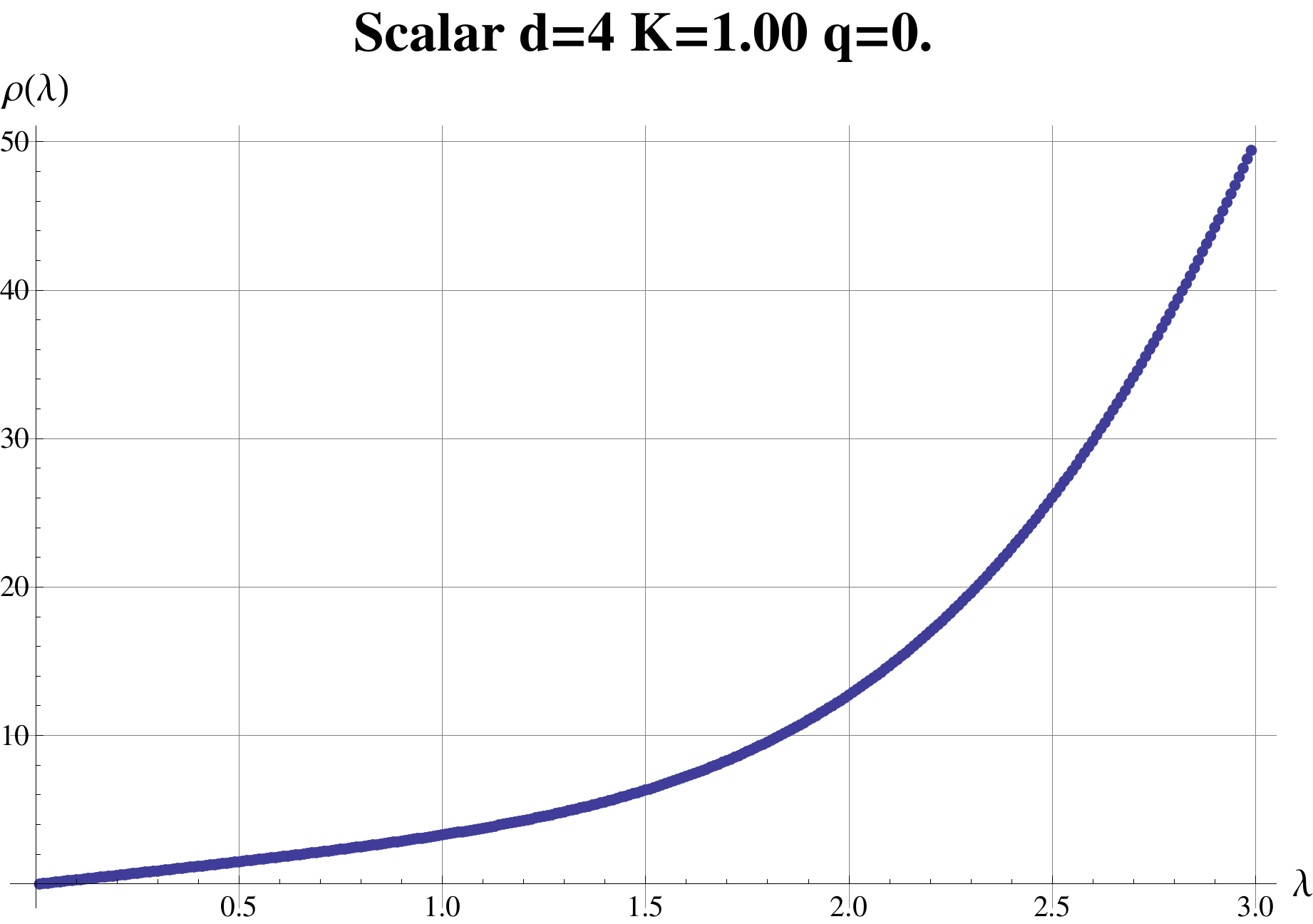}
}
\caption{Numerically calculated ``normalized'' retarded correlation function for the case of scalar
perturbations with $ d=4 $,$ K=1 $ and $ q_\mathbf{s} = 0 $. The plots show:
\protect\subref{fig:correlationfunctionresultspolesscalard4K1q0}
The absolute value of the
``normalized'' correlation function along the analytically calculated asymptotic line of QNM frequencies,
given by the parameters: $ \lambda_0 = -2.7206 - 2.7205 i \quad \Delta\lambda = -1.9691 - 2.3502 i $,
\protect\subref{fig:correlationfunctionresultscirclescalard4K1q0R20}
the real parts of both the
numerically calculated and the analytical asymptotic ``normalized'' correlation functions on a circle in
the complex plane of radius $ |\lambda| = 20 $,
\protect\subref{fig:correlationfunctionresultscontourscalard4K1q0}
a contour plot of the real part
of the numerically calculated ``normalized'' correlation function in the 4-th quadrant of the complex plane,
\protect\subref{fig:correlationfunctionresultsspectralscalard4K1q0}~the spectral function associated with $ G^R $ along the real frequency axis, in units of 
$ -C_s\frac{r_+^{d-1}}{R^d} $.
}
\end{figure}

%
%
%

%
%
%

\item The case of longitudinal vector perturbations in $ d=5 $ bulk dimensions, with $ K=0 $
(flat boundary topology) and $ q_\mathbf{s} = 0.6 $. The results are given in
Figures~\ref{fig:correlationfunctionresultspoleslongvectord5K0q06},
~\ref{fig:correlationfunctionresultscirclelongvectord5K0q06R20},
~\ref{fig:correlationfunctionresultscontourlongvectord5K0q06}
and~\ref{fig:correlationfunctionresultsspectrallongvectord5K0q06}.

\begin{figure}[htbp]
\centering
\subfloat[]{
\label{fig:correlationfunctionresultspoleslongvectord5K0q06}
\includegraphics[width=0.4\textwidth]{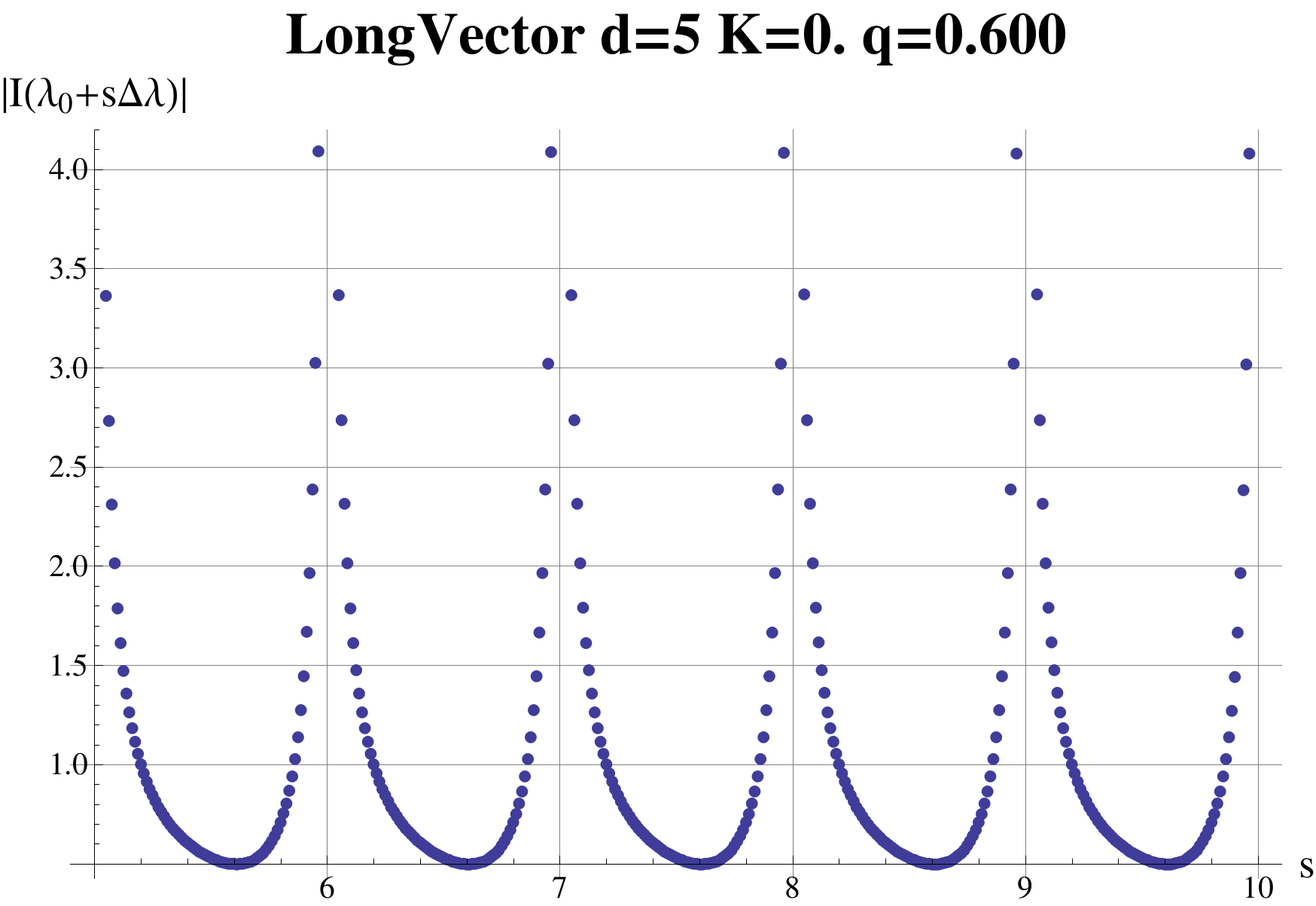}
}
\subfloat[]{
\label{fig:correlationfunctionresultscirclelongvectord5K0q06R20}
\includegraphics[width=0.56\textwidth]{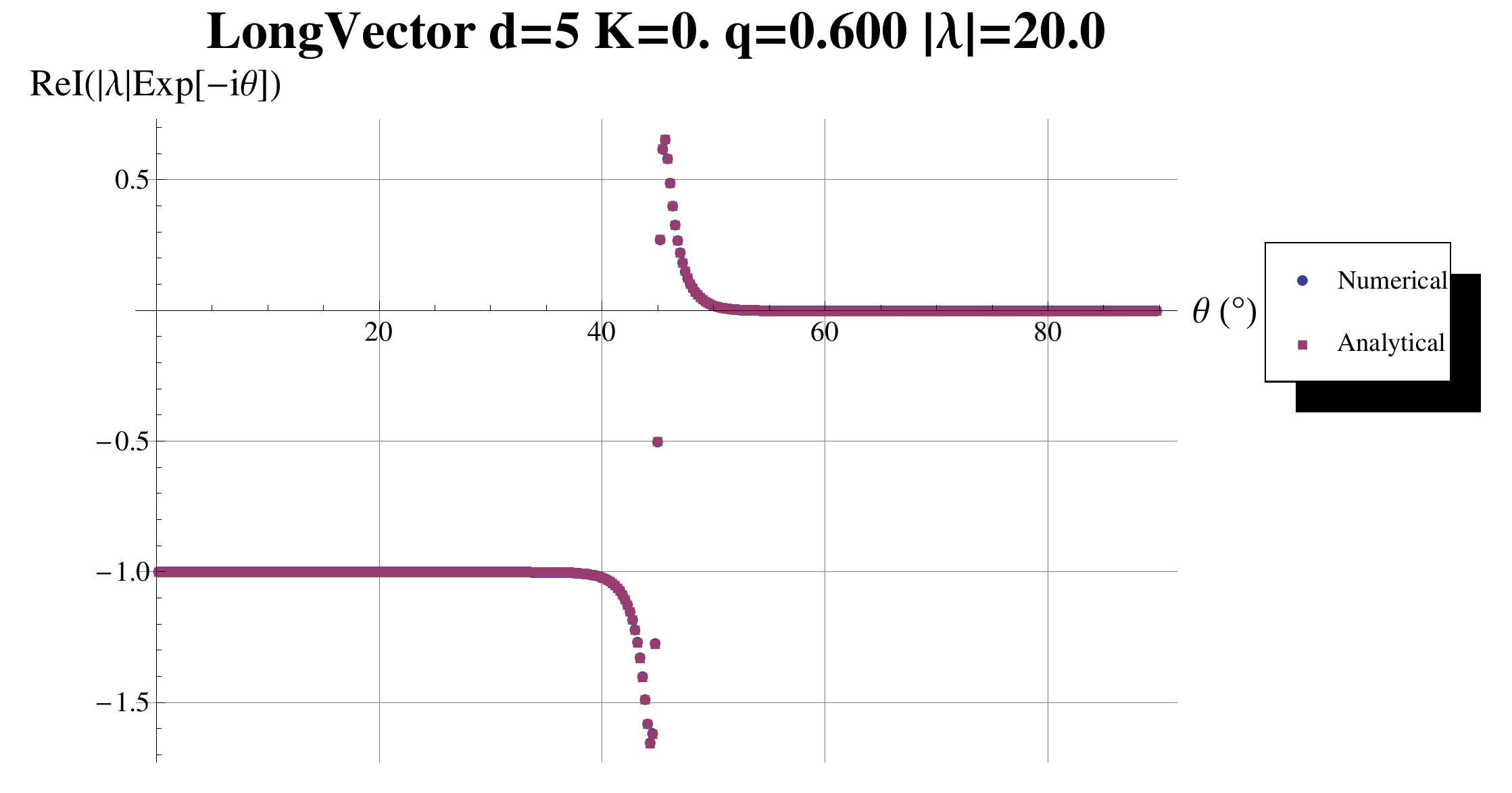}
}\\
\subfloat[]{
\label{fig:correlationfunctionresultscontourlongvectord5K0q06}
\includegraphics[width=0.56\textwidth]{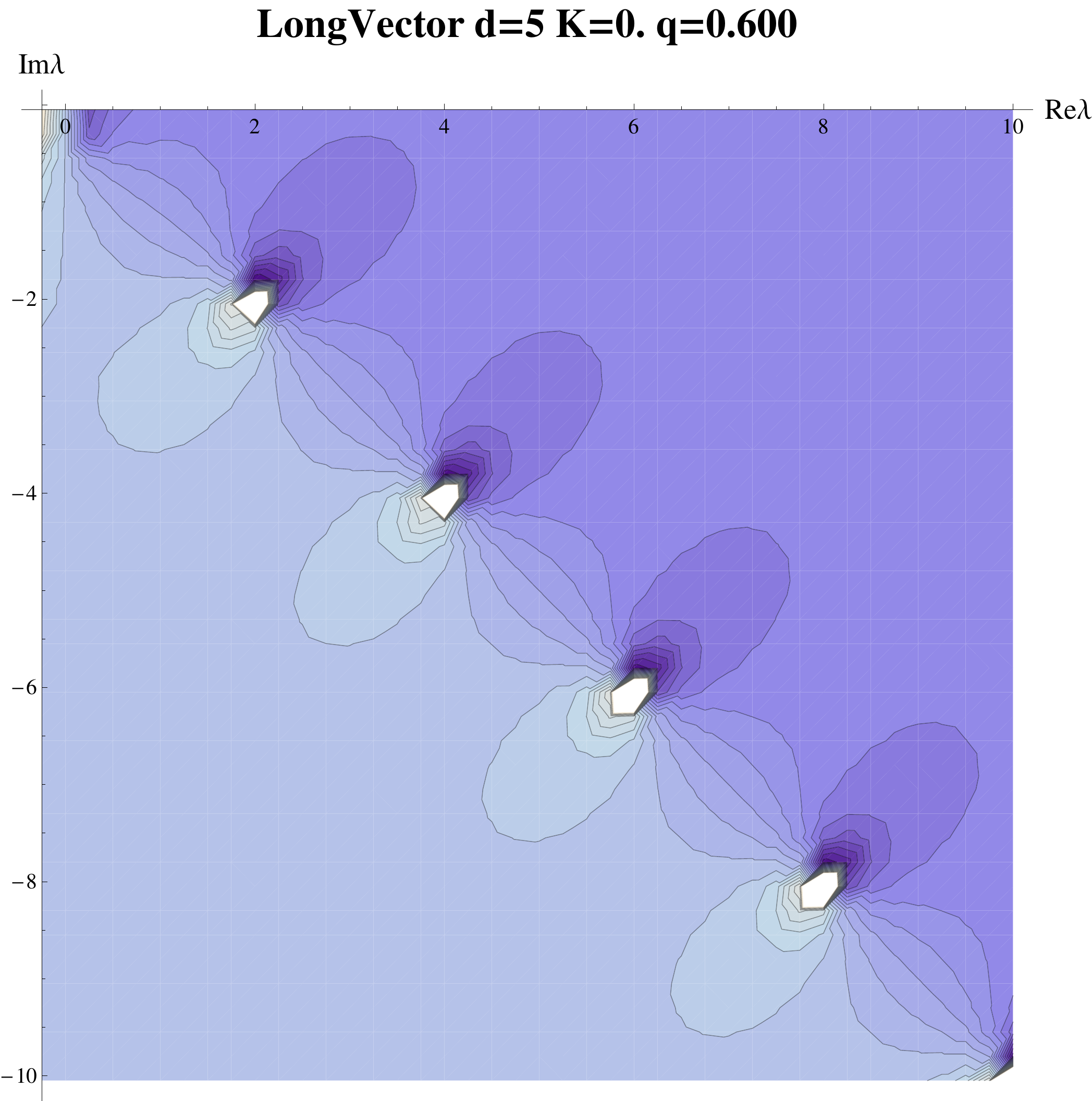}
}
\subfloat[]{
\label{fig:correlationfunctionresultsspectrallongvectord5K0q06}
\includegraphics[width=0.4\textwidth]{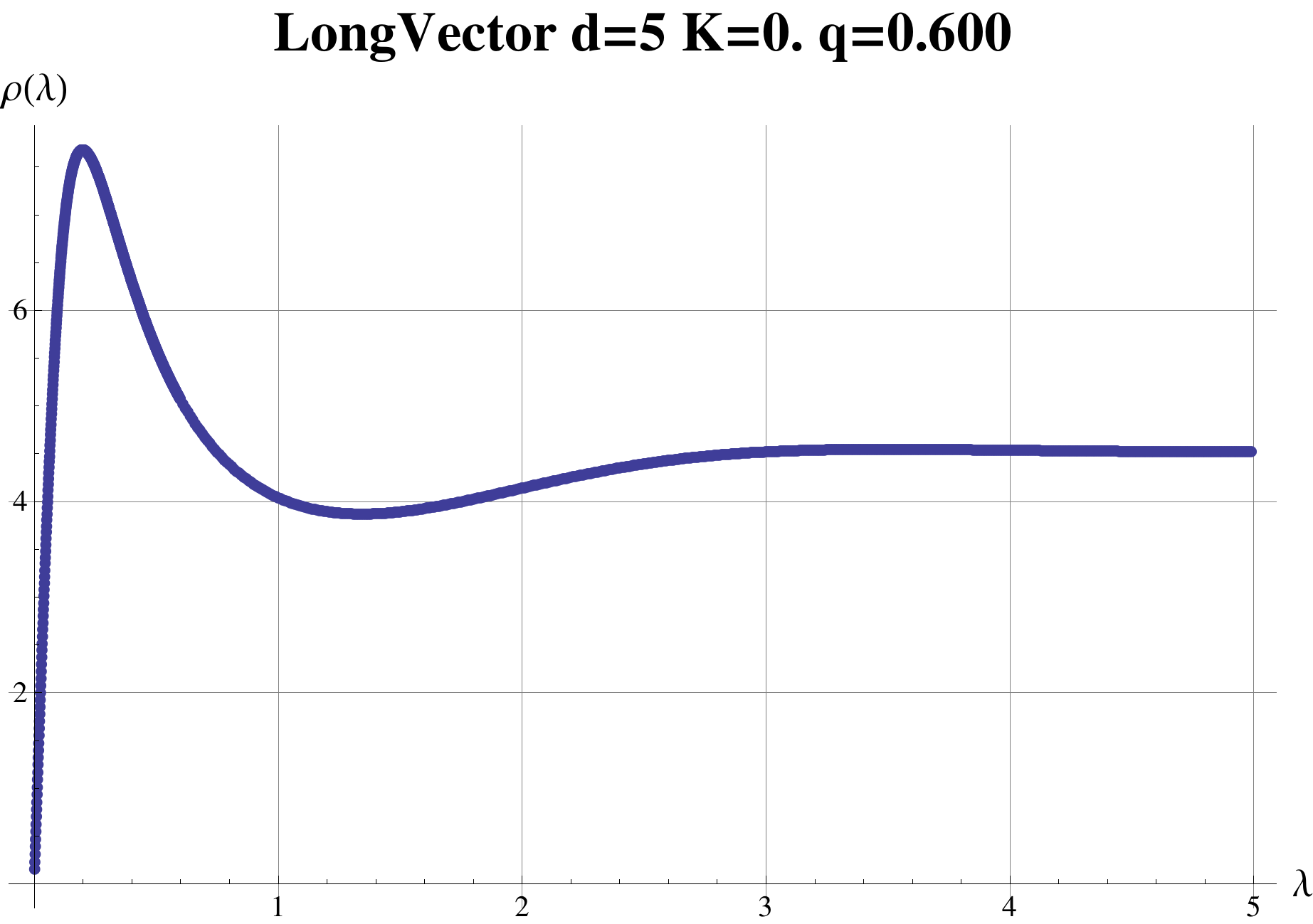}
}
\caption{Numerically calculated ``normalized'' retarded correlation function for the case of longitudinal
vector perturbations with $ d=5 $,$ K=0 $ and $ q_\mathbf{s} = 0.6 $. The plots show:
\protect\subref{fig:correlationfunctionresultspoleslongvectord5K0q06}
The absolute value of
the ``normalized'' correlation function along the analytically calculated asymptotic line of QNM
frequencies, given by the parameters: $ \lambda_0 = 0 \quad \Delta\lambda = -2 - 2 i $,
\protect\subref{fig:correlationfunctionresultscirclelongvectord5K0q06R20}
the real parts of
both the numerically calculated and the analytical asymptotic ``normalized'' correlation functions on a
circle in the complex plane of radius $ |\lambda| = 20 $,
\protect\subref{fig:correlationfunctionresultscontourlongvectord5K0q06}
a contour plot of
the real part of the numerically calculated ``normalized'' correlation function in the 4-th quadrant of
the complex plane,
\protect\subref{fig:correlationfunctionresultsspectrallongvectord5K0q06}~the spectral function associated with $ G_{tt}^R $ along the real frequency axis, in units of 
$ -C_v\frac{r_+^{d-3}}{R^{d-2}} $.}
\end{figure}

%
%
%
%
%

\item The case of transverse vector perturbations in $ d=5 $ bulk dimensions, with $ K=-0.5 $
(hyperbolic boundary topology) and $ q_\mathbf{v} = 1 $. The results are given in
Figures~\ref{fig:correlationfunctionresultspolestransvectord5Km05q1},
~\ref{fig:correlationfunctionresultscircletransvectord5Km05q1R20},
~\ref{fig:correlationfunctionresultscontourtransvectord5Km05q1}
and~\ref{fig:correlationfunctionresultsspectraltransvectord5Km05q1}.

\begin{figure}[htbp]
\centering
\subfloat[]{
\label{fig:correlationfunctionresultspolestransvectord5Km05q1}
\includegraphics[width=0.4\textwidth]{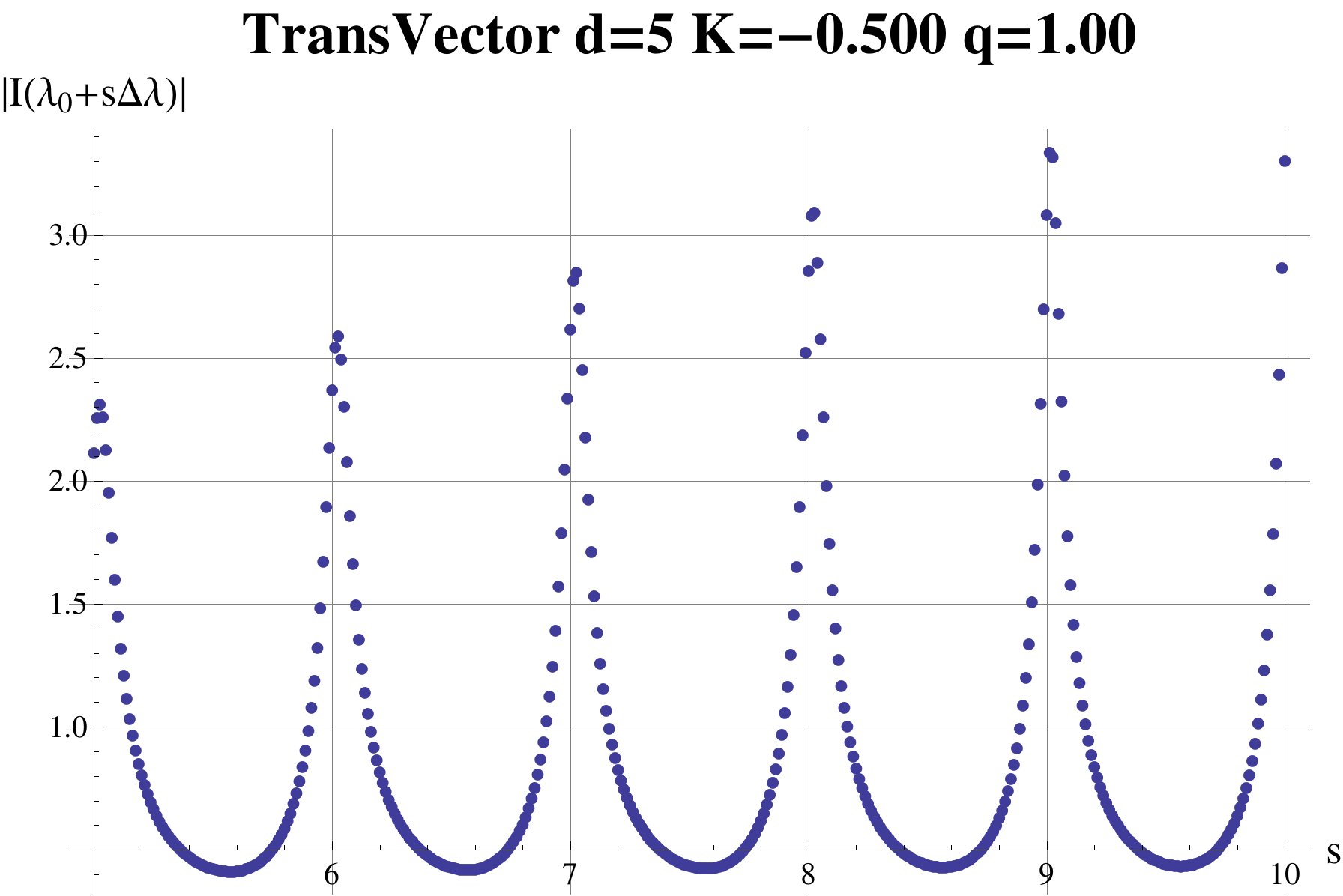}
}
\subfloat[]{
\label{fig:correlationfunctionresultscircletransvectord5Km05q1R20}
\includegraphics[width=0.56\textwidth]{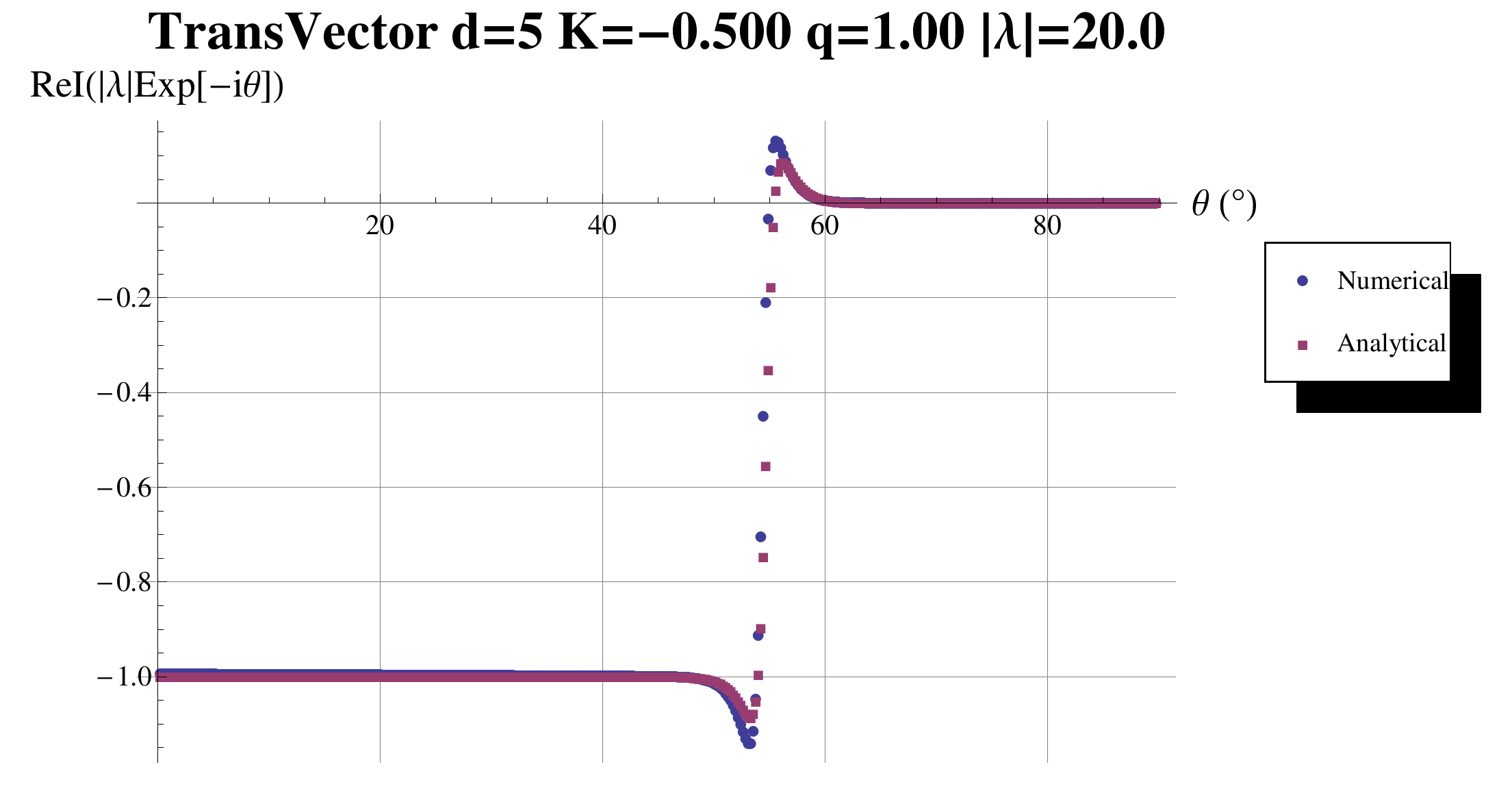}
}\\
\subfloat[]{
\label{fig:correlationfunctionresultscontourtransvectord5Km05q1}
\includegraphics[width=0.56\textwidth]{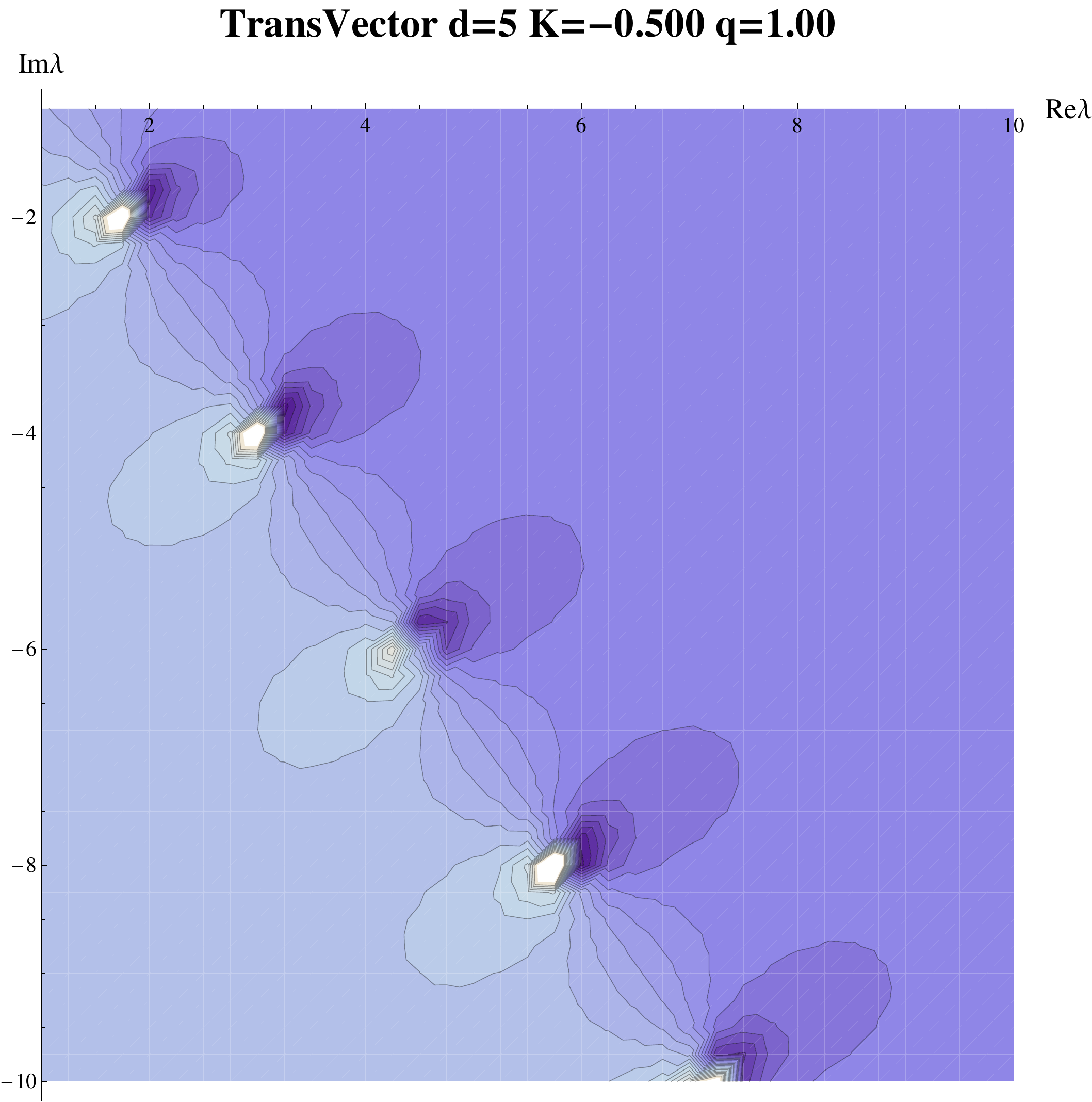}
}
\subfloat[]{
\label{fig:correlationfunctionresultsspectraltransvectord5Km05q1}
\includegraphics[width=0.4\textwidth]{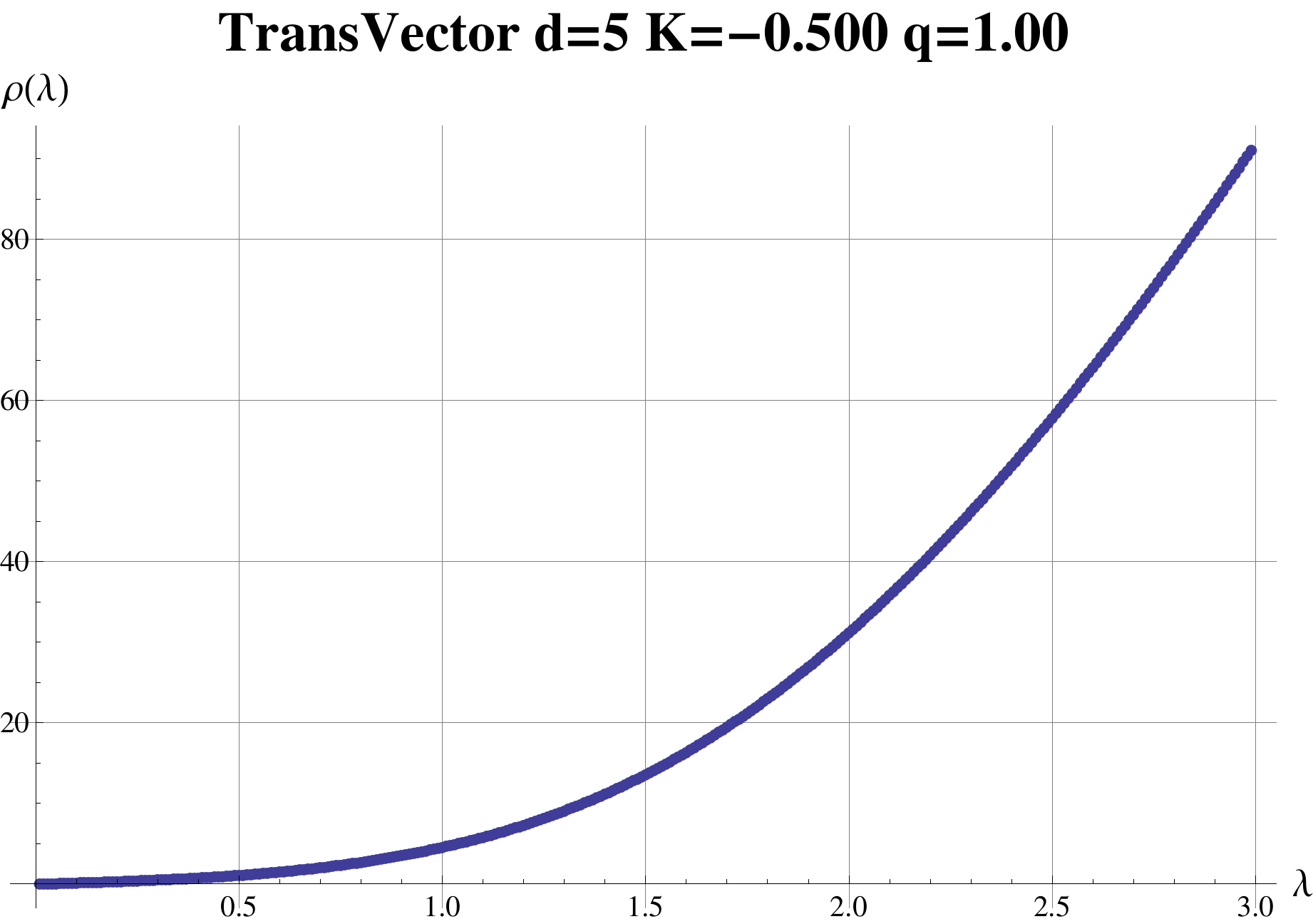}
}
\caption{Numerically calculated ``normalized'' retarded correlation function for the case of transverse
vector perturbations with $ d=5 $,$ K=-0.5 $ and $ q_\mathbf{v} = 1 $. The plots show:
\protect\subref{fig:correlationfunctionresultspolestransvectord5Km05q1}
The absolute value
of the ``normalized'' correlation function along the analytically calculated asymptotic line of QNM
frequencies, given by the parameters:
$ \lambda_0 = -1.4142 - 2.0000 i \quad \Delta\lambda = -1.4142 - 2.0000 i $,
\protect\subref{fig:correlationfunctionresultscircletransvectord5Km05q1R20}
the real parts of
both the numerically calculated and the analytical asymptotic ``normalized'' correlation functions on a
circle in the complex plane of radius $ |\lambda| = 20 $,
\protect\subref{fig:correlationfunctionresultscontourtransvectord5Km05q1}
a contour plot of
the real part of the numerically calculated ``normalized'' correlation function in the 4-th quadrant of
the complex plane,
\protect\subref{fig:correlationfunctionresultsspectraltransvectord5Km05q1}~the spectral function associated with $ G_{\bot\bot}^R $ along the real frequency axis, in units of 
$ -C_v\frac{r_+^{d-3}}{R^{d-2}} $.}
\end{figure}

%
%
%
%
%

\end{itemize}

For the sake of better visual representation of the results, we define the
``normalized'' retarded correlators $ I(\lambda) $ by ``normalizing out'' the power and logarithmic
behaviour of the correlators, keeping only the part of the
correlators that behaves as $ O(1) $ for $ |\lambda| \to \infty $\footnote{See 
Appendix~\ref{app:numericalmethods} for a more precise definition.}. These ``normalized'' correlators
have poles at the QNM frequencies, and approach a constant on each side of the QNM frequencies line for
frequencies far enough away from that line.

Figures~\ref{fig:correlationfunctionresultspolesscalard4K1q0},
~\ref{fig:correlationfunctionresultspoleslongvectord5K0q06}
and~\ref{fig:correlationfunctionresultspolestransvectord5Km05q1} show the value of
$ |I(\lambda_0 + s\Delta\lambda)| $ as a function of $ s $, where
$ \lambda_0 $ and $ \Delta\lambda $ are the analytical asymptotic parameters of the QNM spectrum
as calculated in Subsection~\ref{subsec:qnmasymoptotics} and $ s $ is a positive, real parameter. As
expected, we find the poles of the correlation function approximately at integer values of $ s $.

Figures~\ref{fig:correlationfunctionresultscirclescalard4K1q0R20},
~\ref{fig:correlationfunctionresultscirclelongvectord5K0q06R20}
and~\ref{fig:correlationfunctionresultscircletransvectord5Km05q1R20} show the value of
$ \re I\left(\lambda_R\e^{-i\theta}\right) $ as a function of $ \theta $, where $ \lambda_R $
is some constant radius in the frequency complex plane (in this case $ \lambda_R = 20 $). As expected,
the numerically calculated values are close to the ones of the analytical asymptotic expressions
calculated in Subsection~\ref{subsec:cftasymptoticcorrelatorexpressions}.

Figures~\ref{fig:correlationfunctionresultscontourscalard4K1q0},
~\ref{fig:correlationfunctionresultscontourlongvectord5K0q06}
and~\ref{fig:correlationfunctionresultscontourtransvectord5Km05q1} show a contour plot of
$ \re I(\lambda) $ in the 4-th quadrant of the complex frequency plane. Both the poles at the QNM
frequencies and the step-function-like behaviour of the ``normalized'' correlation function can be
clearly seen in the plots.

Figures~\ref{fig:correlationfunctionresultsspectralscalard4K1q0},
~\ref{fig:correlationfunctionresultsspectrallongvectord5K0q06}
and~\ref{fig:correlationfunctionresultsspectraltransvectord5Km05q1} 
show the values of the appropriate spectral functions on the real frequency axis
($ \rho(\lambda) $ in the case of scalar perturbations, $ \rho_{tt}(\lambda) $ in the case of
longitudinal vector perturbations and $ \rho_{\bot\bot}(\lambda) $ in the case of transverse vector
perturbations). At large frequency values, these functions approach the analytical asymptotic expressions
and are proportional to $ \lambda^s $ (where $ s $ is the appropriate exponent for the perturbation
type), as expected from conformal symmetry considerations. In the longitudinal vector case 
(Figure~\ref{fig:correlationfunctionresultsspectrallongvectord5K0q06}),
the spectral function oscillates at lower frequencies, due to the existence of the hydrodynamic 
diffusion pole.

\clearpage

\section{Discussion}
\label{sec:discussion}

In this work we studied some of the properties of the retarded correlation function of composite
operators of finite temperature gauge field theories with dual gravity descriptions in the large
coupling limit. Of specific interest was the dependence of these
properties on the topology of the space on which the gauge theory is defined, and its temperature.
This goal was achieved by performing exact analytical calculations for specific cases where it was
possible (the hyperbolic case with $ T=T_c $), by performing numerical calculations and by obtaining
approximate expressions for the large frequency limit ($ \frac{L_s^{FT}}{T} = \text{fixed} $ and
$ \frac{\omega}{T} \gg 1 $).

In the large frequency limit we saw that in all cases (all topologies, temperatures and perturbation
types) the correlation functions exhibit the same kind of asymptotic pole structure, with a constant
``gap'' between two consecutive poles. Moreover, we saw that asymptotically these poles are produced by
Bose-Einstein-like factors in the correlation functions, with some complex parameter multiplying the
frequency.

We have also seen that the spatial topology on which the gauge theory is defined is indeed noticeable
in the properties of the correlation function and its poles (corresponding to quasinormal modes in the
dual bulk spacetime), in the case where $ \frac{\sqrt{|R_\Omega^{FT}|}}{T} = \text{fixed} $. In
particular, the asymptotic ``slope'' of the correlator poles encodes information about the spatial
topology and the temperature, but is ``universal'' otherwise - it doesn't depend on the type of operators
considered.
In the spherical (positive curvature) and flat (zero curvature) cases, this slope always has a real part
and an imaginary part. In the spherical case, it depends on the relation between the spatial curvature
and the temperature, while in the flat case it's constant (as expected, since in this case the
temperature is the only scale in the theory, and the correlator poles are necessarily proportional
to it).
In the hyperbolic (negative curvature) case, the slope has a real part for temperatures above $ T_c $,
which goes to $ 0 $ in a non-smooth way for $ T\to T_c $ (In $ d = 4,5 $ dimensions the real part is
also $ 0 $ for temperatures below $ T_c $).
At $ T=T_c $ the slope is completely imaginary, and exact analytical expressions can be found for the
correlation functions and their poles. Interestingly, the poles ``gap'' in this case, and indeed the
general form of the correlation functions, is exactly what one would expect to find for the retarded
correlation function of composite operators in a free thermal field theory (the poles ``gap'' in this
case corresponds to the Matsubara frequency ``gap'' - see~\cite{Hartnoll:2005ju}).

Several open issues are left for further investigation:
\begin{itemize}

\item This work has dealt with uncharged bulk black hole solutions, but we can also expand the scope
and discuss charged black holes, that correspond to gauge theories with non-zero chemical potentials
(see \cite{Son:2006em}).
The background of the perturbations in such solutions contains both a black
hole metric and a non-zero gauge field. We can then ask - how does the charge change the asymptotic
properties of the gauge theory correlation functions and their dependence on the spatial topology?

\item As noted in Subsubsection~\ref{subsubsec:cftasymptoticcorrelatorexpressionsvector}, the method we have
employed for calculating the asymptotic expressions for the correlation functions doesn't work as is
for the vector perturbation modes in $ d=4 $ bulk dimensions. This case requires a different treatment
using complex WKB analysis, although numerical calculations show that our conclusions regarding the
asymptotic ``slope'' of the correlator poles remain true in this case as well.

\item While the asymptotic structure of the correlators (and their poles) has a geometrical meaning
in the gravity (bulk) side, it is unclear if it has an interpretation in the strongly coupled dual
field theory. Weak coupling calculations don't seem to exhibit such a structure (see
~\cite{Hartnoll:2005ju}). The asymptotic poles ``gap'' in the large coupling limit depends only on the
spatial topology and the temperature, suggesting it may have an interpretation involving them, but it
can't be trivially deduced from the symmetries of the theory alone.

\item Considering the properties of the correlation functions in the hyperbolic case for $ T\to T_c $,
does this temperature have any importance in the dual gauge theory? In~\cite{Shen:2007xk}, it is
suggested that in $ d=4 $, the behaviour of the QNM frequencies slope near this temperature is related
to the known phase transition between the topological (hyperbolic) black hole and the MTZ solution
(a hyperbolic black hole with scalar hair - see~\cite{Martinez:2004nb},~\cite{Myung:2008ze}).
It is not clear, however, what is the interpretation of
such a phase transition in the gauge theory side, and whether the argument extends to higher dimensions.

\end{itemize}

\section*{Acknowledgements}
The work is supported in part by the Israeli Science Foundation center
of excellence.

\appendix

\section{List of Notations and Definitions}
\label{app:notations}

Here we list for convenience some of the main notations used throughout this paper.

\begin{itemize}
\item $ d $ is the number of \emph{bulk} dimensions.
\item $ R $ is the AdS radius.
\item $ R_\Omega \equiv R^2 R_\Omega^{FT} $ is the scalar curvature of the $ \Omega_{d-2} $ manifold.
\item $ r_+ $ is the radius of the bulk black hole horizon.
\item $ k \equiv k_{FT}R^2 \equiv \frac{R_\Omega}{(d-2)(d-3)} $.
\item $ \rho \equiv \frac{r_+}{R} $.
\item $ K \equiv \frac{k}{\rho^2} $.
\item $ \omega $ is the frequency of a perturbation mode,
$ L_\mathbf{s/v}^{FT} \equiv \frac{L_\mathbf{s/v}}{R} $
is the Laplace operator eigenvalue of the perturbation mode on the dual field theory side, for scalar
or vector respectively.
\item
$
\lambda \equiv \frac{\omega r_+}{\rho^2} = \frac{\omega R}{\rho}
\qquad
q_\mathbf{s/v} \equiv \frac{L_\mathbf{s/v}}{\rho} = \frac{L_\mathbf{s/v}^{FT}R}{\rho}
$.
\item
$
T_C \equiv \frac{\sqrt{|k|}}{2\pi R} = \frac{1}{2\pi}\sqrt{\frac{|R_\Omega^{FT}|}{(d-2)(d-3)}}
$
(for the case of $ k<0 $).
\item
$
\tilde{g}(z) \equiv 1 + K(1-z)^2 - (1+K)(1-z)^{d-1}
$.
\item
$
\zt_0 = - \sum_{k=1}^{d-1} \gamma_k \ln(1-z_k)
$, where
$
\tilde{g}(z_k)=0
$
and
$
\gamma_k = \frac{1}{\tilde{g}'(z_k)}
$ for $ k = 1, \ldots ,d-1 $.
\item $ C_{s/v} $ is the normalization constant for the bulk action of the scalar/vector field
respectively.
\item
$
\Delta \equiv
\begin{cases}
\frac{d-1}{2} & \text{scalar perturbation,} \\
\frac{d-3}{2} & \text{transverse vector perturbation,} \\
\frac{d-5}{2} & \text{longitudinal vector perturbation.}
\end{cases}
$
\item
$
a,b(\lambda) \equiv
\begin{cases}
-\frac{1}{4}(d-3)-\frac{i\lambda}{2}\pm\frac{i}{2}\sqrt{q_\mathbf{s}^2 - \frac{1}{4}(d-3)^2} &
\text{scalar perturbation,} \\
-\frac{1}{4}(d-5)-\frac{i\lambda}{2}\pm\frac{i}{2}\sqrt{q_\mathbf{v}^2 - \frac{1}{4}(d-5)^2} &
\text{transverse vector perturbation,} \\
-\frac{1}{4}(d-7)-\frac{i\lambda}{2}\pm\frac{i}{2}\sqrt{q_\mathbf{s}^2 - \frac{1}{4}(d-3)^2} &
\text{longitudinal vector perturbation.}
\end{cases}
$\\
$
c(\lambda) \equiv 1-i\lambda
$.
\item
$
\theta_-(\lambda) \equiv
\begin{cases}
\lambda\zt_0 + \frac{\pi}{4}(d-3) & \text{scalar perturbation,} \\
\lambda\zt_0 + \frac{\pi}{4}(d-5) & \text{transverse vector perturbation,} \\
\lambda\zt_0 + \frac{\pi}{4}(d-7) & \text{longitudinal vector perturbation.}
\end{cases}
$ \\
$
\overline{\theta_-}(\lambda) \equiv -\theta_-^*(-\lambda^*) \equiv
\begin{cases}
\lambda\zt_0^* - \frac{\pi}{4}(d-3) & \text{scalar perturbation,} \\
\lambda\zt_0^* - \frac{\pi}{4}(d-5) & \text{transverse vector perturbation,} \\
\lambda\zt_0^* - \frac{\pi}{4}(d-7) & \text{longitudinal vector perturbation.}
\end{cases}
$
\item
$
\theta_+(\lambda) \equiv
\begin{cases}
\lambda\zt_0 - \frac{\pi}{4}(d+1) & \text{scalar perturbation,} \\
\lambda\zt_0 - \frac{\pi}{4}(d-1) & \text{transverse vector perturbation,} \\
\lambda\zt_0 - \frac{\pi}{4}(d-3) & \text{longitudinal vector perturbation.}
\end{cases}
$ \\
$
\overline{\theta_+}(\lambda) \equiv -\theta_+^*(-\lambda^*) \equiv
\begin{cases}
\lambda\zt_0^* + \frac{\pi}{4}(d+1) & \text{scalar perturbation,} \\
\lambda\zt_0^* + \frac{\pi}{4}(d-1) & \text{transverse vector perturbation,} \\
\lambda\zt_0^* + \frac{\pi}{4}(d-3) & \text{longitudinal vector perturbation.}
\end{cases}
$

\end{itemize}

\section{Thermodynamic Quantities}
\label{app:thermodynamicquant}

Given the black hole metric, one may calculate the usual thermodynamic quantities related to the
black hole. In the context of AdS/CFT, these quantities are related to the respective quantities
in the field theory side. Defining:
$
\rho \equiv \frac{r_+}{R}
$
and
$
K\equiv\frac{k}{\rho^2}
$,
we have for the temperature:
\begin{equation}
T = T_{FT} = \frac{f'(r_+)}{4\pi} = \frac{\rho}{4\pi R}\left[2+(K+1)(d-3)\right] \ .
\end{equation}
For the total energy we have:
\begin{equation}
E = E_{FT} 
= \frac{(d-2)Vol(\Omega^{FT}_{d-2})}{16\pi GR}\rho^{d-1}(K+1)+E_0 \ ,
\end{equation}
where $ Vol(\Omega_{d-2}^{FT}) $ is the total volume of the $ \Omega_{d-2}^{FT} $ manifold,
$ G $ is the bulk gravitational constant and $ E_0 $ is independent of the black hole parameters.
Finally the total entropy is given by:
\begin{equation}
S = S_{FT} = \frac{Vol(\Omega_{d-2})}{4G}r_+^{d-2} = \frac{Vol(\Omega_{d-2}^{FT})}{4G}\rho^{d-2} \ .
\end{equation}
For the sake of discussing the field theory correlators and the quasinormal modes, we shall also
define the following dimensionless quantities, for an oscillation mode of frequency $ \omega $
and Laplace operator eigenvalue $ L_s $:
\begin{align}
\lambda &\equiv \frac{\omega r_+}{\rho^2} = \frac{\omega R}{\rho}
= \frac{d-1}{2\pi} \frac{\omega}{T}
\frac{1}{1+\sqrt{1-\frac{d-1}{d-2}\frac{R_\Omega^{FT}}{(2\pi T)^2}}} \\
q_s &\equiv \frac{L_s}{\rho} = \frac{L_s^{FT}R}{\rho}
= \frac{d-1}{2\pi} \frac{L_s^{FT}}{T}
\frac{1}{1+\sqrt{1-\frac{d-1}{d-2}\frac{R_\Omega^{FT}}{(2\pi T)^2}}} \ .
\end{align}
Table~\ref{tab:bulkcftquantities} contains a summary of the physical quantities defined in the
bulk spacetime, the corresponding field theory quantities and their dimensions.

\begin{table}
\caption{A summary of physical quantities in the bulk and field theory spacetimes and their
dimensions}
\label{tab:bulkcftquantities}
\centering
{\renewcommand{\arraystretch}{1.3}
\begin{tabular}{|cc|cc|}
\hline
\multicolumn{2}{|c|}{Bulk Quantity} & \multicolumn{2}{c|}{Field Theory Quantity} \\
\hline
Quantity & Dimension & Quantity & Dimension\\
\hline
$ t $ & $ [L] $ & $ t $ & $ [L] $ \\
$ x^i $ & $ [L]^0 $ & $ x_{FT}^i = R x^i $ & $ [L] $\\
$ r $ & $ [L] $ & & \\
$ \ud s_{bulk}^2 $ & $ [L]^2 $ & $ \ud s_{FT}^2 $ & $ [L]^2 $\\
$ R $ & $ [L] $ & & \\
$ r_+ $ & $ [L] $ & & \\
$ \rho = r_+ R^{-1} $ & $ [L]^0 $ & & \\
$ R_\Omega $ & $ [L]^0 $ & $ R_\Omega^{FT} = R_\Omega R^{-2} $ & $ [L]^{-2} $ \\
$ k $ & $ [L]^0 $ & $ k_{FT} = kR^{-2} $ & $ [L]^{-2} $ \\
$ K = k\rho^{-2} $ & $ [L]^0 $ & & \\
$ T $ & $ [L]^{-1} $ & $ T $ & $ [L]^{-1} $ \\
$ \omega $ & $ [L]^{-1} $ & $ \omega $ & $ [L]^{-1} $\\
$ L_s $ & $ [L]^0 $ & $ L_s^{FT} = L_s R^{-1} $ & $ [L]^{-1} $ \\
$ \lambda = \omega R \rho^{-1} $ & $ [L]^0 $  & & \\
$ q_s = L_s \rho^{-1} $ & $ [L]^0 $  & & \\
$ G $ & $ [L]^{d-2} $ & & \\
$ E $ & $ [L]^{-1} $ & $ E $ & $ [L]^{-1} $ \\
$ S $ & $ [L]^0 $ & $ S $ & $ [L]^0 $ \\
\hline
\end{tabular}}
\end{table}

\begin{figure}[htbp]
\centering
\includegraphics[width=12cm]{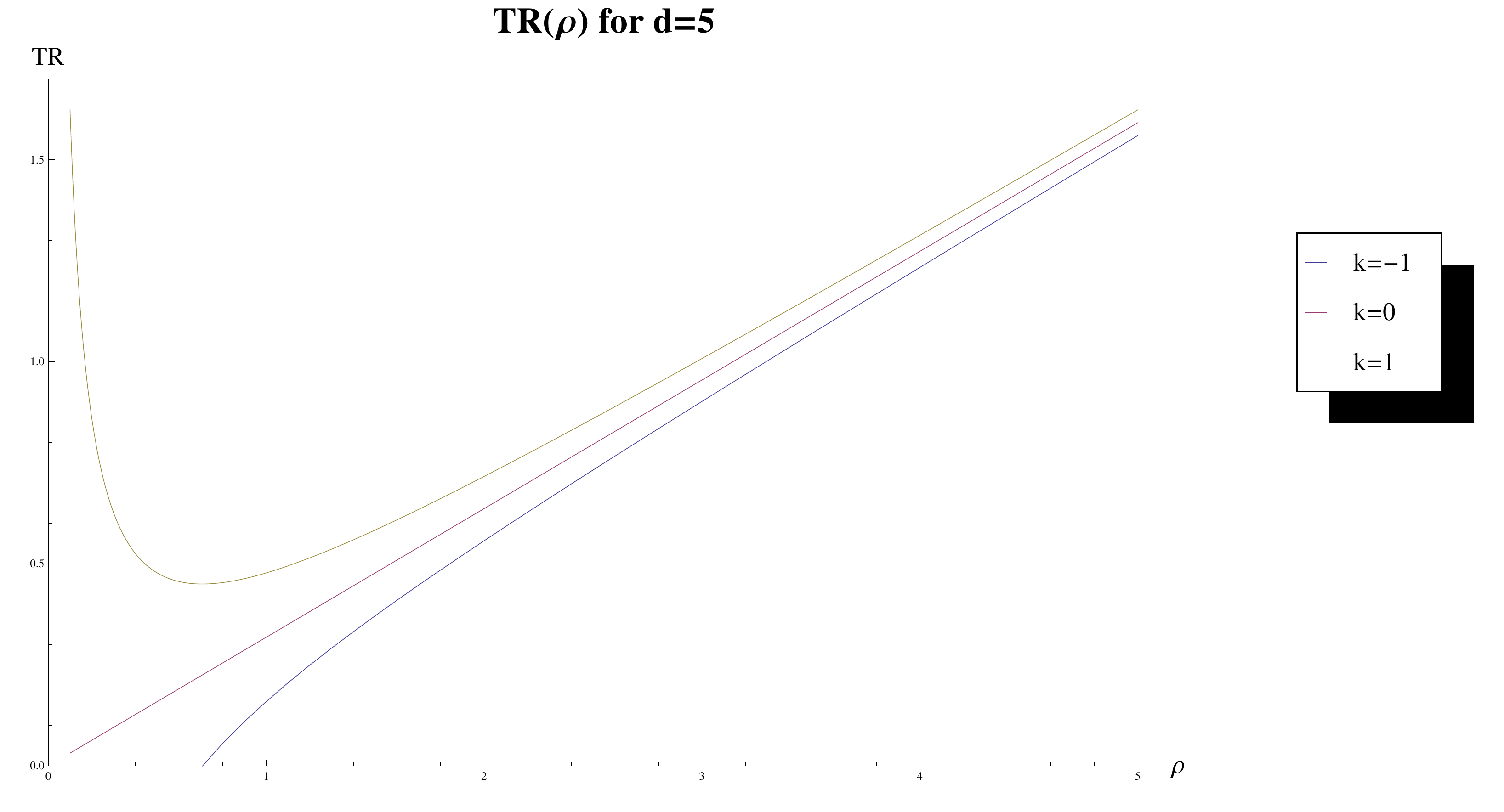}
\caption{$ TR $ as a function of $ \rho $ for the spherical ($ k=1 $), flat ($ k=0 $) and
hyperbolic($ k=-1 $) cases in $ d=5 $ bulk dimensions.}
\label{fig:thermodynamicquantitiestrforrho}
\end{figure}

\section{QNM Equations Derivation}
\label{app:qnmequationsderivation}

\subsection{QNM Scalar Equation}
\label{appsubsec:qnmscalarequationderivation}

The EOM for a minimally coupled, massless scalar field $ \Phi $ are given by
\begin{equation}
\nabla_\mu \nabla^\mu \Phi = 0 \ ,
\end{equation}
or
\begin{equation}
\label{eq:scalareom}
\frac{1}{\sqrt{-g}} \partial_\mu \left( \sqrt{-g}g^{\mu\nu} \partial_{\nu}\Phi \right) = 0
\end{equation}
(where $ g = \det(g_{\mu\nu}) $ ).


To find a QNM, we look for solutions of the form
\begin{equation}
\Phi(t,z,\mathbf{x}) = \psi(z) \e^{-i\omega t} H_{L^2}(\mathbf{x}) \ ,
\end{equation}
where $ H_{L^2}(\mathbf{x}) $ is a (scalar) eigenfunction of the Laplace operator $ \Delta_\Omega $
defined on the manifold $ \Omega_{d-2} $ with corresponding eigenvalue of $ -L_s^2 $. Putting this in, and noticing that
\begin{equation}
\label{eq:scalarlaplaciandef}
\Delta_\Omega H_{L^2}(\mathbf{x}) \equiv \udel \ud H_{L^2}(\mathbf{x})\\
\equiv -\sum_{i,j}\frac{1}{\sqrt{g_\Omega}} \partial_i\left( \sqrt{g_\Omega} g_\Omega^{ij} \partial_j H_{L^2}(\mathbf{x}) \right)
= L_s^2 H_{L^2}(\mathbf{x})
\end{equation}
(where $ \ud $ and $ \udel $ are the exterior derivative and codifferential defined on $ \Omega_{d-2} $, respectively),
we obtain Equation~\ref{eq:scalareq} (scalar QNM equation).

The QNMs in asymptotically AdS spacetimes are defined as the solutions to the equations of motion
obeying the incoming wave boundary condition at the horizon and the Dirichlet boundary condition at
the AdS boundary (at least in the context of AdS/CFT - see Subsection~\ref{subsec:qnmdefinition} and
also~\cite{Son:2002sd},~\cite{Berti:2009kk},~\cite{Nunez:2003eq}). In terms of the defined
variables and coordinates, these translate to
\begin{equation}
\left. \psi \right|_{z=0} \sim z^{-\frac{i\lambda}{C}}
\qquad
\left. \psi \right|_{z=1} = 0 \ ,
\end{equation}
as in Equation~\ref{eq:scalarbc}.

\subsection{QNM Vector Equations}
\label{appsubsec:qnmvectorequationsderivation}

The EOM for a vector (gauge) field $ \mathbf{A} $ are given by
\begin{equation}
\nabla_\mu F^{\mu\nu} = 0 \ ,
\end{equation}
or
\begin{equation}
\label{eq:vectoreom}
\frac{1}{\sqrt{-g}}\partial_\mu \left( \sqrt{-g}g^{\mu\alpha}g^{\nu\beta}F_{\alpha\beta} \right) = 0 \ .
\end{equation}

Putting the metric from Equation~\ref{eq:zmetric} Equation~\ref{eq:vectoreom} we get for the $ \beta $ component of the equation:
\begin{multline}
\label{eq:vectoreom2}
(1-z)^{d-2}\partial_z\left( \frac{\gz}{(1-z)^{d-2}}g^{\beta\gamma}F_{z\gamma} \right)
-\frac{r_+^2}{\rho^4}\frac{1}{\gz}g^{\beta\gamma}\partial_t F_{t\gamma}\\
+\frac{1}{\rho^2}\sum_{i,j} \frac{1}{\sqrt{g_\Omega}}\partial_i \left( \sqrt{g_\Omega}g_\Omega^{ij}g^{\beta\gamma}F_{j\gamma} \right) = 0 \ .
\end{multline}

To find a QNM, we look for solutions proportional to eigenfunctions of the Laplace
operator. The $ t $ and $ z $ components would be proportional to the eigenfunctions of
the scalar Laplace operator:
$
A_t, A_z \propto H_{L^2}(\mathbf{x})
$,
where $ H_{L^2} $ is defined as in Equation~\ref{eq:scalarlaplaciandef}.
The other components would be proportional to the corresponding components of the
eigenfunction of the vector Laplace operator:
$
A_i \propto A_{L^2,i}(\mathbf{x})
$,
where $ A_{L^2} $ is defined as follows:
\begin{equation}
\label{eq:vectoreigenfunctionsdefex}
\Delta_\Omega \mathbf{A}_{L^2}(\mathbf{x})
\equiv \udel \ud \mathbf{A}_{L^2}(\mathbf{x})\\
= \sum_j\frac{1}{\sqrt{g_\Omega}}\partial_j\left( \sqrt{g_\Omega}F_{L^2}^{ij} \right)
= L_v^2 A_{L^2}^i (\mathbf{x}) \ .
\end{equation}

There are now two possibilities that correspond to two different modes:
\begin{enumerate}
\item The ``longitudinal'' mode. This is the case where
$
L_v = 0 \ ,
$
or
$
\udel\ud\mathbf{A}_{0} = 0
$,
from which we can deduce
\begin{equation}
\left< \ud\mathbf{A}_{0}, \ud\mathbf{A}_{0} \right>
= \left< \mathbf{A}_{0} , \udel\ud\mathbf{A}_{0} \right> = 0
\qquad \Rightarrow \qquad
\mathbf{F}_{0} = \ud \mathbf{A}_{0} = 0 \ .
\end{equation}
$ \mathbf{A}_0 $ is, therefore, a closed form and can be written as
$
\mathbf{A}_{0,i} = \partial_i\phi
$,
where $ \phi $ is some (scalar) function\footnote{More precisely, $ \mathbf{A}_0 $ can be written as a
sum of an exact form $ \ud \phi $ and some harmonic form $ \mathbf{B} $. The harmonic component
satisfies $ \udel \mathbf{B} = 0 $, and can therefore be treated as a ``transverse'' mode.}.
To get the equation for a QNM we choose $ \phi $ to be an eigenfunction of the scalar Laplace operator
$ \phi=H_{L^2} $.
The solution will then be of the form
\begin{equation}
\begin{pmatrix}
A_t(t,z,\mathbf{x})\\
A_z(t,z,\mathbf{x})\\
A_1(t,z,\mathbf{x})\\
\vdots\\
A_{d-2}(t,z,\mathbf{x})
\end{pmatrix}
=
\begin{pmatrix}
f(t,z)H_{L^2}(\mathbf{x})\\
h(t,z)H_{L^2}(\mathbf{x})\\
k(t,z)\partial_1 H_{L^2}(\mathbf{x})\\
\vdots\\
k(t,z)\partial_{d-2} H_{L^2}(\mathbf{x})
\end{pmatrix} \ .
\end{equation}
Putting this into Equation~\ref{eq:vectoreom2} we get the following equations:\\
For the $ t $ component-
\begin{equation}
\label{eq:longvectortcomp}
I \equiv \gz (1-z)^{d-4} \pdz\left[ \frac{1}{(1-z)^{d-4}}\left( \partial_t h - \partial_z f \right) \right]
- q_s^2\left( \partial_t k - f \right) = 0 \ .
\end{equation}
For the $ z $ component-
\begin{equation}
\label{eq:longvectorzcomp}
II \equiv -\frac{r_+^2}{\rho^4}\frac{1}{\gz}\partial_t\left( \partial_t h - \partial_z f \right)
-q_s^2 \left( h-\partial_z k \right) = 0 \ .
\end{equation}

From Equations~\ref{eq:longvectortcomp} and~\ref{eq:longvectorzcomp} we can take
$
\partial_z(I) + \partial_t(II) = 0
$,
and defining \\
$
\left( \partial_t h - \partial_z f \right) \equiv \psi(z) \e^{-i\omega t}
$,
we finally obtain Equation~\ref{eq:longvectoreq} (longitudinal vector QNM equation).

As for the boundary conditions, the incoming wave boundary condition at the horizon implies that
\begin{equation}
\left. \psi \right|_{z=0} \sim z^{-\frac{i\lambda}{C}} \ ,
\end{equation}
while the Dirichlet boundary condition at the boundary implies
\begin{equation}
\left. f,h,k \right|_{z=1} = 0 \ .
\end{equation}
Setting $ z=1 $ in Equation~\ref{eq:longvectortcomp} and applying the boundary condition we get for
$ \psi $ the condition:
\begin{equation}
\left. (1-z)^{d-4}\partial_z\left[ \frac{1}{(1-z)^{d-4}} \psi \right] \right|_{z=1} = 0 \ .
\end{equation}

\item The ``transverse'' mode. In this case
$
L_v \ne 0 
$.
Applying the codifferential $ \delta $ to Equation~\ref{eq:vectoreigenfunctionsdefex} we get
\begin{equation}
0=\udel\udel\ud \mathbf{A}_{L^2} = L_v^2 \udel \mathbf{A}_{L^2} \ ,
\end{equation}
and since $ L_v \ne 0 $ we see that
\begin{equation}
\label{eq:transvectorzerodiv}
\udel \mathbf{A}_{L^2} = -\sum_i \frac{1}{\sqrt{g_\Omega}}\partial_i\left( \sqrt{g_\Omega} A_{L^2}^i \right) = 0 \ .
\end{equation}
The solution in this case will be of the form
\begin{equation}
\begin{pmatrix}
A_t(t,z,\mathbf{x})\\
A_z(t,z,\mathbf{x})\\
A_1(t,z,\mathbf{x})\\
\vdots\\
A_{d-2}(t,z,\mathbf{x})
\end{pmatrix}
=
\begin{pmatrix}
0\\
0\\
a(t,z)A_{L^2,1}(\mathbf{x})\\
\vdots\\
a(t,z)A_{L^2,d-2}(\mathbf{x})
\end{pmatrix} \ .
\end{equation}
Putting this into Equation~\ref{eq:vectoreom2}, 
the equations for the $ t $ and $ z $ components are automatically satisfied accroding to 
Equation~\ref{eq:transvectorzerodiv}.\\
Taking into account Equation~\ref{eq:vectoreigenfunctionsdefex}, we get for the i component
\begin{equation}
(1-z)^{d-4} \pdz\left( \frac{\gz}{(1-z)^{d-4}}\pdz a \right)
-\frac{r_+^2}{\rho^4}\frac{1}{\gz}\partial_t^2 a
-\frac{L_v^2}{\rho^2}a = 0 \ .
\end{equation}
Setting $ a(t,z) = \e^{-i\omega t} \psi(z) $ we finally obtain Equation~\ref{eq:transvectoreq} (transverse vector QNM equation).

The boundary conditions are again given by the incoming wave boundary condition at the horizon and the
Dirichlet boundary condition at the boundary, so that:
\begin{equation}
\left.\psi\right|_{z=0} \sim z^{-\frac{i\lambda}{C}}
\qquad
\left.\psi\right|_{z=1} = 0 \ .
\end{equation}
\end{enumerate}	

\section{Derivation of General Formulae For Correlators}
\label{app:corrformulaederivation}

\subsection{Scalar Correlators}

Consider a massless, minimally coupled scalar field in the bulk spacetime (where the bulk
metric is given by Equation~\ref{eq:zmetric}). The action of the scalar field is given by:
\begin{equation}
S[\phi] = C_s\int_0^1\ud z\int_{-\infty}^{\infty}\ud t\int_{\Omega_{d-2}}\ud x^1\ldots\ud x^{d-2}
\sqrt{-g}\nabla_\mu\phi\nabla^\mu\phi \ ,
\end{equation}
where $ C_s $ is the appropriate normalization constant for the bulk scalar field.
Since $ \phi $ is on-shell, it satisfies the EOM:
\begin{equation}
\nabla_\mu\nabla^\mu\phi = 0 \ ,
\end{equation}
so that, using integration by parts:
\begin{multline}
S = C_s\int_0^1\ud z\int_{-\infty}^{\infty}\ud t\int_{\Omega_{d-2}}\ud x^1\ldots\ud x^{d-2}
\sqrt{-g}\left(\nabla_\mu\left(\phi\nabla^\mu\phi\right)-\phi\nabla_\mu\nabla^\mu\phi\right) \\
= C_s\left.\int_{-\infty}^{\infty}\ud t\int_{\Omega_{d-2}}\ud x^1\ldots\ud x^{d-2}
\sqrt{-g}g^{zz}\phi\pdz\phi\right|^{z=1}_{z=0} \ .
\end{multline}
Using the metric from Equation~\ref{eq:zmetric} gives:
\begin{equation}
\sqrt{-g} = \frac{r_+^{d-1}}{(1-z)^d}\sqrt{g_\Omega} \ ,
\end{equation}
so that
\begin{equation}
S[\phi] = C_s\frac{r_+^{d-1}}{(1-z)^d}g^{zz}\left.\int_{-\infty}^{\infty}\ud t\int\ud\Omega_{d-2}
\phi\pdz\phi\right|_{z=0}^{z=1}
\end{equation}
(where $ \ud\Omega_{d-2} \equiv \sqrt{g_\Omega}\ud x^1\ldots\ud x^{d-2} $).

Let $ {H_\mathbf{s}(x)} $ be a complete basis of eigenfunctions of the Laplace
operator defined on the $ \Omega_{d-2} $ manifold, where $ \mathbf{s} $ here stands for all of the
indices (discrete or continuous) required to uniquely identify the eigenfunctions. The
normalization of these functions will be chosen so that:
\begin{equation}
\int\ud\Omega_{d-2} H_{\mathbf{s'}}^*(x)H_{\mathbf{s}}(x) = \delta_{\mathbf{ss'}}\ ,
\end{equation}
where $ \delta_{\mathbf{ss'}} $ here stands for the Kronecker delta in case of a discrete
spectrum and a Dirac delta in case of a continuous spectrum.
One may then expand the field $ \phi $ in terms of these functions and the Fourier components
so that:
\begin{equation}
\label{eq:scalarexpansion}
\phi(z,t,x) = \int\frac{\ud\omega}{2\pi}\sum_\mathbf{s} \e^{-i\omega t}H_{\mathbf{s}}(x)
\psi_{\omega,\mathbf{s}}(z) \ ,
\end{equation}
where $ \sum_\mathbf{s} $ stands here for either a sum in case of a discrete
spectrum or an integral (with the appropriate measure) in case of a continuous spectrum.
Since $ \phi $ is a solution for the EOM, as stated in Subsection~\ref{subsec:cftcorrelatorsfromhol},
$ \psi_{\omega,\mathbf{s}} $ would be a solution of the scalar QNM Equation
(Equation~\ref{eq:scalareq}), with the incoming-wave boundary condition at the horizon:
\begin{equation}
\left.\psi_{\omega,\mathbf{s}}\right|_{z=0} \sim z^{-\frac{i\lambda}{C}} \ .
\end{equation}
As for the boundary, notice that the metric of CFT spacetime is actually given by
Equation~\ref{eq:cftmetric}, so that in the CFT spacetime the normalized
eigenfunctions $ H_{\mathbf{s}}^{FT}(x) $ are given by
\begin{multline}
\int\ud\Omega_{d-2}^{FT} H_{\mathbf{s'}}^{FT*}(x)H_{\mathbf{s}}^{FT}(x)
= R^{d-2}\int\ud\Omega_{d-2} H_{\mathbf{s'}}^{FT*}(x)H_{\mathbf{s}}^{FT}(x)
=\delta_{\mathbf{ss'}} \\
\Rightarrow \qquad H_{\mathbf{s}}^{FT}(x) = R^{-\frac{d-2}{2}}H_{\mathbf{s}}(x) \ .
\end{multline}
As a consequence, on the boundary the field $ \phi $ may be written as:
\begin{equation}
\phi(1,t,x) = \int\frac{\ud\omega}{2\pi}\sum_\mathbf{s} \e^{-i\omega t}
R^{-\frac{d-2}{2}}H_{\mathbf{s}}(x)\phi_0(\omega,\mathbf{s})
\end{equation}
(where $ \phi_0(\omega,\mathbf{s}) $ are the components of $ \phi $ in the CFT spacetime),
which means the boundary condition on $ \psi_{\omega,\mathbf{s}} $ at the AdS boundary is:
\begin{equation}
\left.\psi_{\omega,\mathbf{s}}\right|_{z\to 1} = R^{-\frac{d-2}{2}}\phi_0(\omega,\mathbf{s}) \ .
\end{equation}
Defining a normalized $ \widehat{\psi}_{\omega,\mathbf{s}} $ by
$ \psi_{\omega,\mathbf{s}} \equiv  R^{-\frac{d-2}{2}}\phi_0(\omega,
\mathbf{s})\widehat{\psi}_{\omega,\mathbf{s}} $, the normalized function satisfies
the boundary conditions:
\begin{equation}
\left.\widehat{\psi}_{\omega,\mathbf{s}}\right|_{z=0} \sim z^{-\frac{i\lambda}{C}}
\qquad
\left.\widehat{\psi}_{\omega,\mathbf{s}}\right|_{z\to 1} = 1 \ .
\end{equation}

Putting Equation~\ref{eq:scalarexpansion} into the expression for the action, we have:
\begin{multline}
S[\phi] = C_s\frac{r_+^{d-1}}{(1-z)^d}g^{zz}\int_{-\infty}^{\infty}\ud t\int\ud\Omega_{d-2}
\int\frac{\ud\omega}{2\pi}\int\frac{\ud\omega'}{2\pi}\sum_{\mathbf{s,s'}}\\
\left.\e^{i(\omega'-\omega)t}H_{\mathbf{s'}}^*(x)H_{\mathbf{s}}(x)\psi_{\omega',\mathbf{s'}}^*(z)
\pdz\psi_{\omega,\mathbf{s}}(z)\right|_{z=0}^{z=1} \ .
\end{multline}
Using
\begin{equation}
\int\ud t \,\e^{i(\omega'-\omega)t} = 2\pi\delta(\omega'-\omega)
\end{equation}
and the orthogonality of $ H_{\mathbf{s}}(x) $, the action becomes:
\begin{multline}
S[\phi] = \left.C_s\frac{r_+^{d-1}}{(1-z)^d}g^{zz}\int\frac{\ud\omega}{2\pi}\sum_{\mathbf{s}}
\psi_{\omega,\mathbf{s}}^*(z)\pdz\psi_{\omega,\mathbf{s}}(z)\right|_{z=0}^{z=1} \\
= C_s\frac{r_+^{d-1}}{(1-z)^d}\frac{\gz(1-z)^2}{R^2}\frac{1}{R^{d-2}}
\int\frac{\ud\omega}{2\pi}\sum_{\mathbf{s}}\\
\left.\phi_0^*(\omega,\mathbf{s})
\widehat{\psi}_{\omega,\mathbf{s}}^*(z)\pdz\widehat{\psi}_{\omega,\mathbf{s}}(z)
\phi_0(\omega,\mathbf{s})\right|_{z=0}^{z=1} \ ,
\end{multline}
so that
\begin{equation}
S[\phi] = \left.\int\frac{\ud\omega}{2\pi}\sum_{\mathbf{s}}\phi_0^*(\omega,\mathbf{s})
\mathcal{F}(\omega,\mathbf{s},z)\phi_0(\omega,\mathbf{s})\right|_{z=0}^{z=1} \ ,
\end{equation}
where
\begin{equation}
\mathcal{F}(\omega,\mathbf{s},z) = C_s\frac{r_+^{d-1}}{R^d}\frac{\gz}{(1-z)^{d-2}}
\widehat{\psi}_{\omega,\mathbf{s}}^*(z)\pdz\widehat{\psi}_{\omega,\mathbf{s}}(z) \ .
\end{equation}
Using the prescription described in~\cite{Son:2002sd}, the retarded correlator is
then given by:
\begin{equation}
G^R(\omega,\mathbf{s}) = \left.-2\mathcal{F}(\omega,\mathbf{s},z)\right|_{z\to 1}
= \left.-2C_s\frac{r_+^{d-1}}{R^d}\frac{1}{(1-z)^{d-2}}
\pdz\widehat{\psi}_{\omega,\mathbf{s}}(z)\right|_{z\to 1} \ .
\end{equation}

\subsection{Vector Correlators}

Next consider a vector gauge field in the bulk spacetime. The action of the gauge field is given by:
\begin{equation}
S[A] = C_v\int_0^1\ud z\int_{-\infty}^{\infty}\ud t\int_{\Omega_{d-2}}
\ud x^1\ldots\ud x^{d-2}\sqrt{-g}F_{\mu\nu}F^{\mu\nu} \ ,
\end{equation}
where $ C_v $ is the appropriate normalization constant for the bulk vector field.
Since $ A_\mu $ is on-shell, it satisfies the EOM:
\begin{equation}
\nabla_\mu F^{\mu\nu} = 0 \ ,
\end{equation}
so that
\begin{equation}
F_{\mu\nu} F^{\mu\nu} = 2\nabla_\mu\left(A_\nu F^{\mu\nu}\right) \ ,
\end{equation}
and using integration by parts:
\begin{equation}
S = \left.2C_v\int_{-\infty}^{\infty}\ud t\int_{\Omega_{d-2}}\ud x^1\ldots\ud x^{d-2}
\sqrt{-g}A_\nu F^{z\nu} \right|_{z=0}^{z=1} \ .
\end{equation}
Using the metric from Equation~\ref{eq:zmetric} gives:
\begin{equation}
S = 2C_v\frac{r_+^{d-1}}{(1-z)^d}g^{zz}\left.\int_{-\infty}^{\infty}\ud t\int\ud\Omega_{d-2}
\left[g^{tt}A_t F_{zt} + g^{ij}A_i F_{zj}\right]\right|_{z=0}^{z=1}
\end{equation}
(where $ \ud\Omega_{d-2} \equiv \sqrt{g_\Omega}\ud x^1\ldots\ud x^{d-2} $).

Let $ H_\mathbf{s}(x) $ again be a complete basis of eigenfunctions of the Laplace operator on the
$ \Omega_{d-2} $ manifold, normalized so that:
\begin{equation}
\label{eq:orthogonality1}
\int\ud\Omega_{d-2} H_{\mathbf{s'}}^*(x) H_\mathbf{s}(x) = \delta_{\mathbf{ss'}} \ .
\end{equation}
Let us further define $ \mathbf{\tilde{A}_v}(x) $ as a set of vector eigenfunctions of the
Laplace operator, such that:
\begin{equation}
\udel\ud\mathbf{\tilde{A}_v} = L_\mathbf{v}^2 \mathbf{\tilde{A_v}} \qquad L_\mathbf{v} \ne 0 \ ,
\end{equation}
and $ \udel\mathbf{\tilde{A}_v} = 0 $. Here $ \mathbf{v} $ stands for all of the indices required to
uniquely identify the eigenfunctions.
The normalization of these functions will be chosen so that:
\begin{equation}
\label{eq:orthogonality2}
\int\ud\Omega_{d-2}\, g_\Omega^{ij} \tilde{A}_{\mathbf{v},i}^*(x)\tilde{A}_{\mathbf{v'},j}(x) =
\delta_{\mathbf{vv'}} \ .
\end{equation}
Note that $ \mathbf{\tilde{A}}_v $ along with $ \frac{\partial_i H_{\mathbf{s}}(x)}{L_s} $
form a complete orthonormal basis.
One may expand the field $ \mathbf{A} $ in terms of these functions and the Fourier components:
\begin{align}
A_t(z,t,x) &= \int\frac{\ud\omega}{2\pi}\sum_{\mathbf{s}}\e^{-i\omega t}H_{\mathbf{s}}(x)
f_{\omega,\mathbf{s}}(z) \\
A_z(z,t,x) &= \int\frac{\ud\omega}{2\pi}\sum_{\mathbf{s}}\e^{-i\omega t}H_{\mathbf{s}}(x)
h_{\omega,\mathbf{s}}(z) \\
A_i(z,t,x) &= \int\frac{\ud\omega}{2\pi}\sum_{\mathbf{s}}\e^{-i\omega t}\partial_i H_{\mathbf{s}}(x)
k_{\omega,\mathbf{s}}(z) +
\int\frac{\ud\omega}{2\pi}\sum_{\mathbf{v}}\e^{-i\omega t}\tilde{A}_{\mathbf{v},i}(x)
a_{\omega,\mathbf{v}}(z) \ .
\end{align}
Since $ \mathbf{A} $ is a solution for the EOM, as stated in Subsection~\ref{subsec:cftcorrelatorsfromhol},
$ f_{\omega,\mathbf{s}} $, $ h_{\omega,\mathbf{s}} $, $ k_{\omega,\mathbf{s}} $ and
$ a_{\omega,\mathbf{v}} $ would be solutions of the vector QNM Equations (see
Appendix~\ref{app:qnmequationsderivation}) with incoming-wave boundary conditions
at the horizon. As for the boundary, since the field theory metric is given by
Equation~\ref{eq:cftmetric}, in the field theory side the normalized eigenfunctions are
given by:
\begin{align}
H_\mathbf{s}^{FT}(x) &= R^{-\frac{d-2}{2}}H_\mathbf{s}(x) \\
\mathbf{\tilde{A}_v}^{FT}(x) &= R^{-\frac{d-4}{2}}\mathbf{\tilde{A}_v} \ .
\end{align}
As a consequence, on the boundary the field $ \mathbf{A} $ may be written as:
\begin{eqnarray}
A_t(1,t,x) &=& \int\frac{\ud\omega}{2\pi}\sum_{\mathbf{s}}\e^{-i\omega t}R^{-\frac{d-2}{2}}
H_\mathbf{s}(x)A_t^0(\omega,\mathbf{s}) \\
A_i(1,t,x) &=& \int\frac{\ud\omega}{2\pi}\sum_{\mathbf{s}}\e^{-i\omega t}R^{-\frac{d-4}{2}}
\frac{\partial_i H_\mathbf{s}(x)}{L_\mathbf{s}} A_\|^0(\omega,\mathbf{s}) \nonumber\\
&& + \int\frac{\ud\omega}{2\pi}\sum_{\mathbf{v}}\e^{-i\omega t}R^{-\frac{d-4}{2}}
\tilde{A}_{\mathbf{v},i}(x) A_\bot^0(\omega,\mathbf{v})
\end{eqnarray}
(where $ A_t^0(\omega,\mathbf{s}) $, $ A_\|^0(\omega,\mathbf{s}) $ and $ A_\bot^0(\omega,\mathbf{v}) $
are the components of $ \mathbf{A} $ in the field theory spacetime). The boundary conditions
at the AdS boundary are therefore:
\begin{align}
\label{eq:fboundarycondition}
\left.f_{\omega,\mathbf{s}}\right|_{z\to 1} &= R^{-\frac{d-2}{2}}A_t^0(\omega,\mathbf{s}) \\
\label{eq:kboundarycondition}
\left.k_{\omega,\mathbf{s}}\right|_{z\to 1} &= R^{-\frac{d-4}{2}}L_\mathbf{s}^{-1}
A_\|^0(\omega,\mathbf{s}) \\
\label{eq:aboundarycondition}
\left.a_{\omega,\mathbf{v}}\right|_{z\to 1} &= R^{-\frac{d-4}{2}}A_\bot^0(\omega,\mathbf{v}) \ .
\end{align}

Next, one may derive expansions for the $ F_{\mu\nu} $ tensor:
\begin{equation}
F_{zt} = -\int\frac{\ud\omega}{2\pi}\sum_{\mathbf{s}}\e^{-i\omega t}H_{\mathbf{s}}(x)
\left[-i\omega h_{\omega,\mathbf{s}}-\pdz f_{\omega,\mathbf{s}}\right]
\end{equation}
\begin{equation}
F_{zj} = \int\frac{\ud\omega}{2\pi}\sum_{\mathbf{s}}\e^{-i\omega t}\partial_j H_\mathbf{s}(x)
\left[\pdz k_{\omega,\mathbf{s}}-h_{\omega,\mathbf{s}}\right] \\
+ \int\frac{\ud\omega}{2\pi}\sum_{\mathbf{v}}\e^{-i\omega t}\tilde{A}_{\mathbf{v},j}(x)
\pdz a_{\omega,\mathbf{v}} \ .
\end{equation}
We proceed by inserting these expressions into the expression for the action. Using the
orthogonality properties of Equations~\ref{eq:orthogonality1} and~\ref{eq:orthogonality2},
and also the following derived properties:
\begin{equation}
\int\ud\Omega_{d-2}\, g_\Omega^{ij}\partial_i H_\mathbf{s'}^*(x) \partial_j H_\mathbf{s}(x)
= \left\langle\ud H_\mathbf{s'},\ud H_\mathbf{s}\right\rangle = \left\langle H_\mathbf{s'},\udel\ud
H_\mathbf{s}\right\rangle = L_\mathbf{s}^2 \delta_\mathbf{ss'}
\end{equation}
\begin{equation}
\int\ud\Omega_{d-2}\, g_\Omega^{ij}\partial_i H_\mathbf{s'}^*(x) \tilde{A}_{\mathbf{v},j}(x)
= \left\langle \ud H_\mathbf{s'} , \mathbf{\tilde{A}_v} \right\rangle
= \left \langle H_\mathbf{s'} , \udel \mathbf{\tilde{A}_v} \right\rangle = 0 \ .
\end{equation}
We see that the action decomposes into a longitudinal part $ S_\| $, that involves
$ f_{\omega,\mathbf{s}} $, $ h_{\omega,\mathbf{s}} $ and $ k_{\omega,\mathbf{s}} $,
and a transverse part $ S_\bot $ that involves only $ a_{\omega,\mathbf{v}} $, so that:
\begin{equation}
S = S_\| + S_\bot \ .
\end{equation}

For the longitudinal part of the action, we get the following expression:
\begin{multline}
S_\| = 2C_v\frac{r_+^{d-1}}{(1-z)^d}g^{zz} \int\frac{\ud\omega}{2\pi}\sum_\mathbf{s}
\left[\frac{(1-z)^2}{\rho^2\gz} f_{\omega,\mathbf{s}}^*
\left(-i\omega h_{\omega,\mathbf{s}}-\pdz f_{\omega,\mathbf{s}}\right) \right.\\
\left.\left.+\frac{(1-z)^2}{R^2}q_\mathbf{s}^2 k_{\omega,\mathbf{s}}^*
\left(\pdz k_{\omega,\mathbf{s}}-h_{\omega,\mathbf{s}}\right)\right]\right|_{z=0}^{z=1}
\end{multline}
(where the definition $ q_\mathbf{s}\equiv \frac{L_\mathbf{s}}{\rho} $ was used).
Since $ f_{\omega,\mathbf{s}} $, $ h_{\omega,\mathbf{s}} $ and $ k_{\omega,\mathbf{s}} $
satisfy the longitudinal vector QNM Equations, we may use Equation~\ref{eq:longvectorzcomp} to get
the relation:
\begin{equation}
q_\mathbf{s}^2\left(\pdz k_{\omega,\mathbf{s}} - h_{\omega,\mathbf{s}}\right)
= -\frac{r_+^2}{\rho^4\gz}i\omega\left(-i\omega h_{\omega,\mathbf{s}}
-\pdz f_{\omega,\mathbf{s}}\right) \ .
\end{equation}
Using this relation, along with the definitions:
\begin{align}
\psi_{\omega,\mathbf{s}} &\equiv -i\omega h_{\omega,\mathbf{s}} -\pdz f_{\omega,\mathbf{s}} \\
\chi_{\omega,\mathbf{s}} &\equiv -i\omega k_{\omega,\mathbf{s}} -f_{\omega,\mathbf{s}} \ ,
\end{align}
we obtain the expression:
\begin{equation}
S_\| =-2C_v \frac{r_+^{d-3}}{(1-z)^{d-4}} \left.\int\frac{\ud\omega}{2\pi}\sum_\mathbf{s}
\chi_{\omega,\mathbf{s}}^*(z)\psi_{\omega,\mathbf{s}}(z)\right|_{z=0}^{z=1} \ .
\end{equation}
Looking at Equations~\ref{eq:longvectortcomp} and~\ref{eq:longvectoreq}, we see that
$ \chi_{\omega,\mathbf{s}} $ and $ \psi_{\omega,\mathbf{s}} $ satisfy the following
system of differential equations:
\begin{align}
\label{eq:longvectorchipsiequation1}
\chi_{\omega,\mathbf{s}} &= \frac{1}{q_\mathbf{s}^2}\gz(1-z)^{d-4}\pdz
\left[\frac{1}{(1-z)^{d-4}}\psi_{\omega,\mathbf{s}}\right] \\
\label{eq:longvectorchipsiequation2}
\psi_{\omega,\mathbf{s}} &= -\frac{q_\mathbf{s}^2\gz}{\lambda^2-q_\mathbf{s}^2\gz}\pdz
\chi_{\omega,\mathbf{s}} \ .
\end{align}
From Equations~\ref{eq:kboundarycondition} and~\ref{eq:fboundarycondition} we can derive the boundary
condition for $ \chi_{\omega,\mathbf{s}} $:
\begin{equation}
\left.\chi_{\omega,\mathbf{s}}\right|_{z\to 1} = -i\omega R^{-\frac{d-4}{2}}L_\mathbf{s}^{-1}
A_\|^0(\omega,\mathbf{s}) - R^{-\frac{d-2}{2}}A_t^0(\omega,\mathbf{s}) \ .
\end{equation}
Next, define the normalized $ \widehat{\chi}_{\omega,\mathbf{s}} $ and
$ \widehat{\psi}_{\omega,\mathbf{s}} $ by:
\begin{align}
\chi_{\omega,\mathbf{s}} &\equiv \left[-i\omega R^{-\frac{d-4}{2}}L_\mathbf{s}^{-1}
A_\|^0(\omega,\mathbf{s}) - R^{-\frac{d-2}{2}}A_t^0(\omega,\mathbf{s})\right]
\widehat{\chi}_{\omega,\mathbf{s}}\\
\psi_{\omega,\mathbf{s}} &\equiv \left[-i\omega R^{-\frac{d-4}{2}}L_\mathbf{s}^{-1}
A_\|^0(\omega,\mathbf{s}) - R^{-\frac{d-2}{2}}A_t^0(\omega,\mathbf{s})\right]
\widehat{\psi}_{\omega,\mathbf{s}} \ ,
\end{align}
so that $ \widehat{\chi}_{\omega,\mathbf{s}} $ and $ \widehat{\psi}_{\omega,\mathbf{s}} $
also satisfy Equations~\ref{eq:longvectorchipsiequation1} and~\ref{eq:longvectorchipsiequation2}, with
the boundary condition:
\begin{equation}
\left.\widehat{\chi}_{\omega,\mathbf{s}}\right|_{z\to 1} = 1 \ .
\end{equation}
Using these definitions in the expression for the action gives:
\begin{multline}
S_\| = \int\frac{\ud\omega}{2\pi}\sum_\mathbf{s} \\
\left.\left[i\frac{\lambda}{q_\mathbf{s}}A_\|^0(\omega,\mathbf{s})+A_t^0(\omega,\mathbf{s})\right]^*
\mathcal{F}_\|(\omega,\mathbf{s},z)
\left[i\frac{\lambda}{q_\mathbf{s}}A_\|^0(\omega,\mathbf{s})+A_t^0(\omega,\mathbf{s})\right]
\right|_{z=0}^{z=1} \ ,
\end{multline}
where:
\begin{equation}
\mathcal{F}_\|(\omega,\mathbf{s},z) = -2C_v\frac{r_+^{d-3}}{R^{d-2}}\frac{1}{(1-z)^{d-4}}
\widehat{\chi}_{\omega,\mathbf{s}}^*\widehat{\psi}_{\omega,\mathbf{s}} \ .
\end{equation}
Using the prescription described in \cite{Son:2002sd}, we can then get the components of the retarded
correlators that correspond to the longitudinal mode:
\begin{multline}
G_{tt}^R(\omega,\mathbf{s}) = \left.-2\mathcal{F}_\|(\omega,\mathbf{s},z)\right|_{z\to 1}
= \left.4C_v\frac{r_+^{d-3}}{R^{d-2}}\frac{1}{(1-z)^{d-4}}
\widehat{\chi}_{\omega,\mathbf{s}}^*\widehat{\psi}_{\omega,\mathbf{s}}\right|_{z\to 1}\\
= \left.4C_v\frac{r_+^{d-3}}{R^{d-2}}\frac{1}{(1-z)^{d-4}}
\frac{\psi_{\omega,\mathbf{s}}}{\chi_{\omega,\mathbf{s}}}\right|_{z\to 1}
\end{multline}
\begin{align}
G_{t\|}^R(\omega,\mathbf{s}) &= i\frac{\lambda}{q_\mathbf{s}}G_{tt}^R(\omega,\mathbf{s}) \\
G_{\|\|}^R(\omega,\mathbf{s}) &= \frac{\lambda^2}{q_\mathbf{s}^2}G_{tt}^R(\omega,\mathbf{s}) \ .
\end{align}
Note that from Equation~\ref{eq:longvectorchipsiequation1} we have:
\begin{equation}
\left.\frac{\psi_{\omega,\mathbf{s}}}{\chi_{\omega,\mathbf{s}}}\right|_{z\to 1}
= \left.q_\mathbf{s}^2\frac{\psi_{\omega,\mathbf{s}}}
{(1-z)^{d-4}\pdz\left[\frac{1}{(1-z)^{d-4}}\psi_{\omega,\mathbf{s}}\right]}\right|_{z\to 1} \ .
\end{equation}

Turning to the transverse part of the action, we get the following expression:
\begin{equation}
S_\bot = \left.2C_v\frac{r_+^{d-3}}{(1-z)^{d-4}}\frac{\gz}{R^2}
\int\frac{\ud\omega}{2\pi}\sum_\mathbf{v} a_{\omega,\mathbf{v}}^*(z)\pdz a_{\omega,\mathbf{v}}(z)
\right|_{z=0}^{z=1} \ .
\end{equation}
Considering the boundary condition in Equation~\ref{eq:aboundarycondition} we define the normalized
$ \widehat{a}_{\omega,\mathbf{v}} $ by:
\begin{equation}
a_{\omega,\mathbf{v}} \equiv R^{-\frac{d-4}{2}}A_\bot^0(\omega,\mathbf{v})
\widehat{a}_{\omega,\mathbf{v}} \ ,
\end{equation}
so that:
\begin{equation}
\left.\widehat{a}_{\omega,\mathbf{v}}\right|_{z\to 1} = 1 \ .
\end{equation}
Inserting this definition into the expression for the action, we get:
\begin{equation}
S_\bot = \left.\int\frac{\ud\omega}{2\pi}\sum_\mathbf{v}
A_\bot^{0*}(\omega,\mathbf{v})\mathcal{F}_\bot(\omega,\mathbf{v},z)A_\bot^{0}(\omega,\mathbf{v})
\right|_{z=0}^{z=1} \ ,
\end{equation}
where:
\begin{equation}
\mathcal{F}_\bot(\omega,\mathbf{v},z)
= 2C_v\frac{r_+^{d-3}}{R^{d-2}}\frac{\gz}{(1-z)^{d-4}}
\widehat{a}_{\omega,\mathbf{v}}^*(z)\pdz\widehat{a}_{\omega,\mathbf{v}}(z) \ .
\end{equation}
The components of the retarded correlators corresponding to the transverse mode are then:
\begin{multline}
G_{\bot\bot}^R(\omega,\mathbf{v})
= \left.-2\mathcal{F}_\bot(\omega,\mathbf{v},z)\right|_{z\to 1}\\
= \left.-4C_v\frac{r_+^{d-3}}{R^{d-2}}\frac{1}{(1-z)^{d-4}}
\widehat{a}_{\omega,\mathbf{v}}^*(z)\pdz\widehat{a}_{\omega,\mathbf{v}}(z)\right|_{z\to 1}\\
= \left.-4C_v\frac{r_+^{d-3}}{R^{d-2}}\frac{1}{(1-z)^{d-4}}
\frac{\pdz a_{\omega,\mathbf{v}}(z)}{a_{\omega,\mathbf{v}}(z)}\right|_{z\to 1} \ .
\end{multline}

\section{Derivation of Exact Solutions for the \texorpdfstring{$ K=-1 $}{K=-1} Case}
\label{app:exactsolutionsfortcderivation}

In the $K=-1$ case, both the massless scalar and the transverse vector equations can be written in the
form:
\begin{equation}
(1-z)^\alpha\pdz\left[\frac{\gz}{(1-z)^\alpha}\pdz\psi\right]+\left[\frac{\lambda^2}{\gz}-q^2\right]\psi = 0 \ ,
\end{equation}
or
\begin{equation}
\gz\partial_z^2\psi+\left[\tilde{g}'(z)+\frac{\alpha\gz}{1-z}\right]\pdz\psi
+\left[\frac{\lambda^2}{\gz}-q^2\right]\psi = 0 \ ,
\end{equation}
where
\begin{equation}
\alpha=
\begin{cases}
d-2 & \text{for scalar},\\
d-4 & \text{for transverse vector}
\end{cases} \ ,
\end{equation}
with the boundary conditions:
\begin{equation}
\left.\psi\right|_{z=0} \sim z^{-\frac{i\lambda}{C}}
\qquad
\left.\psi\right|_{z=1}=0 \ .
\end{equation}
In the case $K=-1$ we have
\begin{equation}
\gz=1-(1-z)^2 \ .
\end{equation}
In order to solve the equation, first make the transformation:
\begin{equation}
w\equiv\gz=1-(1-z)^2=z(2-z) \ ,
\end{equation}
so that
\begin{equation}
\pdz = \frac{\partial w}{\partial z}\partial_w = \tilde{g}'(z)\partial_w = 2(1-z)\partial_w
= 2\sqrt{1-w}\partial_w \ .
\end{equation}
The equation then becomes:
\begin{equation}
w(1-w)\partial_w^2\psi
+\left[1+\frac{1}{2}(\alpha-3)w\right]\partial_w\psi
+\left[\frac{\lambda^2}{4w}-\frac{1}{4}q^2\right]\psi = 0 \ ,
\end{equation}
with the boundary conditions:
\begin{equation}
\left.\psi\right|_{w=0} \sim w^{-\frac{i\lambda}{2}}
\qquad
\left.\psi\right|_{w=1} = 0 \ .
\end{equation}
Next, define
\begin{equation}
\psi \equiv w^\gamma \phi \ ,
\end{equation}
where $\gamma=-i\frac{\lambda}{2}$, and get an equation for $\phi$:
\begin{multline}
w(1-w)\partial_w^2\phi
+\left[(1+2\gamma)+\left(\frac{1}{2}(\alpha-3)-2\gamma\right)w\right]\partial_w\phi\\
+\left[\gamma-\gamma^2+\frac{1}{2}\gamma(\alpha-3)-\frac{1}{4}q^2\right]\phi = 0 \ ,
\end{multline}
with the boundary conditions:
\begin{equation}
\label{eq:hypergeometriceq1}
\left.\phi\right|_{w=0}=1
\qquad
\left.\phi\right|_{w=1}=0 \ .
\end{equation}
Equation~\ref{eq:hypergeometriceq1} is known as the hypergeometric equation.
The solution to this equation that also satisfies the boundary condition at $w=0$ is
$\sb{2}F_1\left(a,b;c;w\right)$, where:
\begin{align}
\label{eq:scalartransexactsolutionfortcparams1}
c &= 1-i\lambda \\
\label{eq:scalartransexactsolutionfortcparams2}
a,b &= -\frac{1}{4}(\alpha-1)-i\frac{\lambda}{2} \pm \frac{1}{2}\sqrt{\frac{1}{4}(\alpha-1)^2-q^2} \ .
\end{align}

The longitudinal vector equation can be written in the form:
\begin{equation}
\pdz\left[\gz(1-z)^\alpha\pdz\left(\frac{1}{(1-z)^\alpha}\psi\right)\right]
+\left[\frac{\lambda^2}{\gz}-q^2\right]\psi = 0 \ ,
\end{equation}
or
\begin{equation}
\gz\partial_z^2\psi
+\left[\tilde{g}'(z)+\frac{\alpha\gz}{1-z}\right]\pdz\psi
+\left[\pdz\left(\frac{\alpha\gz}{1-z}\right)+\frac{\lambda^2}{\gz}-q^2\right]\psi = 0 \ ,
\end{equation}
where
\begin{equation}
\alpha=d-4 \ ,
\end{equation}
with the boundary conditions:
\begin{equation}
\left.\psi.\right|_{z=0} \sim z^{-\frac{i\lambda}{C}}
\qquad
\left.(1-z)^\alpha\pdz\left[\frac{1}{(1-z)^\alpha}\psi\right]\right|_{z=1} = 0 \ .
\end{equation}
Defining again:
\begin{equation}
w \equiv \gz = 1-(1-z)^2 \ ,
\end{equation}
the equation becomes:
\begin{multline}
w(1-w)\partial_w^2\psi
+\left[1+\frac{1}{2}(\alpha-3)w\right]\partial_w\psi\\
+\left[\frac{\lambda^2}{4w}-\frac{1}{4}q^2+\frac{1}{2}\alpha+\frac{\alpha w}{4(1-w)}\right]\psi = 0 \ .
\end{multline}
Next, define
\begin{equation}
\psi \equiv w^\gamma (1-w)^\delta \phi \ ,
\end{equation}
where $ \gamma = -i\frac{\lambda}{2} $ and $ \delta=\frac{1}{2} $, and get an equation for $ \phi $:
\begin{multline}
w(1-w)\partial_w^2\phi
+\left[(1+2\gamma)+\left(\frac{1}{2}(\alpha-3)-2\gamma-2\delta\right)w\right]\partial_w\phi\\
+\left[\gamma-\gamma^2+\frac{1}{2}\gamma(\alpha-3)-\frac{1}{4}q^2
+\frac{1}{4}\alpha-\delta(\delta-1)+\left(\frac{1}{2}(\alpha-3)-2\gamma\right)\delta\right]\phi = 0 \ .
\end{multline}
As for the boundary conditions, at $ w=0 $ we get:
\begin{equation}
\left.\phi\right|_{z=0}=0 \ ,
\end{equation}
while at $ w=1 $ we have
\footnote{Since the exponents of the equation at $ w=1 $ are $0$ and $\frac{\alpha-1}{2}$,
and since $ d\geq 4 $, near $w=1$ the solution $ \phi=o((1-w)^{-1}) $.}
\begin{multline}
\left.(1-z)^\alpha\pdz\left[\frac{1}{(1-z)^\alpha}\psi\right]\right|_{z=1}
= \left.2(1-w)^{\frac{\alpha+1}{2}}\partial_w\left[\frac{w^\gamma}{(1-w)^\frac{\alpha-1}
{2}}\phi\right]\right|_{w=1} \\
= \left.2w^\gamma(1-w)^{\frac{\alpha+1}{2}}\partial_w\left[\frac{1}{(1-w)^\frac{\alpha-1}
{2}}\phi\right] + 2\gamma(1-w)w^{\gamma-1}\phi\right|_{w=1} \\
\sim \left.2(1-w)^{\frac{\alpha+1}{2}}\partial_w\left[\frac{1}{(1-w)^\frac{\alpha-1}
{2}}\phi\right]\right|_{w=1} = 0 \ .
\end{multline}

The solution is once again $ \sb{2}F_1(a,b;c;w) $, where:
\begin{align}
\label{eq:longvectorexactsolutionfortcparams1}
c &= 1-i\lambda\\
\label{eq:longvectorexactsolutionfortcparams2}
a,b &= -\frac{1}{4}(\alpha-3)-i\frac{\lambda}{2}\pm\frac{1}{2}\sqrt{\frac{1}{4}(\alpha+1)^2-q^2} \ .
\end{align}

\section{Derivation of Vector Correlators for the \texorpdfstring{$ K=-1 $}{K=-1} Case}
\label{app:exactvectorcorrelatorsderivationfortc}

For the longitudinal vector mode, we continue from the transformation defined in
Appendix~\ref{app:exactsolutionsfortcderivation}.
The expression in Equation~\ref{eq:longvectorcftcorrelator} may be written in terms of $ \phi $
and $ w $:
\begin{equation}
\label{eq:longvectorcorrelatorinw}
G_{tt}^R(\omega,\mathbf{s})
=\left.4C_v\frac{r_+^{d-3}}{R^{d-2}}\frac{1}{(1-w)^\frac{\alpha}{2}}\frac{q_\mathbf{s}^2}{2}
\frac{(1-w)^\frac{1}{2}\phi_{\omega,\mathbf{s}}}
{(1-w)^\frac{\alpha+1}{2}\partial_w\left[\frac{1}{(1-w)^\frac{\alpha-1}{2}}\phi_{\omega,\mathbf{s}}\right]
+\gamma\frac{1-w}{w}\phi_{\omega,\mathbf{s}}}\right|_{w\to 1} \ .
\end{equation}

As explained in Appendix~\ref{app:exactsolutionsfortcderivation}, the solution to the
EOM with an incoming-wave boundary condition at the horizon is
$ \phi_{\omega,\mathbf{s}} = {}_2F_1\left(a,b;c;w\right)  $,
where $ a $, $ b $ and $ c $ are given by Equations~\ref{eq:longvectorexactsolutionfortcparams1}
and~\ref{eq:longvectorexactsolutionfortcparams2}.
Define:
\begin{equation}
\Delta \equiv c-a-b = \frac{\alpha-1}{2} = \frac{d-5}{2} \ .
\end{equation}

There are now several possible cases:
\begin{enumerate}
\item $ d $ and is even ($ \Delta $ is non-integer). In this case, the connection
formula in Equation~\ref{eq:hypergeometricevenconnectionformula} holds, so that:
\begin{multline}
\phi_{\omega,\mathbf{s}} = {}_2F_1\left(a,b;c;w\right)
= A \,{}_2F_1\left(a,b;-\Delta+1;1-w\right) \\
+ B (1-w)^\Delta\,{}_2F_1\left(c-a,c-b;\Delta+1;1-w\right)
\end{multline}
\begin{multline}
(1-w)^{\frac{\alpha+1}{2}}\partial_w\left[\frac{1}{(1-w)^\frac{\alpha-1}
{2}}\phi_{\omega,\mathbf{s}}\right]
= \frac{(c-a)(c-b)}{c}{}_2F_1(a,b;c+1;w) \\
= \tilde{A} \,{}_2F_1\left(a,b;-\Delta;1-w\right) \\
+ \tilde{B} (1-w)^{\Delta+1}\,{}_2F_1\left(c-a+1,c-b+1;\Delta+2;1-w\right) \ ,
\end{multline}
where:
\begin{align}
A &= \frac{\Gamma(c)\Gamma(\Delta)}{\Gamma(c-a)\Gamma(c-b)} \\
B &= \frac{\Gamma(c)\Gamma(-\Delta)}{\Gamma(a)\Gamma(b)} \\
\tilde{A} &= \frac{(c-a)(c-b)}{c}\frac{\Gamma(c+1)\Gamma(\Delta+1)}{\Gamma(c-a+1)\Gamma(c-b+1)}
= \Delta A \\
\tilde{B} &= \frac{(c-a)(c-b)}{c}\frac{\Gamma(c+1)\Gamma(-\Delta-1)}{\Gamma(a)\Gamma(b)} \ .
\end{align}
We proceed by calculating the limit:
\begin{equation}
\label{eq:longvectorexactcorrelatorlimitcalculation}
\left.\frac{\psi_{\omega,\mathbf{s}}}{\chi_{\omega,\mathbf{s}}}\right|_{w=1-\epsilon}
\approx \frac{q_\mathbf{s}^2}{2\Delta}\frac{B}{A}\epsilon^\frac{\alpha}{2} \ ,
\end{equation}
where contact terms have been dropped. Putting this into
Equation~\ref{eq:longvectorcorrelatorinw} we get:
\begin{equation}
\boxed{
G_{tt}^R(\omega,\mathbf{s})
= \frac{4C_v}{d-5}\frac{r_+^{d-3}}{R^{d-2}}q_\mathbf{s}^2
\frac{\Gamma(-\Delta)}{\Gamma(\Delta)}\frac{\Gamma(a+\Delta)}{\Gamma(a)}
\frac{\Gamma(b+\Delta)}{\Gamma(b)} \ .
}
\end{equation}
The other components of the longitudinal mode correlator can be calculated from
Equations~\ref{eq:longvectorcftcorrelator2} and~\ref{eq:longvectorcftcorrelator3}.

\item $ d>5 $ and is odd ($ \Delta>0 $ is an integer). In this case, the connection
formula in Equation~\ref{eq:hypergeometricoddconnectionformula} holds, so that:
\begin{equation}
\left.\phi_{\omega,\mathbf{s}}\right|_{w\to 1} = \left.{}_2F_1\left(a,b;c;w\right)\right|_{w\to 1}
= A+\ldots+B(1-w)^\Delta+\ldots
\end{equation}
\begin{multline}
\left.(1-w)^{\frac{\alpha+1}{2}}\partial_w\left[\frac{1}{(1-w)^\frac{\alpha-1}
{2}}\phi_{\omega,\mathbf{s}}\right]\right|_{w\to 1}\\
= \left.\frac{(c-a)(c-b)}{c}{}_2F_1(a,b;c+1;w)\right|_{w\to 1} \\
= \tilde{A}+\ldots+\tilde{B}(1-w)^{\Delta+1}+\ldots \ ,
\end{multline}
where:
\begin{align}
A &=  \frac{\Gamma(c)\Gamma(\Delta)}{\Gamma(c-a)\Gamma(c-b)} \\
B &=  \frac{\Gamma(c)}{\Gamma(a)\Gamma(b)}\frac{(-1)^{\Delta+1}}{(\Delta)!}
\left[\psi(a+\Delta)+\psi(b+\Delta)\right]\\
\tilde{A} &= \frac{(c-a)(c-b)}{c}\frac{\Gamma(c+1)\Gamma(\Delta+1)}
{\Gamma(c-a+1)\Gamma(c-b+1)} = \Delta A \\
\tilde{B} &=  \frac{(c-a)(c-b)}{c}\frac{\Gamma(c+1)}{\Gamma(a)\Gamma(b)}
\frac{(-1)^{\Delta+2}}{(\Delta+1)!}
\left[\psi(a+\Delta+1)+\psi(b+\Delta+1)\right] \ .
\end{align}

Proceeding in the same manner as in Equation~\ref{eq:longvectorexactcorrelatorlimitcalculation}
above, we have:
\begin{equation}
\left.\frac{\psi_{\omega,\mathbf{s}}}{\chi_{\omega,\mathbf{s}}}\right|_{w=1-\epsilon}
\approx \frac{q_\mathbf{s}^2}{2\Delta}\frac{B}{A}\epsilon^\frac{\alpha}{2} \ ,
\end{equation}
so that:
\begin{equation}
\boxed{
G_{tt}^R(\omega,\mathbf{s})
= 2C_v\frac{r_+^{d-3}}{R^{d-2}}q_\mathbf{s}^2\frac{(-1)^{\Delta+1}}{\Gamma^2(\Delta+1)}
(a)_\Delta(b)_\Delta\left[\psi(a+\Delta)+\psi(b+\Delta)\right] \ .
}
\end{equation}
The other components of the longitudinal mode correlator can be calculated from
Equations~\ref{eq:longvectorcftcorrelator2} and~\ref{eq:longvectorcftcorrelator3}.

\item $ d=5 $ ($ \Delta=0 $). In this case the following connection formula holds:
\begin{multline}
\phi_{\omega,\mathbf{s}} = {}_2F_1(a,b;c;w)
= -\frac{\Gamma(c)}{\Gamma(a)\Gamma(b)}\sum_{n=0}^{\infty} \frac{(a)_n(b)_n}{(n!)^2}(1-w)^n \ln(1-w)\\
 -\frac{\Gamma(c)}{\Gamma(a)\Gamma(b)}\sum_{n=0}^{\infty} \frac{(a)_n(b)_n}{(n!)^2}
 \left[\psi(a+n)+\psi(b+n)-2\psi(n+1)\right](1-w)^n
\end{multline}
(while for $ {}_2F_1(a,b;c+1;w) $ the formula in Equation~\ref{eq:hypergeometricoddconnectionformula}
still holds), so that:
\begin{equation}
\left.\phi_{\omega,\mathbf{s}}\right|_{w\to 1} = \left.{}_2F_1\left(a,b;c;w\right)\right|_{w\to 1}
= A\ln(1-w)+\ldots+B+\ldots
\end{equation}
\begin{multline}
\left.(1-w)^{\frac{\alpha+1}{2}}\partial_w\left[\frac{1}{(1-w)^\frac{\alpha-1}
{2}}\phi_{\omega,\mathbf{s}}\right]\right|_{w\to 1}\\
= \left.\frac{(c-a)(c-b)}{c}{}_2F_1(a,b;c+1;w)\right|_{w\to 1} \\
= \tilde{A}+\ldots+\tilde{B}(1-w)+\ldots \ ,
\end{multline}
where:
\begin{align}
A &=  -\frac{\Gamma(c)}{\Gamma(a)\Gamma(b)} \\
B &=  -\frac{\Gamma(c)}{\Gamma(a)\Gamma(b)}\left[\psi(a)+\psi(b)\right]\\
\tilde{A} &= \frac{(c-a)(c-b)}{c}\frac{\Gamma(c+1)}
{\Gamma(c-a+1)\Gamma(c-b+1)} = -A \\
\tilde{B} &=  \frac{(c-a)(c-b)}{c}\frac{\Gamma(c+1)}{\Gamma(a)\Gamma(b)}
\left[\psi(a+1)+\psi(b+1)\right] \ .
\end{align}

Calculating the limit, we get:
\begin{equation}
\left.\frac{\psi_{\omega,\mathbf{s}}}{\chi_{\omega,\mathbf{s}}}\right|_{w=1-\epsilon}
\approx -\frac{q_\mathbf{s}^2}{2}\frac{B}{A}\epsilon^\frac{1}{2} \ ,
\end{equation}
where contact terms have been dropped. Putting this into
Equation~\ref{eq:longvectorcorrelatorinw} we get:
\begin{equation}
G_{tt}^R(\omega,\mathbf{s})
= -2C_v\frac{r_+^{2}}{R^{3}}q_\mathbf{s}^2 \left[\psi(a)+\psi(b)\right] \ .
\end{equation}

This result also matches the general result for $ d>5 $ in Equation~\ref{eq:longvectorexactcorrelatorintcodd}, so that one may extend that result
to $ d=5 $ as well.
The other components of the longitudinal mode correlator can be calculated from
Equations~\ref{eq:longvectorcftcorrelator2} and~\ref{eq:longvectorcftcorrelator3}.

\end{enumerate}

As for the transverse vector mode, the calculations are very similar to the scalar case.
Writing the expression in Equation~\ref{eq:transvectorcftcorrelator} in terms of $ \phi $
and $ w $ we get:
\begin{equation}
G_{\bot\bot}^R(\omega,\mathbf{v})
=\left.-8C_v\frac{r_+^{d-3}}{R^{d-2}}\frac{1}{(1-w)^\frac{d-5}{2}}
\frac{\partial_w\phi_{\omega,\mathbf{v}}}{\phi_{\omega,\mathbf{v}}}\right|_{w\to 1} \ .
\end{equation}
The solution to the EOM with an incoming-wave boundary condition at the horizon is again
$ \phi_{\omega,\mathbf{s}} = {}_2F_1\left(a,b;c;w\right)  $,
where $ a,b $ and $ c $ are given by Equations~\ref{eq:scalartransexactsolutionfortcparams1}
and~\ref{eq:scalartransexactsolutionfortcparams2}. The rest of the calculations
are identical to the scalar case, except with:
\begin{equation}
\Delta \equiv c-a-b = \frac{\alpha+1}{2} = \frac{d-3}{2} \ ,
\end{equation}
so that we get for even $ d $:
\begin{equation}
\boxed{
G_{\bot\bot}^R(\omega,\mathbf{v})
= 4(d-3)C_v \frac{r_+^{d-3}}{R^{d-2}}\frac{\Gamma(-\Delta)}{\Gamma(\Delta)}
\frac{\Gamma(a+\Delta)}{\Gamma(a)}\frac{\Gamma(b+\Delta)}{\Gamma(b)} \ ,
}
\end{equation}
and for odd $ d $:
\begin{equation}
\boxed{
G_{\bot\bot}^R(\omega,\mathbf{v})
= 8C_v\frac{r_+^{d-3}}{R^{d-2}}\frac{(-1)^{\Delta+1}}{\Gamma^2(\Delta)}(a)_\Delta(b)_\Delta
\left[\psi(a+\Delta)+\psi(b+\Delta)\right] \ .
}
\end{equation}

\section{Effective Potentials and Their Asymptotics}
\label{app:effectivepotentials}

In order to obtain an asymptotic expression for the QNM frequencies and corresponding correlation
functions, the Equations~\ref{eq:scalareq},~\ref{eq:longvectoreq} and~\ref{eq:transvectoreq} must first
be put into the Schr\"odinger form:
\begin{equation}
\label{eq:schrodingerform}
\partial_{\zt}^2\phi + \left[ \lambda^2-V(\zt) \right]\phi = 0 \ .
\end{equation}

Both the scalar equation and the ``transverse'' vector equation are of the form
\begin{equation}
\label{eq:scalartransvectoreq}
(1-z)^\alpha\partial_z\left[ \frac{\tilde{g}(z)}{(1-z)^\alpha} \partial_z \psi \right] +
\left[ \frac{\lambda^2}{\tilde{g}(z)} - q^2 \right] \psi = 0 \ ,
\end{equation}
where
\begin{equation}
\alpha=
\begin{cases}
d-2 & \text{for scalar},\\
d-4 & \text{for transverse vector}
\end{cases} \ .
\end{equation}
In these cases we can bring the equation to the form in Equation~\ref{eq:schrodingerform} by defining
\begin{equation}
\psi(z) = (1-z)^{\frac{\alpha}{2}} \phi(z) \ .
\end{equation}
After substituting $ \psi $ in Equation~\ref{eq:scalartransvectoreq} and some algebra we get the equation
\begin{equation}
\gz\pdz\left[\gz\pdz\phi\right] + \left[\lambda^2-V_1(z)\right] = 0 \ ,
\end{equation}
where
\begin{multline}
V_1(z) = \frac{\alpha}{2}\gz(1-z)^{\frac{\alpha}{2}}\pdz\left(\frac{\gz}{(1-z)^{\frac{\alpha}{2}+1}} \right) + q^2\gz\\
= \frac{\alpha}{2}\frac{\gz\pdz\gz}{1-z} + \frac{\alpha}{2}\left(\frac{\alpha}{2}+1\right)\frac{\tilde{g}^2(z)}{(1-z)^2} + q^2\gz \ .
\end{multline}
The boundary conditions are accordingly
\begin{equation}
\left.\phi\right|_{z=0} \sim z^{-\frac{i\lambda}{C}}
\qquad
\left.(1-z)^{\frac{\alpha}{2}}\phi\right|_{z=1} = 0 \ .
\end{equation}

Next we define a new coordinate as follows:
\begin{equation}
\zt \equiv \int \frac{1}{\gz} \ud z \ ,
\end{equation}
so that $ \gz\pdz = \pdzt $. Changing to this new coordinate we obtain the form of
Equation~\ref{eq:schrodingerform} with $ V_1\left(z(\zt)\right) $ as the effective potential.

The ``longitudinal'' vector equation is:
\begin{equation}
\label{eq:longvectoreq_withalpha}
\partial_z \left[ \tilde{g}(z)(1-z)^{\alpha} \partial_z \left( \frac{1}{(1-z)^{\alpha}} \psi \right) \right] + \left[ \frac{\lambda^2}{\tilde{g}(z)} - q_s^2 \right] \psi = 0 \ ,
\end{equation}
where
\begin{equation}
\alpha = d-4 \ .
\end{equation}
Define again
\begin{equation}
\psi(z) = (1-z)^{\frac{\alpha}{2}} \phi(z) \ .
\end{equation}
Substituting into Equation~\ref{eq:longvectoreq_withalpha} we again get
\begin{equation}
\gz\pdz\left[\gz\pdz\phi\right] + \left[\lambda^2-V_2(z)\right]\phi = 0 \ ,
\end{equation}
only with the effective potential
\begin{multline}
V_2(z) = -\frac{\alpha}{2}\gz(1-z)^{-\frac{\alpha}{2}}\pdz\left((1-z)^{\frac{\alpha}{2}-1}\gz\right)\\
= -\frac{\alpha}{2}\frac{\gz\pdz\gz}{1-z}+\frac{\alpha}{2}\left(\frac{\alpha}{2}-1\right)\frac{\tilde{g}^2(z)}{(1-z)^2} + q^2 \gz \ .
\end{multline}
The boundary conditions are accordingly
\begin{equation}
\left.\phi\right|_{z=0} \sim z^{-\frac{i\lambda}{C}}
\qquad
\left.(1-z)^{\alpha}\pdz\left[(1-z)^{-\frac{\alpha}{2}}\phi\right]\right|_{z=1} = 0 \ .
\end{equation}

Transforming again to $ \zt $ we obtain the form of Equation~\ref{eq:schrodingerform} with $ V_2\left(z(\zt)\right) $ as the effective potential.

Looking at the effective potentials, we can see their behaviour near the horizon, near the boundary and at $ z\to\infty $ (or $ \zt=\zt_0 $).

At the horizon ($ z\to 0 $, $ \zt\to -\infty $) we have:
\begin{align}
V_1(z) &\to 0\\
V_2(z) &\to 0 \ .
\end{align}

At the boundary ($ z\to 1 $, $ \zt\to 0 $) we have:
\begin{align}
V_1(z) &\approx \frac{\alpha}{2}\left(\frac{\alpha}{2}+1\right) \frac{1}{(1-z)^2} \approx \frac{\alpha}{2}\left(\frac{\alpha}{2}+1\right)\frac{1}{\zt^2} = \frac{j_1^2-1}{4\zt^2}\\
V_2(z) &\approx \frac{\alpha}{2}\left(\frac{\alpha}{2}-1\right) \frac{1}{(1-z)^2} \approx \frac{\alpha}{2}\left(\frac{\alpha}{2}-1\right)\frac{1}{\zt^2} = \frac{j_1^2-1}{4\zt^2} \ ,
\end{align}
so that, generally
\begin{equation}
V(z) \approx \frac{j_1^2-1}{4\zt^2}
\qquad
j_1 =
\begin{cases}
d-1 & \text{for scalar},\\
d-3 & \text{for transverse vector},\\
|d-5| & \text{for longitudinal vector}
\end{cases} \ .
\end{equation}

At the limit $ z\to\infty $ ($ \zt\to\zt_0 $) we have:
\begin{equation}
V_1(z)\approx\frac{\alpha}{2}(1+K)^2\left[\frac{\alpha}{2}-d+2\right]z^{2d-4}\\
= \frac{1}{(d-2)^2}\frac{\alpha}{2}\left[\frac{\alpha}{2}-d+2\right]\frac{1}{(\zt-\zt_0)^2}
\end{equation}
\begin{equation}
V_2(z) \approx \frac{\alpha}{2}(1+K)^2\left[\frac{\alpha}{2}+d-2\right]z^{2d-4}\\
= \frac{1}{(d-2)^2}\frac{\alpha}{2}\left[\frac{\alpha}{2}+d-2\right]\frac{1}{(\zt-\zt_0)^2} \ ,
\end{equation}
so that, generally
\begin{equation}
V(z) \approx \frac{j_{\infty}^2-1}{4(\zt-\zt_0)^2}
\qquad
j_\infty =
\begin{cases}
0 & \text{for scalar,}\\
\frac{2}{d-2} & \text{for transverse vector,}\\
2-\frac{2}{d-2} & \text{for longitudinal vector}
\end{cases} \ .
\end{equation}

\section{Asymptotic Correlators for the Vector Case}
\label{app:cftasymptoticcorrelatorexpressionsvector}

Proceeding from the definitions in Subsection~\ref{subsec:qnmasymoptotics}, the expressions in
Equation~\ref{eq:longvectorcftcorrelator} and~\ref{eq:transvectorcftcorrelator} can be written in
terms of $ \phi $ and $ \zt $:
\begin{equation}
\label{eq:longvectorcftcorrelatorinzt}
G_{tt}^R(\omega,\mathbf{s})
= \left.4C_v\frac{r_+^{d-3}}{R^{d-2}}\frac{(-1)^d}{\zt^{d-4}}q_\mathbf{s}^2
\frac{\phi_{\omega,\mathbf{s}}}{\zt^\frac{\alpha}{2}\pdzt\left[\frac{1}{\zt^\frac{\alpha}{2}}
\phi_{\omega,\mathbf{s}}\right]}\right|_{\zt\to 0}
\end{equation}
\begin{equation}
\label{eq:transvectorcftcorrelatorinzt}
G_{\bot\bot}^R(\omega,\mathbf{v})
= \left.-4C_v\frac{r_+^{d-3}}{R^{d-2}}\frac{(-1)^d}{\zt^{d-4}}
\frac{\pdzt \phi_{\omega,\mathbf{v}}}{\phi_{\omega,\mathbf{v}}}\right|_{\zt\to 0} \ .
\end{equation}

We again continue
with the method applied in Subsection~\ref{subsec:qnmasymoptotics} and
Subsubsection~\ref{subsubsec:cftasymptoticcorrelatorexpressionsscalar}.
In this case, the solution around
$ \zt\to\zt_0 $ ($ z\to\infty $) is given by Equation~\ref{eq:monoasympinftyplus} in direction (1)
and Equation~\ref{eq:monoasympinftymin} in direction (2).
From the boundary condition at the horizon ($ z=0 $, $ \zt\to -\infty $) we have:
\begin{equation}
\phi \approx \e^{-i\lambda\zt} \ .
\end{equation}

We again distinguish between two possible cases:
\begin{enumerate}

\item $ d $ is even. In this case the solution around $ \zt\to 0 $ is approximately:
\begin{equation}
\phi \approx B_+ P_+(\zt) + B_- P_-(\zt)\\
= B_+\sqrt{2\pi\lambda\zt} J_{\frac{j_1}{2}}(\lambda\zt)
+ B_-\sqrt{2\pi\lambda\zt} J_{-\frac{j_1}{2}}(\lambda\zt) \ ,
\end{equation}
where:
\begin{equation}
j_1 =
\begin{cases}
d-3 & \text{for transverse vector,}\\
|d-5| & \text{for longitudinal vector}
\end{cases}
\end{equation}
(so that $ \frac{j_1}{2} > 0 $ is non-integer). Replacing the Bessel functions
with their asymptotic form, we have:
\begin{equation}
\phi
\approx \left[B_+\e^{-i\beta_+} + B_-\e^{-i\beta_-}\right]\e^{i\lambda\zt} +
 \left[B_+\e^{i\beta_+} + B_-\e^{i\beta_-}\right]\e^{-i\lambda\zt} \ ,
\end{equation}
where $ \beta_\pm \equiv \frac{\pi}{4}(1 \pm j_1)$.
We now match the solutions on lines (1) and (2). We have for line (1):
\begin{equation}
\label{eq:cftmonoeq9}
A_+\e^{i(-\lambda\zt_0-\alpha_+)} + A_-\e^{i(-\lambda\zt_0-\alpha_-)} = 0 \ .
\end{equation}
For the section (2) we have:
\begin{align}
\label{eq:cftmonoeq10}
A_+\e^{i(-\lambda\zt_0+3\alpha_+)} + A_-\e^{i(-\lambda\zt_0+3\alpha_-)}
&= B_+\e^{-i\beta_+} + B_-\e^{-i\beta_-} \\
\label{eq:cftmonoeq11}
A_+\e^{i(\lambda\zt_0+\alpha_+)} + A_-\e^{i(\lambda\zt_0+\alpha_-)}
&= B_+\e^{i\beta_+} + B_-\e^{i\beta_-} \ .
\end{align}
Equations~\ref{eq:cftmonoeq9},~\ref{eq:cftmonoeq10} and~\ref{eq:cftmonoeq11} form a linear system
of equations.
Solving for $ B_\pm $, $ A_- $ we can get an expression for $ \frac{B_+}{B_-} $:
\begin{equation}
\frac{B_+}{B_-}
= -i^{j_1}\frac{\e^{2i\theta_-}-2\cos\left(\frac{\pi}{2}j_\infty\right)}
{\e^{2i\theta_-}+2\cos\left(\frac{\pi}{2}j_\infty\right)} \ ,
\end{equation}
where $ \theta_- \equiv \lambda\zt_0 + \frac{\pi}{4}(j_1 -2) $.

Next we evaluate the correlators. Developing $ \phi $ around $ \zt=0 $, we again have:
\begin{equation}
\phi = A\zt^{-\Delta+\frac{1}{2}}+\ldots+B\zt^{\Delta+\frac{1}{2}}+\ldots \ ,
\end{equation}
with:
\begin{align}
A &= \frac{\sqrt{2\pi}\lambda^{-\Delta+\frac{1}{2}}}
{2^{-\Delta}\Gamma(-\Delta+1)} B_- \\
B &= \frac{\sqrt{2\pi}\lambda^{\Delta+\frac{1}{2}}}
{2^{\Delta}\Gamma(\Delta+1)} B_+
\end{align}
(where $ \Delta\equiv \frac{j_1}{2} $).
For the longitudinal vector mode (for which $j_1 = d-5 = \alpha-1 = 2\Delta  $), we have:
\begin{equation}
\left.\frac{\psi_{\omega,\mathbf{s}}}{\chi_{\omega,\mathbf{s}}}\right|_{\zt=\epsilon}
\approx -\frac{q_\mathbf{s}^2}{j_1} \frac{B}{A}\epsilon^\alpha \ .
\end{equation}
Putting this into Equation~\ref{eq:longvectorcftcorrelatorinzt}, and using the fact that for
the longitudinal vector mode $ j_\infty = 2-\frac{2}{d-2} $ and the symmetry properties of the retarded
correlator (as in Subsubsection~\ref{subsubsec:cftasymptoticcorrelatorexpressionsscalar}),
we obtain the correlator:
\begin{empheq}[box=\fbox]{multline}
G_{tt}^R(\omega,\mathbf{s})
\approx \frac{4C_v}{d-5}R^{d-4}
(L_\mathbf{s}^{FT})^2\frac{\Gamma(-\Delta)}{\Gamma(\Delta)}\left(\frac{i\omega}{2}\right)^{d-5}\\
\frac{\e^{2i\theta_-}+2\cos\left(\frac{\pi}{d-2}\right)}
{\e^{2i\theta_-}-2\cos\left(\frac{\pi}{d-2}\right)}
\frac{\e^{2i\overline{\theta_-}}+2\cos\left(\frac{\pi}{d-2}\right)}
{\e^{2i\overline{\theta_-}}-2\cos\left(\frac{\pi}{d-2}\right)} \ ,
\end{empheq}
where:
\begin{equation}
\overline{\theta_-}(\lambda) \equiv -\theta_-^*(-\lambda^*)
= \lambda\zt_0^*-\frac{\pi}{4}(d-7) \ .
\end{equation}
The other components of the longitudinal mode correlator can be calculated from
Equations~\ref{eq:longvectorcftcorrelator2} and~\ref{eq:longvectorcftcorrelator3}.

For the transverse vector mode (for which $ j_1 = d-3 = \alpha+1 = 2\Delta $), we have:
\begin{equation}
\left.\frac{\pdzt\phi_{\omega,\mathbf{v}}}{\phi_{\omega,\mathbf{v}}}\right|_{\zt=\epsilon}
= j_1 \frac{B}{A}\epsilon^\alpha \ .
\end{equation}
Putting this into Equation~\ref{eq:transvectorcftcorrelatorinzt}, and using the fact that for
the transverse vector mode $ j_\infty = \frac{2}{d-2} $ and the symmetry properties of the retarded
correlator , we obtain the result:

\begin{empheq}[box=\fbox]{multline}
G_{\bot\bot}^R(\omega,\mathbf{v})
\approx 4(d-3)C_v R^{d-4}
\frac{\Gamma(-\Delta)}{\Gamma(\Delta)}\left(\frac{i\omega}{2}\right)^{d-3}\\
\frac{\e^{2i\theta_-}-2\cos\left(\frac{\pi}{d-2}\right)}
{\e^{2i\theta_-}+2\cos\left(\frac{\pi}{d-2}\right)}
\frac{\e^{2i\overline{\theta_-}}-2\cos\left(\frac{\pi}{d-2}\right)}
{\e^{2i\overline{\theta_-}}+2\cos\left(\frac{\pi}{d-2}\right)} \ ,
\end{empheq}
where:
\begin{equation}
\overline{\theta_-}(\lambda) \equiv -\theta_-^*(-\lambda*)
= \lambda\zt_0^*-\frac{\pi}{4}(d-5) \ .
\end{equation}

\item $ d $ is odd (so that $ \frac{j_1}{2} \geq 0 $ is an integer). In this case the solution around
$ \zt\to 0 $ is approximately:
\begin{equation}
\phi \approx B_+ P_+(\zt) + B_- P_-(\zt)\\
= B_+\sqrt{2\pi\lambda\zt} J_{\frac{j_1}{2}}(\lambda\zt)
+ B_-\sqrt{2\pi\lambda\zt} Y_{\frac{j_1}{2}}(\lambda\zt) \ .
\end{equation}

Replacing the Bessel functions
with their asymptotic form, we have:
\begin{equation}
\phi
\approx \left[(B_+-iB_-)\e^{-i\beta_+}\right]\e^{i\lambda\zt} +
\left[(B_++iB_-)\e^{i\beta_+}\right]\e^{-i\lambda\zt} \ ,
\end{equation}
where $ \beta_+ = \frac{\pi}{4}(1+j_1) $.
We now match the solutions on lines (1) and (2). We have for line (1):
\begin{equation}
\label{eq:cftmonoeq12}
A_+\e^{i(-\lambda\zt_0-\alpha_+)} + A_-\e^{i(-\lambda\zt_0-\alpha_-)} = 0 \ .
\end{equation}
For the section (2) we have:
\begin{align}
\label{eq:cftmonoeq13}
A_+\e^{i(-\lambda\zt_0+3\alpha_+)} + A_-\e^{i(-\lambda\zt_0+3\alpha_-)}
&= (B_+ -iB_-)\e^{-i\beta_+} \\
\label{eq:cftmonoeq14}
A_+\e^{i(\lambda\zt_0+\alpha_+)} + A_-\e^{i(\lambda\zt_0+\alpha_-)}
&= (B_+ +iB_-)\e^{i\beta_+} \ .
\end{align}
Equations~\ref{eq:cftmonoeq12},~\ref{eq:cftmonoeq13} and~\ref{eq:cftmonoeq14} form a linear system
of equations.
Solving for $ B_\pm $, $ A_- $ we get an expression for $ \frac{B_+}{B_-} $:
\begin{equation}
\frac{B_+}{B_-}
= 2i\frac{\e^{2i\theta_+}}{\e^{2i\theta_+} - 2\cos\left(\frac{\pi}{2}j_\infty\right)} - i \ ,
\end{equation}
where $ \theta_+ \equiv \lambda\zt_0 - \frac{\pi}{4}(j_1 +2) $.

Next we evaluate the correlators.
We first assume that $ j_1>0 $ (this is satisfied for the transverse mode with $ d\geq 5 $, or the
longitudinal mode with $ d>5 $).
Developing $ \phi $ around $ \zt=0 $, we again have:
\begin{equation}
\phi = A\zt^{-\Delta+\frac{1}{2}}+\ldots+B\zt^{\Delta+\frac{1}{2}}+\ldots \ ,
\end{equation}
with:
\begin{equation}
A = -\frac{\sqrt{2\pi}\lambda^{-\Delta+\frac{1}{2}}}{\pi 2^{-\Delta}}
(\Delta-1)!\,B_-
\end{equation}
\begin{multline}
B = \frac{2\sqrt{2\pi}\lambda^{\Delta+\frac{1}{2}}}
{\pi 2^\Delta\Gamma(\Delta+1)}\ln\left(-\frac{\lambda}{2}\right)\,B_- \\
-\frac{\sqrt{2\pi}\lambda^{\Delta+\frac{1}{2}}}{\pi 2^\Delta(\Delta)!}
\left[\psi(1)+\psi(\Delta+1)\right]\,B_-
+\frac{\sqrt{2\pi}\lambda^{\Delta+\frac{1}{2}}}{2^\Delta \Gamma(\Delta+1)}\,B_+
\end{multline}
(where $ \Delta\equiv \frac{j_1}{2} $).
For the longitudinal vector mode (for which $j_1 = d-5 = \alpha-1 = 2\Delta  $), we again have:
\begin{equation}
\left.\frac{\psi_{\omega,\mathbf{s}}}{\chi_{\omega,\mathbf{s}}}\right|_{\zt=\epsilon}
\approx -\frac{q_\mathbf{s}^2}{j_1} \frac{B}{A}\epsilon^\alpha\ .
\end{equation}
Putting this into Equation~\ref{eq:longvectorcftcorrelatorinzt}, and using the fact that for
the longitudinal vector mode $ j_\infty = 2-\frac{2}{d-2} $ and the symmetry properties of the
correlator, we obtain the correlator:
\begin{empheq}[box=\fbox]{multline}
G_{tt}^R(\omega,\mathbf{s})
\approx 2C_v R^{d-4}(L_\mathbf{s}^{FT})^2\frac{(-1)^{\Delta+1}}{\Gamma^2(\Delta+1)}
\left(\frac{i\omega}{2}\right)^{d-5} \\
\left[2\pi i\frac{\e^{2i\theta_+}}{\e^{2i\theta_+} + 2\cos\left(\frac{\pi}{d-2}\right)}
- 2\pi i\frac{\e^{2i\overline{\theta_+}}}{\e^{2i\overline{\theta_+}} + 2\cos\left(\frac{\pi}{d-2}\right)}
+ 2\ln\left(\frac{i\lambda}{2}\right)\right] \ ,
\end{empheq}
where:
\begin{equation}
\overline{\theta_-}(\lambda) \equiv -\theta_-^*(-\lambda*)
= \lambda\zt_0^*+\frac{\pi}{4}(d-3) \ .
\end{equation}
The other components of the longitudinal mode correlator can be calculated from
Equations~\ref{eq:longvectorcftcorrelator2} and~\ref{eq:longvectorcftcorrelator3}.

Similarly, for the transverse vector mode (for which $ j_1 = d-3 = \alpha+1 = 2\Delta $), we get:
\begin{empheq}[box=\fbox]{multline}
G_{\bot\bot}^R(\omega,\mathbf{v})
\approx 8C_v R^{d-4}
\frac{(-1)^{\Delta+1}}{\Gamma^2(\Delta)}
\left(\frac{i\omega}{2}\right)^{d-3}\\
\left[2\pi i\frac{\e^{2i\theta_+}}{\e^{2i\theta_+} - 2\cos\left(\frac{\pi}{d-2}\right)}
- 2\pi i\frac{\e^{2i\overline{\theta_+}}}{\e^{2i\overline{\theta_+}} - 2\cos\left(\frac{\pi}{d-2}\right)}
+ 2\ln\left(\frac{i\lambda}{2}\right)\right] \ ,
\end{empheq}
where:
\begin{equation}
\overline{\theta_-}(\lambda) \equiv -\theta_-^*(-\lambda*)
= \lambda\zt_0^*+\frac{\pi}{4}(d-1) \ .
\end{equation}

Next we turn to the case of $ j_1=0 $, which is satisfied for the longitudinal mode with $ d=5 $.
In this case, Equation~\ref{eq:asymptoticcorrelatorsphiaround0odd} is still valid with
$ \Delta = \frac{j_1}{2} = 0 $, so that:
\begin{equation}
\phi = A\zt^\frac{1}{2}\ln(-\zt)+\ldots+B\zt^\frac{1}{2}+\ldots \ ,
\end{equation}
with:
\begin{equation}
A = \frac{2\sqrt{2\pi}\lambda^\frac{1}{2}}{\pi}B_-
\end{equation}
\begin{equation}
B
= \frac{2\sqrt{2\pi}\lambda^\frac{1}{2}}{\pi}\left[\ln\left(-\frac{\lambda}{2}\right)-\psi(1)\right]B_-
+ \sqrt{2\pi}\lambda^\frac{1}{2}B_+ \ .
\end{equation}
This case is restricted for the longitudinal case, for which we have:
\begin{equation}
\left.\frac{\psi_{\omega,\mathbf{s}}}{\chi_{\omega,\mathbf{s}}}\right|_{\zt=\epsilon}
= \left.q_\mathbf{s}^2 \frac{\phi_{\omega,\mathbf{s}}}{\zt^\frac{1}{2}\pdzt\left[\frac{1}
{\zt^\frac{1}{2}}\phi_{\omega,\mathbf{s}}\right]}\right|_{\zt=\epsilon}\\
= q_\mathbf{s}^2\left[\epsilon\ln(-\epsilon)+\ldots+\frac{B}{A}\epsilon+\ldots\right]
\approx q_\mathbf{s}^2 \frac{B}{A}\epsilon \ ,
\end{equation}
where contact terms were dropped in the last equality.
Putting $ \frac{B}{A} $ into this expression we get:
\begin{equation}
\left.\frac{\psi_{\omega,\mathbf{s}}}{\chi_{\omega,\mathbf{s}}}\right|_{\zt=\epsilon}
= \frac{q_\mathbf{s}^2}{2}\left[\pi\frac{B_+}{B_-}+2\ln\left(-\frac{\lambda}{2}\right)\right] \epsilon
\ .
\end{equation}
This result is consistent with the corresponding expression for $ d>5 $, so that we may
extend the rest of the results for $ d>5 $ to $ d=5 $ as well.

\end{enumerate}

\section{Numerical Methods}
\label{app:numericalmethods}

Here we outline the numerical methods used for the exact calculation of the QNM spectrum and the
retarded correlation functions associated with the massless scalar and vector perturbation modes.

\subsection{QNM spectrum calculation}
\label{appsubsec:qnmnumericalmethod}

For each of the QNM Equations~\ref{eq:scalareq},~\ref{eq:longvectoreq} and~\ref{eq:transvectoreq}
and their corresponding boundary conditions, given the values of the parameters $ d $, $ K $ and $ q $,
one may use the Frobenius method to calculate the exact values of $ \lambda $ for which the eqautions
have non-trivial solutions. The stages of the calculation are as follows:

\begin{enumerate}

\item We write the QNM equation in the form:
\begin{equation}
\label{eq:qnmequationpolynomform}
P_1(z)\pdz^2\psi + P_2(z)\pdz\psi + P_3(z)\psi = 0 \ ,
\end{equation}
where $ P_1(z) $, $ P_2(z) $ and $ P_3(z) $ are polynomials in $ z $ that depend on the values of
$ d $, $ K $, $ q $ and $ \lambda $. We write the boundary condition at $ z=1 $ (the AdS boundary) in a
similar form:
\begin{equation}
\left.Q_1(z)\pdz\psi + Q_2(z)\psi\right|_{z\to 1} = 0 \ ,
\end{equation}
where $ Q_1(z) $ and $ Q_2(z) $ are polynomials in $ z $\footnote{This boundary condition should be
interpreted at the first order of $ 1-z $ that doesn't trivially vanish.}.
For example, for the scalar QNM perturbation we have:
\begin{align}
P_1(z) &= \tilde{g}^2(z)(1-z) \\
P_2(z) &= \gz\left[\tilde{g}'(z)(1-z)+(d-2)\gz\right] \\
P_3(z) &= (1-z)\left[\lambda^2-q^2\gz\right] \\
Q_1(z) &= 0 \\
Q_2(z) &= 1 \ .
\end{align}

\item We calculate the exponents of the QNM equation at $ z=0 $ and $ z=1 $ via its indicial equation.
At $ z=0 $ we have:
\begin{equation}
\left.\psi\right|_{z=0} \sim z^{\pm\frac{i\lambda}{C}} \ ,
\end{equation}
where $ C=\tilde{g}'(0) $. In accordance with the incoming-wave boundary condition at the horizon,
we always choose the exponent $ \gamma = -\frac{i\lambda}{C}$.
At $ z=1 $ we get the real exponents $ \beta_1 $ and $ \beta_2 $, and we define:
\begin{equation}
\beta \equiv \min(\beta_1,\beta_2) \ .
\end{equation}

\item We make the transformation:
\begin{equation}
\label{eq:qnmnumericaltransform}
\psi = z^\gamma(1-z)^\beta\phi \ ,
\end{equation}
so that $ \phi $ is regular and finite at $ z=0 $. Re-writing the equation in terms of $ \phi $
we get:
\begin{equation}
\label{eq:numericalnormalizedqnmequation}
\widetilde{P}_1(z)\pdz^2\phi + \widetilde{P}_2(z)\pdz\phi + \widetilde{P}_3(z)\phi = 0 \ ,
\end{equation}
with the boundary condition at the horizon:
\begin{equation}
\left.\widetilde{Q}_1(z)\pdz\phi + \widetilde{Q}_2(z)\phi\right|_{z\to 1} = 0 \ ,
\end{equation}
where $ \widetilde{P}_1(z) $, $ \widetilde{P}_2(z) $, $ \widetilde{P}_3(z) $, $ \widetilde{Q}_1(z) $ and
$ \widetilde{Q}_2(z) $ are polynomials in $ z $. We assume that $ \widetilde{P}_1 $, $ \widetilde{P}_2 $
and $ \widetilde{P}_3 $ are given by:
\begin{align}
\widetilde{P}_1(z) &= \sum_{k=0}^{n_1} c_k^1 z^k \\
\widetilde{P}_2(z) &= \sum_{k=0}^{n_2} c_k^2 z^k \\
\widetilde{P}_3(z) &= \sum_{k=0}^{n_3} c_k^3 z^k \ .
\end{align}

\item In order to increase the numerical stability of the calculation and decrease the time required for
calculation we minimize the degree of the polynomials in the equation by dividing  $ \widetilde{P}_1 $,
$ \widetilde{P}_2 $ and $ \widetilde{P}_3 $ by their greatest common divisor (as polynomials in $ z $).

\item We solve the equation using the Frobenius method:
We develop $ \phi $ as a power series around $ z=0 $ up to the $ N $-th degree:
\begin{equation}
\label{eq:numericalexpansion}
\phi(z) = \sum_{k=1}^{N+1} a_{k-1}z^{k-1}
\end{equation}
We then put the series into the Equation~\ref{eq:numericalnormalizedqnmequation} and
obtain an equation for each power of $ z $ up to the $ (N-1) $-th degree. The homogeneous system
of equations reads:
\begin{equation}
S\mathbf{a} = 0 \ ,
\end{equation}
where
\begin{equation}
\mathbf{a} =
\begin{pmatrix}
a_0 \\
a_1 \\
\vdots \\
a_N
\end{pmatrix} \ ,
\end{equation}
and
\begin{equation}
\label{eq:numericallinearequations}
S_{ij} = c_{i-j+2}^1 (j-1)(j-2) + c_{i-j+1}^2 (j-1) + c_{i-j}^3
\end{equation}
(for $ 1 \leq i \leq N $, $ 1 \leq i \leq N+1 $ and $ -2 \leq i-j \leq \max(n_1-2,n_2-1,n_3) $).
The boundary condition at $ z=1 $ gives another equation:
In the case of $ \beta_1 = \beta_2 $, the series in Equation~\ref{eq:numericalexpansion} doesn't
converge at $ z=1 $ for $ N\to\infty $, since one of the solutions to
Equation~\ref{eq:numericalnormalizedqnmequation} goes like $ \log(1-z) $ near $ z=1 $, while the other
is finite at $ z=1 $. In order to choose the finite solution (as dictated by the boundary condition in
this case), we choose the following condition that is necessary for convergence:
\begin{equation}
a_N = 0 \ .
\end{equation}
In the case of $ \beta_1 \neq \beta_2 $, we have the boundary condition:
\begin{equation}
b_1\pdz\phi + b_2\phi = 0 \ ,
\end{equation}
where:
\begin{equation}
b_1 = \left.\frac{\widetilde{Q}_1(z)}{(1-z)^s}\right|_{z\to 1}
\qquad
b_2 = \left.\frac{\widetilde{Q}_2(z)}{(1-z)^s}\right|_{z\to 1} \ ,
\end{equation}
and $ s $ is the minimal exponent such that $ b_1 \neq 0 $ or $ b_2 \neq 0 $.
In terms of the series expansion this gives the equation:
\begin{equation}
\sum_{j=1}^{N+1} S_{N+1,j}a_{j-1} = 0 \ ,
\end{equation}
where
\begin{equation}
S_{N+1,j} = b_1(j-1) + b_2 \ .
\end{equation}
We end up with a linear system of $ N+1 $ equations, where the coefficients are polynomials in
$ \lambda $:
\begin{equation}
S_{ij}(\lambda,d,K,q) = \sum_{l=0}^M S_{ij}^{(l)}(d,K,q)\lambda^l \ .
\end{equation}
Finally we find all the values of $ \lambda $ for which a non-trivial solution exists to the system
of equations, by solving the generalized eigenvalue problem:
\begin{equation}
\det(S) = \det\left(\sum_{l=0}^M S_{ij}^{(l)}\lambda^l\right) = 0 \ .
\end{equation}

\item We filter out all of the values of $ \lambda $ that don't converge as we take $ N\to\infty $,
by performing the above calculation for the $ N $-th degree and for the $ (N+1) $-th degree,
and including only values of $ \lambda $ that satisfy the condition:
\begin{equation}
\left| \lambda_k^{(N+1)} - \lambda_k^{(N)} \right| < \epsilon \ ,
\end{equation}
where $ \epsilon $ is some convergence threshold.

\end{enumerate}

\subsection{Correlators calculation}

For each type of perturbation discussed in this chapter, and given the values of the parameters $ d $,
$ K $, $ q $ and $ \lambda $, we outline a numerical method based on the Frobenius solution to calculate
the exact value of the retarded correlation function of the dual gauge theory operators. The
calculation of the correlators require the evaluation of the expressions in
Equations~\ref{eq:scalarcftcorrelator},~\ref{eq:longvectorcftcorrelator}
and~\ref{eq:transvectorcftcorrelator}. The stages of the calculation are as follows:

\begin{enumerate}

\item As in Subsection~\ref{appsubsec:qnmnumericalmethod}, we write the relevant QNM equation in the
form of Equation~\ref{eq:qnmequationpolynomform}.

\item We again calculate the exponents of the QNM equation at $ z=0 $ and $ z=1 $ via the indicial
equation. At the horizon we choose the the exponent $ \gamma = -\frac{i\lambda}{C} $ corresponding to
the incoming-wave condition (and the retarded correlator in the gauge theory). At $ z=1 $ we get the
real exponents $ \beta_1 $ and $ \beta_2 $, and we define:
\begin{equation}
\beta \equiv \min(\beta_1,\beta_2)
\qquad
\beta' \equiv \max(\beta_1,\beta_2) - \min(\beta_1,\beta_2) \ .
\end{equation}

\item We again perform the transformation in Equation~\ref{eq:qnmnumericaltransform}, and get the
Equation~\ref{eq:numericalnormalizedqnmequation}.

\item We divide the polynomials $ \widetilde{P}_1 $, $ \widetilde{P}_2 $ and $ \widetilde{P}_3 $ by
their greatest common divisor.

\item We solve the equation around $ z=0 $ using the Frobenius method: We develop $ \phi $ as a power
series around $ z=0 $ up to the N-th degree:
\begin{equation}
\phi_1(z) = \sum_{k=1}^{N+1} a_{k-1}^{(1)} z^{k-1} \ .
\end{equation}
We put the expansion into the equation \ref{eq:numericalnormalizedqnmequation} and obtain an equation
for each power of $ z $ up to the $ (N-1) $-th degree. We end up with the system of equations:
\begin{equation}
S\mathbf{a} = 0 \ ,
\end{equation}
where $ S $ is given by Equation~\ref{eq:numericallinearequations}. To get a single solution a
normalization equation needs to be added, and we choose:
\begin{equation}
a_0^{(1)} = 1 \ .
\end{equation}
Solving this system of equations for specific values of $ d $, $ K $, $ q $ and $ \lambda $, we obtain
the solution $ \phi _1 $ (which is regular at $ z=0 $).

\item We find the two independent solutions of the equation, $ \phi_2 $ and $ \phi_3 $ around $ z=1 $
using the Frobenius method:
First, we assume that:
\begin{equation}
\left.\phi_2\right|_{z=1} \approx (1-z)^{\beta'} \ .
\end{equation}
We define:
\begin{equation}
\label{eq:qnmnumericaltransform2}
\phi_2 = (1-z)^{\beta'}\widetilde{\phi}_2 \ ,
\end{equation}
and again re-write the equation in terms of $ \widetilde{\phi}_2 $. We develop $ \widetilde{\phi}_2 $
as a power series around $ z=1 $ up to the $ N $-th degree:
\begin{equation}
\widetilde{\phi}_2(z) = \sum_{k=1}^{N+1} a_{k-1}^{(2)} (z-1)^{k-1} \ .
\end{equation}
Putting the expansion into the equation, we again end up with a system of $ N $ linear equations for the
coefficients $ a_k^{(2)} $ (similar to Equations~\ref{eq:numericallinearequations}, but here the
parameters $ c_k^i $ are the coefficients of the polynomials in the QNM equation after the
transformation given in Equation~\ref{eq:qnmnumericaltransform2}, developed around $ z=1 $).
Adding the normalization equation:
\begin{equation}
a_0^{(2)} = 1 \ ,
\end{equation}
we solve the system of equations to get the solution $ \widetilde{\phi}_2 $ and then transform back to
$ \phi_2 $.
The other solution around $ z=1 $ takes the form:
\begin{equation}
\phi_3 = \widetilde{\phi}_3 + c\ln(1-z)\phi_2 \ ,
\end{equation}
where:
\begin{equation}
\left.\widetilde{\phi}_3\right|_{z=1} \approx 1 \ .
\end{equation}
We develop $ .\widetilde{\phi}_3 $ as a power series around $ z=1 $ up to the $ N $-th degree:
\begin{equation}
\widetilde{\phi}_3(z) = \sum_{k=1}^{N+1} a_{k-1}^{(3)} (z-1)^{k-1} \ ,
\end{equation}
and put $ \phi_3 $ into the equation. We end up with a system of $ N $ linear equations for the
coefficients $ a_k^{(3)} $ and the coefficient $ c $. Two equations must now be added:
A normalization equation, and another equation to fix the extra degree of freedom of adding
to $ \phi_3 $ a function that is proportional to $ \phi_2 $.
In the case where $ \beta' > 0 $, we choose the following two equations:
\begin{align}
a_0^{(3)} &= 1 \\
a_{\beta'}^{(3)} &= 0 \ .
\end{align}
If $ \beta' = 0 $, we choose:
\begin{align}
c &= 1 \\
a_0^{(3)} &= 0 \ .
\end{align}
Solving this system of $ N+2 $ equations we get the solution $ \phi_3 $.

\item We extract the connection coefficients $ A,B $ linearly relating the solution $ \phi_1 $ to the
solutions $ \phi_2 $, $ \phi_3 $:
\begin{equation}
\label{eq:numericalconnectionrelation}
\phi_1(z) = A\phi_3(z) + B\phi_2(z) \ .
\end{equation}
This can be accomplished by choosing a value $ 0 < z_0 < 1 $\footnote{$ z_0 $ should be chosen so that
all three series solutions are well converged in the vicinity of $ z_0 $.}, and setting $ z=z_0 $ both
in Equation~\ref{eq:numericalconnectionrelation} and its derivative with respect to $ z_0 $, thereby
obtaining 2 linear equations for $ A,B $:
\begin{align}
\phi_1(z_0) &= A\phi_3(z_0) + B\phi_2(z_0) \\
\phi_1'(z_0) &= A\phi_3'(z_0) + B\phi_2'(z_0) \ .
\end{align}
Solving these equations, we get the values of $ A,B $, and may then calculate the relation
$ \frac{B}{A} $ that enters  into the expressions for the retarded correlators (as explained in
Subsection~\ref{subsec:generalformulaeforcorrelators}).

\item For the sake of better visual representation of the results, we define and calculate the
``normalized'' retarded correlators $ I(\lambda) $ in the following way:
In all cases we may write the retarded correlator in the form:
\begin{equation}
G^R(\lambda) = N(\lambda) \left[I(\lambda) + M(\lambda)\right] \ ,
\end{equation}
where $ |I(\lambda)| = O(1) $ for $ |\lambda|\to\infty $, $ N(\lambda) $ is proportional to
$ \lambda^s $ ($ s $ being the appropriate exponent for perturbation type) and $ M(\lambda) $
is proportional to $ \ln(a\lambda) $.
For example, for the scalar perturbation mode (see
Subsubsection~\ref{subsubsec:cftasymptoticcorrelatorexpressionsscalar}):
If $ d $ is even-
\begin{equation}
N(\lambda) = -2(d-1)C_s\frac{r_+^{d-1}}{R^d}\frac{\Gamma(-\Delta)}{\Gamma(\Delta)}\left(\frac{i\lambda}
{2}\right)^{d-1}
\qquad
M(\lambda) = 0 \ .
\end{equation}
If $ d $ is odd-
\begin{equation}
N(\lambda) = 8\pi iC_s\frac{r_+^{d-1}}{R^d} \frac{(-1)^{\Delta+1}}{\Gamma^2(\Delta)}\left(\frac{i\lambda}
{2}\right)^{d-1}
\qquad
M(\lambda) = -\frac{i}{\pi}\ln\left(\frac{i\lambda}{2}\right) \ .
\end{equation}
Using the numerically calculated value of $ \frac{B}{A} $ and the values of $ N $ and $ M $ appropriate
for each perturbation type we may extract a numerical value for $ I(\lambda) $.

\end{enumerate}

\section*{Acknowledgements}

The work is supported in part by the Israeli Science Foundation center
of excellence.

\end{document}